\documentclass{aa} 
\usepackage{tabularx}
\usepackage{booktabs}
\usepackage{makecell}
\usepackage{natbib,twoopt}
\usepackage{graphicx}
\usepackage{wrapfig}
\usepackage{float}
\usepackage{txfonts}
\usepackage{multirow}
\usepackage{tablefootnote}
\usepackage{amsmath}
\usepackage{siunitx}
\usepackage{indentfirst}
\usepackage[compatibility=false]{caption}
\usepackage{subfig}
\usepackage{blindtext}
\usepackage[T1]{fontenc}
\usepackage[utf8]{inputenc}
\usepackage[dvipsnames]{xcolor}
\usepackage{xcolor}
\definecolor{darkgoldenrod}{rgb}{0.72, 0.53, 0.04}
\usepackage[colorlinks = true,
            linkcolor = RoyalBlue,
            urlcolor = darkgoldenrod,
            citecolor = RoyalBlue,
            anchorcolor = RoyalBlue,
            draft = false]{hyperref}

\newcommand{\comment}[1]{}

\makeatletter
\let\linenumbers\relax
\let\nolinenumbers\relax
\let\modulolinenumbers\relax

\let\runninglinenumbers\relax
\@ifundefined{makeLineNumber}{}{
  \renewcommand\makeLineNumber[2]{}
}
\makeatother

\begin{document} 
   \title{NGC 3521 as the Milky Way near-twin:\\UV-to-radio-decameter spectral energy distribution}

  \author{
Kompaniiets, O.V.$^{1}$\thanks{E-mail: kompaniets@mao.kiev.ua}
Vavilova, I.B.$^{1}$\thanks{E-mail: irivav@mao.kiev.ua}
Vasylkivskyi, Y.V. $^{2}$\thanks{E-mail: vasylkivskyi@rian.kharkov.ua}
Konovalenko, O.O.$^{2}$
Pastoven, O.S.$^{1}$ 
Izviekova, I.O.$^{1,3}$
Junais,$^{5}$
Dmytrenko, A.M. $^{1}$
Dobrycheva, D.V.$^{1}$ 
Fedorov, P.N.$^{4}$
Khramtsov, V.P.$^{1,4}$
Sergijenko, O.$^{1}$
Vasylenko, A.A.$^{1}$
}

\institute{
$^{1}$Main Astronomical Observatory of the NAS of Ukraine, 27, Akademik Zabolotnyi St., Kyiv 03143, Ukraine\\
$^{2}$ Institute of Radio Astronomy of the National Academy of Sciences of Ukraine, 4, Mystetstv St., Kharkiv, 61002, Ukraine\\
$^{3}$ International Center for Astronomical, Medical, and Ecological Research, National Academy of Sciences of Ukraine, 27, Akademik Zabolotnyi St., Kyiv 03123, Ukraine\\
$^{4}$Institute of Astronomy, V.N. Karazin Kharkiv National University, 35, Sumska St., Kharkiv, 61022, Ukraine \\
$^{5}$Instituto de Astrofísica de Canarias (IAC), Tenerife, Spain
}
\titlerunning{NGC 3521: SED from UV to Radio decameter ramges}
\authorrunning{Kompaniiets O.V.,  Vavilova I.B., Vasylkivskyi Y.V. et al.}

   \date{Received December 15, 2025; accepted April 01,2026}
 
\abstract
{
Context. {Milky Way analogs (MWAs) are defined on the basis of structural, kinematic, and global physical parameters. Our advanced approach to MWAs search also includes such principal features of the Milky Way (MW) as its isolated position in the local cosmic web, the absence or weak nuclear activity, and the small mass of the supermassive black hole. We forward the suggestion that spectral energy distribution (SED) shape and such properties as star-formation rate, stellar and dust masses should be similar within the expected parametric scatter for MW and MWAs, which are at the same co-evolutionary cosmological scale. However, constructing a reliable SED reference of MWAs remains challenging due to heterogeneous photometry and the lack of constraints across the overall spectral range. NGC~3521 is one of the closest known MWAs and an excellent candidate for extending SED-based similarity criteria.}

Aims.{We aim to construct a homogeneous, aperture photometry-based SED of NGC~3521 from the UV to the radio decameter range and to assess whether the integrated SED can serve as an additional indicator for the search for and validation of MWAs.}

Methods. {We report the SED model for NGC 3521 based on the measurements across the UV to radio wavelength ranges, exploiting, for the first time, both the data at the decameter range and aperture photometry for GALEX, SDSS, WISE, Spitzer/MIPS, Herschel/PACS, SPIRE, and VLA images. To constrain the decameter emission and derive an upper limit in the 24--32 MHz band, we present observational data obtained in Jan-Feb 2022 using the Ukrainian T-shaped radio telescope (UTR-2). The observed SED was modeled with \textsc{CIGALE}, in which we developed a radio prescription (radio\_extra) module that accounts for emission and absorption effects in the radio meter and decameter domains. The nuclear activity of NGC 3521 was analyzed with ZTF and NEOWISE archival data for the 2014-2025 period.}

Results. {The SED measurements contain 27 photometric points. The preferred SED model from UV-to-decameter ranges yields $M_{\star}\simeq6.0\times10^{10} M_{\odot}$, ${\rm SFR}\simeq1.65 M_{\odot} {\rm yr}^{-1}$, $M_{\rm dust}\simeq1.3\times10^{8} M_{\odot}$, and an effective dust temperature of $\sim23$ K. We found genuine variability of the NGC 3521 central region. The optical trends, measured with ZTF PSF-fit photometry on a seeing-limited scale of $\lesssim3^{\prime\prime}$ (FWHM), primarily trace the compact nuclear region and are consistent with a variable nuclear continuum superimposed on a relatively stable stellar component. The same behavior is observed in the MIR, but due to the larger effective scale of the NEOWISE PSF-fit photometry ($\sim7.5^{\prime\prime}$) it reflects a combination of nuclear variations and the contribution of warm dust in the central region.}

Conclusions. {Considering NGC 3521 as the MW near-twin galaxy, we present, for the first time, its SED model from UV to radio decameter ranges. We show that the integrated SED, especially when extended below 100 MHz, provides a complementary, physically motivated diagnostic. We demonstrate that the SEDs model of our Galaxy and NGC~3521 are similar, with the most noticeable residual offsets occurring in the FUV and near the far-IR dust peak. The decameter constraint provides an upper limit $\log L_\nu(28 \mathrm{MHz}) \approx 30.19$ of the spectral luminosity at 28 MHz, which is consistent with the extrapolated luminosity of a Milky Way of $\log L_\nu(28 \mathrm{MHz} \approx 30.17$ placed at 10.7 Mpc. An exceptional consistency of SED shapes of our Galaxy and NGC~3521 allows us to conclude that SED shape may serve as an additional indicator for the search of MWAs. In turn, it helps to extrapolate how MW properties appear to an extragalactic observer. }
}

   \maketitle

\section{Introduction}
\label{sec:Intro}
\noindent 
NGC 3521 is a nearby spiral flocculent galaxy in the constellation Leo (RAJ = 11h05m48.8s, DecJ = $-00^\circ02'13''$, a visible diameter of $\approx$26 kpc, an angular size of $\approx$$10'$, and $z = 0.002672$). The morphological type SAB(rs) of this galaxy, according to the de Vaucouleurs classification, implies the existence of an intermediate bar, a faint inner ring, and spiral arms that are moderate to weakly loose. Its nuclear emission is consistent with a composite H~II/LINER type \citep{Das2003, Pastoven2024}. Due to its high inclination angle ($i = 72.7^{\circ}$), subtracting the diffuse, non-star-forming background emission is more complicated, which can affect the inferred relations, especially at spatial scales larger than 8 arcseconds \citep{Liu2011}.

One of the first dynamical analyses of NGC 3521 by \cite{Burbidge1964} reveals rotation curve parameters and a mass from the H$\alpha$ and [N~II] emission lines of $\approx$ $8\times10^{10} M_{\odot}$ within $\sim170''$ from the center. Later, \cite{Casertano1991} verified that the rotation curve declines toward the galaxy's outskirts. Photometric observations by \cite{Dettmar1993} indicated that this behavior at low surface brightness levels may suggest a past interaction or minor merger. Further spectroscopic studies confirmed kinematic asymmetries and the presence of counter-rotating stellar and gaseous components along both major and minor axes \citep{Zeilinger2001}, which can be interpreted within the framework of an axisymmetric dynamo scenario, where density waves generate spiral magnetic fields with a radial component contributing up to $40$–$60\%$ of the azimuthal field \citep{Knapik2000}. 
Over the last few decades, NGC 3521 has been included in many multiwavelength sky surveys and has also been observed individually to obtain more precise parameters. This allows for the construction of its spectral energy distribution in a wide spectral range (see Section \ref{sec:Multiwavelength_Data}). 

Since the seminal work by \cite{deVaucouleurs1978}, the basic structural and photometric parameters for the search of the Milky Way galaxies-analogs (MWAs) were adopted (morphology, size, bar and bulge sizes, isophotal and effective radii, luminosity in the B-filter, optical color, inner ring, etc.). Later on, such parameters of our Galaxy as stellar mass, star formation rate, bulge-to-disk ratio, bulge-to-total ratio, disk scale length, rotation velocity were considered in various combinations altogether with structural and photometric parameters (e.g. \cite{Mutch2011, Licquia2015, Fraser2019, Boardman2020a, Boardman2020b, Tuntipong2024}) to compile a list of MWAs  \cite{Pilyugin2019, Pilyugin2023} found that the low metallicity at the periphery of the Milky Way can also serve as an indicator for MWAs' search. In fact, these authors exploited a minimal set of parameters when selecting MWAs by key indicators such as stellar mass, rotation curve. metallicity, etc.

We consider as many as possible parameters-indicators of galaxies belonging to the MWA class. The more parameters of the selected MWA galaxy coincide within a certain spread of the same parameters of our Galaxy, the more confidently we can speak of a multi parametric offset and consider such a MWA galaxy to be a Milky Way near-twin. In this context, we proposed \citep{Vavilova2024} that an advanced approach to MWAs search should also include such principal features of the Milky Way as its isolated position in the local cosmic web, the absence or weak nuclear activity, and the small mass of the supermassive black hole (SMBH). We adopted the isolation criteria proposed for catalogs of isolated galaxies by \cite{Karachentseva1973, Karachentseva2010} with some modifications for the 3D cosmic web \citep{Dobrycheva2025,Kompaniiets2025}. We verified the type of spectral activity of NGC 3521 as a LINER \citep{Kompaniiets2025} using the Baldwin–Phillips–Terlevich (BPT) diagnostic diagram \citep{Baldwin1981, Kewley2001, Kauffmann2003, Schawinski2007}. The SMBH mass is $2.6\times10^{6}M_{\odot}$ \citep{Zhang2009} estimated by Sersic index relation \citep{Graham2007}.

The galaxy NGC 3521 (=~KIG~461) is classified as an isolated system according to the criteria of the Catalog of Isolated Galaxies \citep{Karachentseva1973}. NGC~3521 belongs to the diffuse elongated Leo~Spur structure of the Local Volume, together with the systems surrounding NGC 3115 and NGC 2784 \citep{Kourkchi2017}. Analyzing the Subaru Hyper Suprime Camera data, \cite{Muller2025} discovered new candidates to the dwarf satellites around NGC 3521, pointing out their similar luminosity function to that of the Milky Way. \cite{Karachentsev2022} identified four new low-surface-brightness satellites around NGC 3521 using DECaLS imaging, covering a projected area of $750\times750$ kpc. Their estimates of the total mass, $M_{\mathrm{T}} =(0.90\pm0.42)\times10^{12} M_{\odot}$, and the mass-to-light ratio, $M_{\mathrm{T}}/L_{K}=(7\pm3) M_{\odot}/L_{\odot}$, together with the declining rotation curve \citep{Dettmar1993}, indicate a relatively shallow potential well and a low-mass dark halo. 

The Milky Way can be considered an isolated galaxy for at least 7-10 Gyr during its evolution. The last major merger of our Galaxy with Gaia-Sausage-Enceladus (GSE) was at $z \approx$ 2 (see the results of the high-resolution N-body simulations by \cite{Naidu2021}). As for the possible collision of the MW---M31 system, \cite{Sawala2024} has demonstrated that the neighboring small galaxies M33 and the Large Magellanic Cloud (LMC) can make this merger less likely because the LMC orbit runs perpendicular to the orbit of MW--M31 system. No less critical here is the fact of more precise distance moduli \citep{DiValentino, H0DN} and dynamic (masses) data for Local Group galaxies and the MW-M31 orbital geometry \citep{Elyiv2020, Banik2022}, the position of Local Void lying adjacent to the Local Group and the MW moving away from this void \citep{Tully1987, Lindner1995, Mazurenko2024}. The TNG50 simulations of the Magellanic Clouds analogs show that such systems may form through distinct processes \citep{Haslbauer2024}. Recent Gaia DR3-based studies provide new constraints on the Milky Way kinematics: the spatial orientation of the local velocity ellipsoids indicates departures from a simple axisymmetric disc \citep{Dmytrenko2025a}, while an independent analysis of Gaia DR3 RGB stars delivers an updated Galactic rotation curve and suggests azimuth-dependent variations at large radii \citep{Fedorov2025}.

The physical parameters of NGC~3521 and NGC~3115 are often compared with those of the MW--M31 system. We studied the location of NGC 3521 within the local 3D Cosmic Web \citep{Kompaniiets2025} and concluded that the environmental densities of NGC 3521 and MW are comparable. The projected surface density derived with the Voronoi tessellation method is $\approx$$0.7~\mathrm{gal~Mpc^{-2}}$, while the fifth-nearest-neighbor estimator yields $\approx 0.4~\mathrm{gal~Mpc^{-2}}$. In comparison, the Milky Way has an environmental density of $\Sigma_{5} \approx 0.13~\mathrm{gal~Mpc^{-3}}$, when estimated with the fifth-nearest-neighbor method after excluding dwarf companions such as the Magellanic Clouds. 

The comparable baryonic mass, disk scale length, and rotation curve allowed \cite{Mast2006, McGaugh2016} to suggest the similarity of NGC 3521 with the Milky Way. \cite{Pilyugin2023} reinforced this similarity, considering oxygen abundance gradients and kinematic profiles. Earlier, the asymmetries in the outer stellar and gaseous distributions along the major axis were revealed by \cite{VilaVilaro2015} based on the data of CO and H~I observations. The IFU observations with VIRUS-W by \cite{Fabricius2015} and subsequent modeling by \cite{Coccato2018} explained the counter-rotating stellar components, e.g., elevated velocity dispersion in the bulge and slower central rotation relative to the disc. They identified three main stellar populations: an old ($\gtrsim7$~Gyr), an intermediate-age ($\sim3$~Gyr ago, formed during a secondary star formation burst or merger event), and a young ($\lesssim1$~Gyr, associated with recent disc activity). 

In the context of our advanced approach \citep{Vavilova2024}, we forward the suggestion that spectral energy distribution shape and global physical properties, such as star-formation rate, $M_\star$, $M_{\rm dust}$ should be similar within the expected parametric scatter for MW and MWA, which are at the same co-evolutionary cosmological scale. For verifying this, we first selected NGC 3521 as the nearby isolated galaxy with SB morphology, having similar structural and physical parameters, HII/LINER nuclear activity, rotation curve, and metallicity. Importantly, this galaxy has a satisfactory series of multiwavelength observations from X-ray to radio meter ranges. All this makes NGC 3521 an ideal candidate for SED modeling and verifying our suggestion. So, secondly, we decided to supplement this series with our own observations in the radio decameter range, and, thirdly, to exploit homogeneous aperture photometry for broad-band SED measurements. We applied the CIGALE software to construct an SED model. This allowed us to derive global properties of its star-dust evolutionary history for comparison with the SED shape and global properties of the Milky Way. Additionally, by analyzing the SEDs of MWA twins, we obtain an additional astroinformation channel for the SED of our Galaxy. 

\cite{Zibetti2011} analyzed the SED of NGC 3521 in pixels of seven nearby galaxies, including NGC 3521, in the optical and infrared (IR) ranges. NGC 3521 proved to be the most anomalous: its SED shows high variability on small scales, suggesting a complex relationship among star formation, stellar age, and dust distribution parameters. The SED of NGC 3521 galaxy with various aims was modeled by \cite{Dale2012} in IR bands, \cite{Brown2014} from UV to mid-IR, \cite{Hunt2019} from FUV to sub-mm, \cite{Chang2020} from IR to sub-mm, \cite{Pastoven2024} from UV to radio cm ranges as well as included by \cite{DESI2023} in the DESI Siena Galaxy Atlas. \cite{Pastoven2024} performed a preliminary analysis of flux radiation from the central region of NGC~3521 and constructed a baseline model describing its emission from the ultraviolet (UV) to radio cm ranges. 

Our research aims to construct an SED model for NGC 3521 across the UV to radio wavelength ranges, exploiting, for the first time, both decameter-range data and aperture photometry across all wavelength ranges. Throughout this article, we distinguish between (i) the observed UV--to--radio SED, constructed from homogeneous photometric measurements, and (ii) the physical parameters inferred from it via SED modeling. While the measured SED is model-independent, the derived quantities (e.g. $M_\star$, SFR, $M_{\rm dust}$) depend on assumptions about the star-formation history, dust attenuation/emission, and energy balance (as implemented in \textsc{CIGALE}). We therefore use SED fitting primarily as an interpretative layer, to place the MW--NGC~3521 SED comparison into a physical context.

The broader goal of this work is to connect the integrated UV--to--radio SED of Milky Way analogues with their evolutionary state.  By constructing a homogeneous UV--radio decameter SED for NGC~3521, we provide a first benchmark for testing SED-level similarity and illustrate, in this case study, which parts of the UV--to--radio decameter SED exhibit the most prominent offsets between the MW and a MW near-twin.

\begin{table*}
\centering
\small
\caption{Multiwavelength photometry data for NGC 3521 from UV to radio decameter ranges}
\label{tab:NGC3521_fluxes}
\setlength{\tabcolsep}{4pt}
\begin{tabular}{lrrrrrrrrrrrr}
\hline
 Survey & \multicolumn{12}{c}{Fluxes [mJy]} \\
\hline
     GALEX & FUV   & $\sigma_{\rm FUV}$ & NUV      & $\sigma_{\rm NUV}$  & & & & & & & &\\
     (this work) & 13.24 & 0.05               & 23.62    & 0.02                & & & & & & & & \\[3pt]\hline

     SDSS& $u$    & $\sigma_u$         & g     & $\sigma_g$         & r       & $\sigma_r$          & $i$     & $\sigma_i$ & $z$     & $\sigma_z$ & & \\
     (this work)      & 153.80 & 0.07               & 546.07  & 0.03               & 1039.61   & 0.05                & 1489.46 & 0.08       & 2023.05 & 0.28       & & \\[3pt]\hline

      WISE& W1      & $\sigma_{\rm W1}$  & W2      & $\sigma_{\rm W2}$  & W3       & $\sigma_{\rm W3}$   & W4     & $\sigma_{\rm W4}$ & & & \\
      (this work)& 1388.11 & 0.04               & 1012.55 & 0.05               & 5666.23  & 0.28                & 6156.08 & 0.55       &     & & & \\[3pt]\hline

       MIPS& MIPS1   & $\sigma_{\rm M1}$ & MIPS2    & $\sigma_{\rm M2}$   & MIPS3     & $\sigma_{\rm M3}$ & & & & & & \\
       (this work)& 5264.77 & 0.59              & 51909.76 & 22.55               & 177009.98 & 187.26            & & & & & & \\[3pt]\hline

Herschel & PACS$_{70}$  & $\sigma_{70}$   & PACS$_{100}$& $\sigma_{100}$  & PACS$_{160}$        & $\sigma_{160}$ &  PSW      & $\sigma_{\rm PSW}$ & PMW      & $\sigma_{\rm PMW}$ & PLW     & $\sigma_{\rm PLW}$  \\
(this work)& 79564.06     & 46.02           & 150037.12   & 47.83           & 205026.54           & 50.54          & 100503.96 & 51.98              & 41635.80 & 59.17              & 14858.65 & 54.83             \\[3pt]\hline

      VLA& VLA$_{\rm L}$ & $\sigma_{\rm L}$ & & & & & & & & & & \\
  (this work)& 374.77        & 3.78             & & & & & & & & & & \\[3pt]\hline

NRAO\tablefootmark{a}  & 750 MHz & upper & & & & & & & & & & \\
                       & 1100    & limit & & & & & & & & & & \\[3pt]\hline

Molonglo Radio & 408 MHz & $\sigma_{\rm 408 MHz}$ & & & & & & & & & & \\
 Telescope \tablefootmark{b}                                        & 1460    & 0.5                     & & & & & & & & & & \\[3pt]\hline

Culgoora circular  & 160 MHz & upper & 80 MHz & upper & & & & & & & & \\
  array\tablefootmark{c} & 3900    & limit & 7000   & limit & & & & & & & & \\[3pt]\hline

Clark Lake\tablefootmark{d} & 57.5 MHz & $\sigma_{\rm 57.5 MHz}$ & & & & & & & & & & \\
TPT array    & 13000    & 2000                      & & & & & & & & & & \\[3pt]\hline

UTR-2\tablefootmark{e} & 28 MHz & $\sigma_{\rm 28 MHz}$ & upper & & & & & & & & & \\
(this work)  & 11220  & 0.42                    & limit & & & & & & & & & \\
\hline
\end{tabular}

\tablefoot{\tablefoottext{a}{Fluxes: NRAO 750 MHz measurement by \cite{Heeschen1964};}
\tablefoottext{b}{Molonglo Radio Telescope 408 MHz measurement by \cite{Large1981};}
\tablefoottext{c}{Culgoora circular array 60 and 80 MHz (upper limits) by \cite{Slee1995};}
\tablefoottext{d}{Clark Lake Radio Telescope TPT array 57.5 MHz measurement by \cite{Israel1990};}
\tablefoottext{e}{Ukrainian T-shape Radio Telescope (UTR-2) by own observations in Jan-Feb, 2022 (see Section 4) 28 MHz measurement.}
}
\end{table*}

We describe the multiwavelength data sources for SED models in the UV to radio decameter ranges as well as the X-ray data in Section \ref{sec:Multiwavelength_Data}. The aperture photometry in the decameter range, including the specificity of observations and sensitivity measurements, is justified in Section \ref{sec:APT_UV_VLA} and Appendix \ref{APT}. Aperture photometry, including measurement technique, flux conversion, and extinction correction, performed for CIGALE across UV to radio ranges, is presented in Section \ref{sec:sed_UV-Decameter} and Appendix \ref{sec:sed_UV-VLA}.  The results on SED model covering UV to radio decameter ranges are given in Section \ref{sec:sed_results}. Nuclear emission behavior and luminosity curves based on the ZTF data are analyzed in Section \ref{sec:nucleus}. The photometric extraction, angular resolution, and PSF-related considerations for ZTF and NEOWISE data are described in Appendix ~\ref{sec:ZTF_optical} and \ref{sec:NEOWISE_data}, respectively. The multiwavelength properties of NGC 3521, as follows from the SED modeling, are discussed in Section \ref{sec:discussion}, and conclusions are in Section \ref{sec:conclusions}.

\section{Multiwavelength data sources for SED models}
\label{sec:Multiwavelength_Data}

Our study of isolated galaxies at $z<0.1$ confirms that they typically exhibit lower continuum luminosities in multiple bands compared to galaxies of similar stellar mass in denser environments --- most notably in the radio continuum \citep{Pulatova2023, Vasylenko2020, Vavilova2021, Vavilova2022, Kompaniiets2024} --- and, on average, show weaker nuclear activity than galaxies in denser environments \citep{Volvach2011, Melnyk2015, Pulatova2015, Dobrycheva2018}. However, our new results indicate that the X-ray luminosities of isolated galaxies hosting AGN are not suppressed and are instead typical of Seyfert galaxies \citep{Kompaniiets2023IAU, Kompaniiets2025arxiv}. These trends complicate the assembly of homogeneous, truly panchromatic datasets for isolated systems (including MWAs) over the full spectral range. In this respect, NGC 3521 is an exceptionally well-covered case: extensive archival observations exist from the X-ray down to the meter-wave radio domain, while the decameter measurements are provided by our observations (see Section \ref{sec:utr2}).

NGC 3521 has the LINER type of nuclear activity \citep{Das2003, Pastoven2024} and the X-ray Eddington ratio as $L_{0.3-8}/L_{Edd}$ = $2.2\times10^{-7}$ \citep{Zhang2009}. For this reason, we did not use X-ray data to measure SED models, analyzing nuclear activity separately (see Section \ref{sec:nucleus}). 

\subsection{UV to IR range}

NGC~3521 has been included in several major sky surveys, as the THINGS \citep{Walter2008}, the James~Clerk~Maxwell Telescope Nearby Galaxies Legacy Survey \citep{Warren2010}, SDSS \citep{SDSS}, the Spitzer Infrared Nearby Galaxies Survey \citep{Spitzer}, and others. To obtain integrated fluxes from an aperture encompassing the entire galaxy, and to enable subsequent multiwavelength SED model analysis, we used calibrated FITS images retrieved from available public archives and surveys \citep{Vavilova2020}:

- Ultraviolet (NUV and FUV) data from the GALEX All Sky Imaging Surveys (AIS; GR2/GR3, \citep{Galex}); 

- Optical data from the Sloan Digital Sky Survey (SDSS DR17,\citep{SDSS}).

- Infrared data from the AllWISE Image Atlas \citep{Wise}, the Spitzer Infrared Nearby Galaxies Survey (SINGS) \citep{Spitzer}, and the Herschel Far-infrared and Submillimeter Photometry for the KINGFISH Sample of nearby Galaxies \citep{Hersel}.

\begin{figure*}
\centering
\begin{minipage}[t]{0.24\textwidth}\centering
  \includegraphics[width=\linewidth]{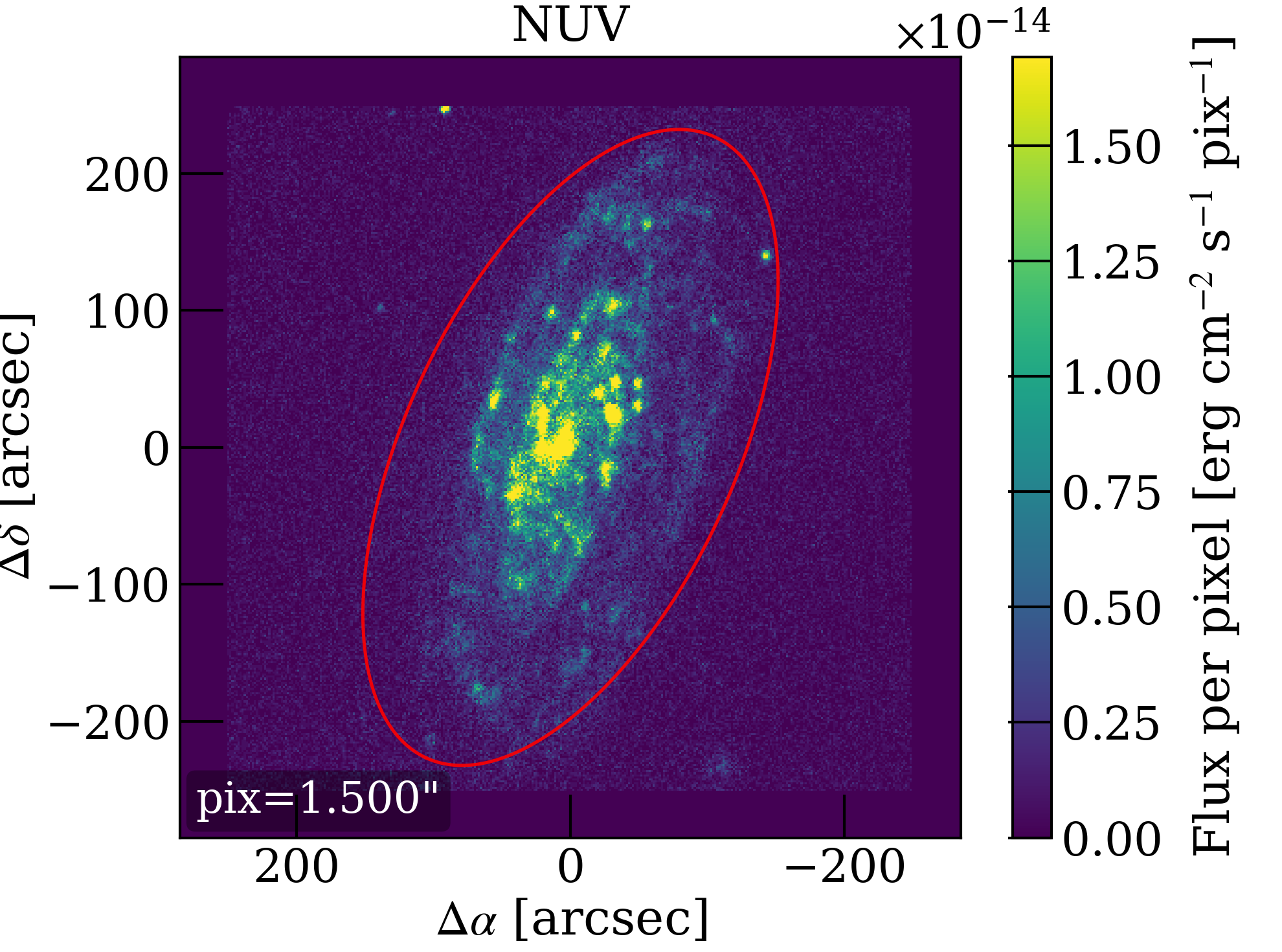}\\[0pt]
  {\small (a) GALEX NUV}
\end{minipage}\hfill%
\begin{minipage}[t]{0.24\textwidth}\centering
  \includegraphics[width=\linewidth]{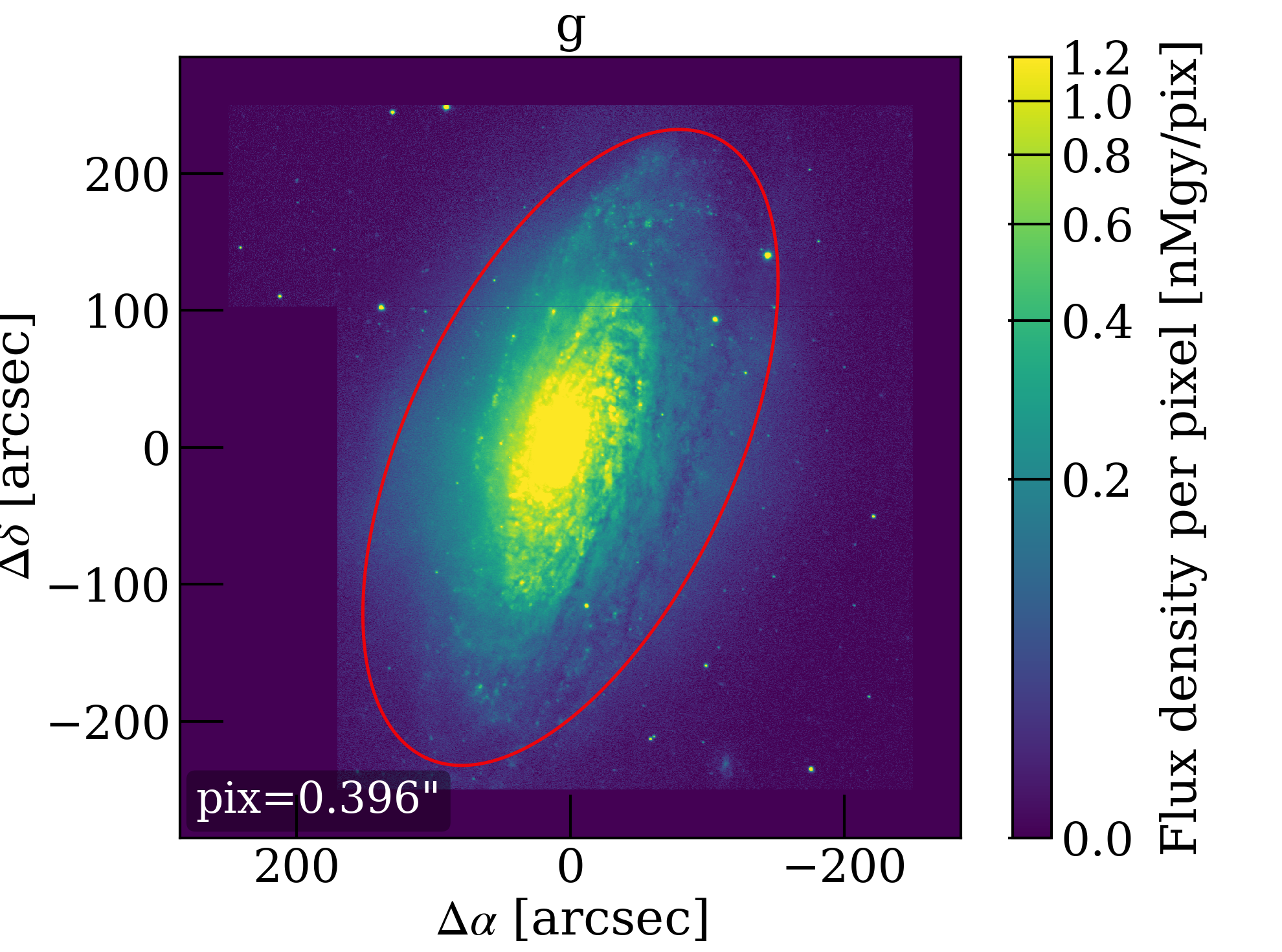}\\[0pt]
  {\small (b) SDSS g--band}
\end{minipage}
\begin{minipage}[t]{0.24\textwidth}\centering
  \includegraphics[width=\linewidth]{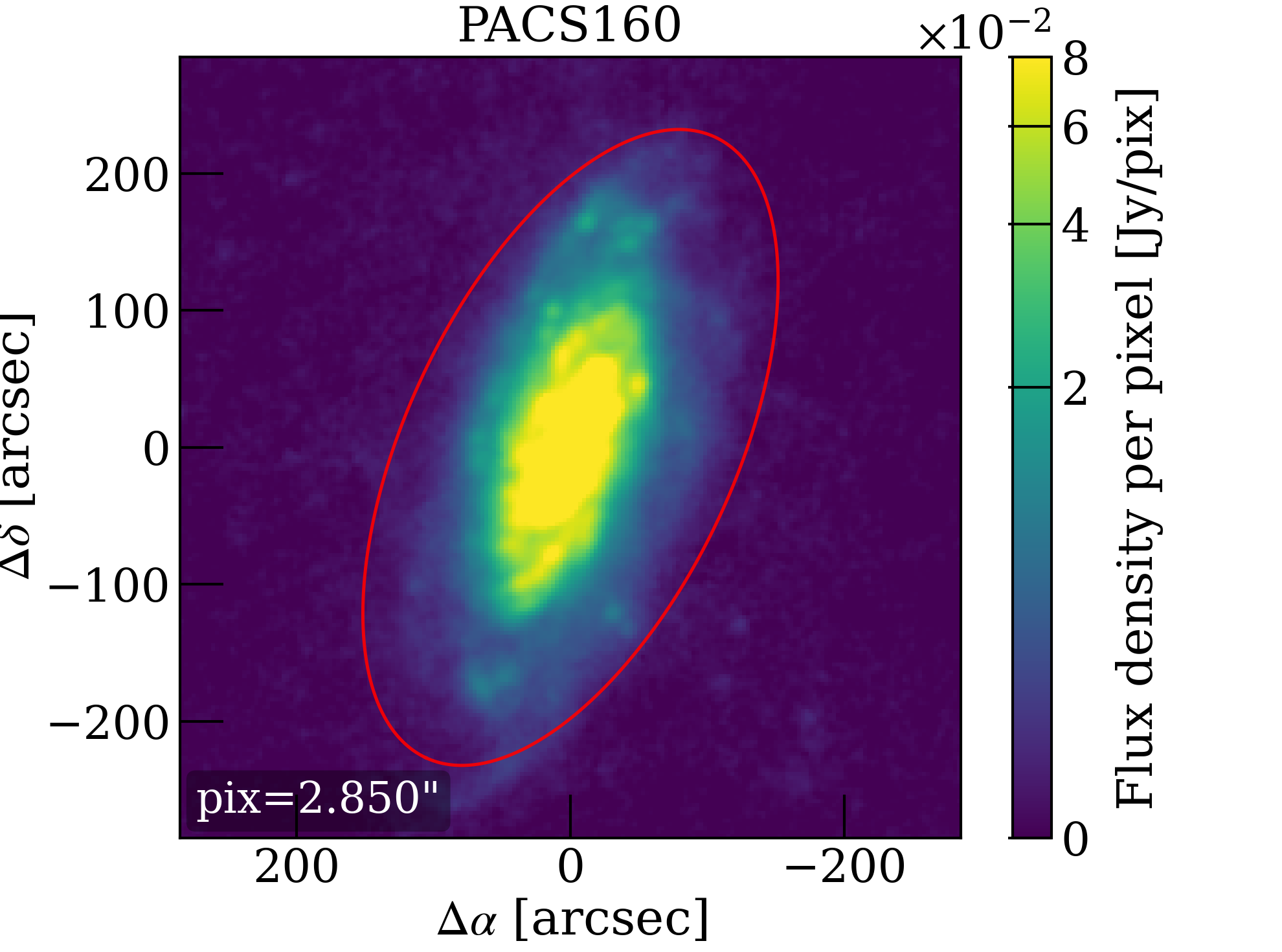}\\[0pt]
  {\small (c) PACS 160 $\mu$m}
\end{minipage}\hfill%
\begin{minipage}[t]{0.24\textwidth}\centering
  \includegraphics[width=\linewidth]{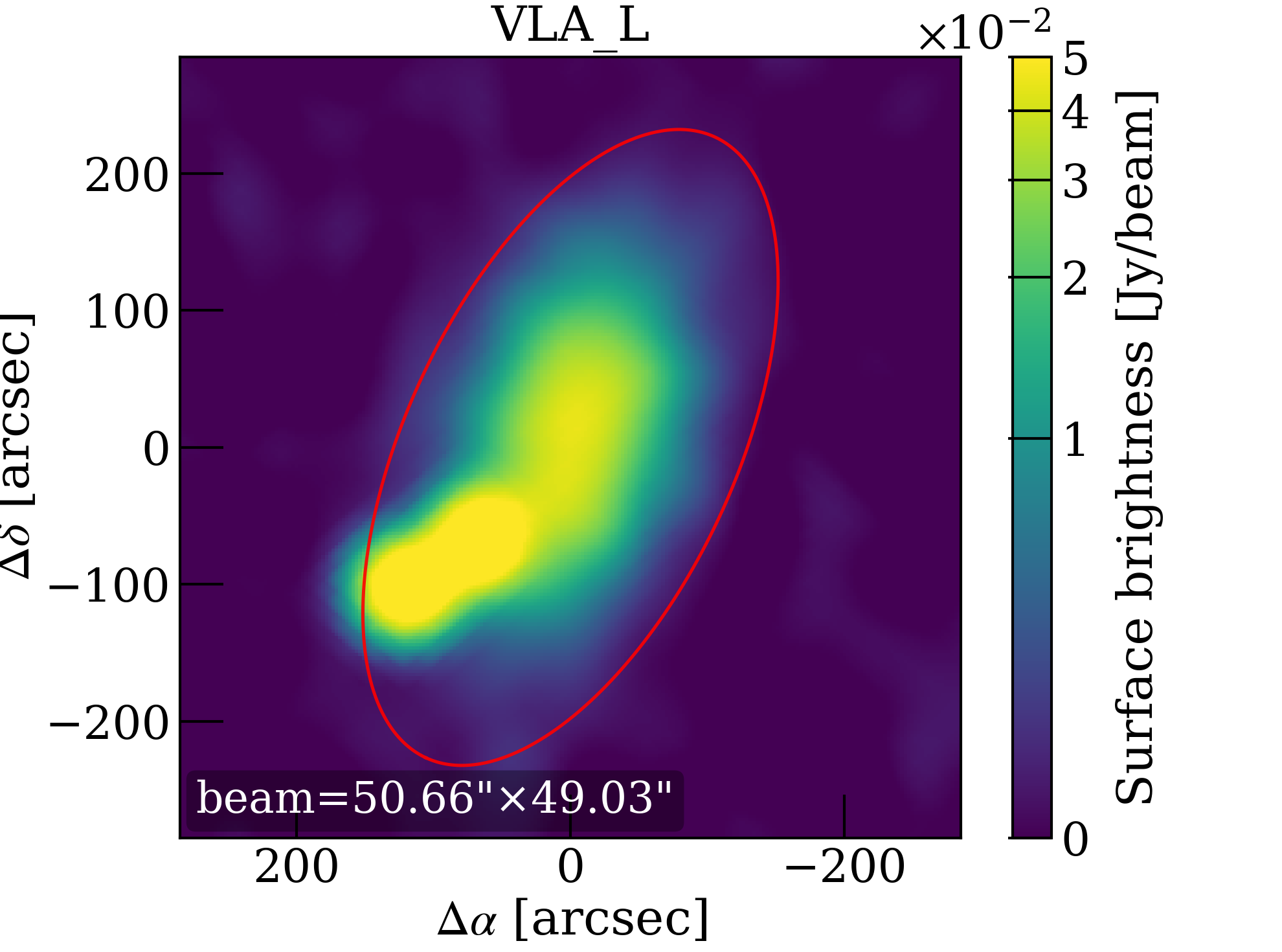}\\[-2pt]
  {\small (d) VLA L--band}
\end{minipage}\hfill%
 \caption{Examples of multiwavelength images of NGC 3521 used for aperture photometry and SED construction:
(a) GALEX NUV, (b) SDSS g--band, (c) Herschel/PACS 160 $\mu$m, and (d) VLA L--band radio continuum.
All panels show the same field of view ($570''\times570''$) centered on the galaxy; the axes are angular offsets from the center in arcsec.
The color scales are shown in commonly used map units, as labeled on each color-bar:
for GALEX NUV, we convert the count-rate map to a band-integrated flux per pixel in $\mathrm{erg cm^{-2} s^{-1} pix^{-1}}$ using the standard GALEX photometric calibration and the effective band pass width; for SDSS g--band, we display flux density per pixel (nanomaggies$/$pix);
for PACS 160 $\mu$m, the map is shown in Jy$/$pix, and for the VLA L--band in Jy$/$beam.
The pixel scale (and, for the VLA map, the beam size BMAJ$\times$BMIN) is indicated in each image.
To preserve faint emission while avoiding saturation in bright regions, all images are displayed with an inverse-hyperbolic-sine (asinh) stretch.
The red ellipse marks the adopted isophotal aperture used for integrated flux measurements.
}
\label{fig:ngc3521_multiwavelength}
\end{figure*}

\subsection{Radio range}
\label{sec:utr2}
 
The estimates of flux density in the radio cm range were performed based on the data from the NRAO VLA Sky Survey (NVSS) at frequency $f$=1.4 GHz \citep{NVSS}. Fluxes at lower frequencies were taken from several works: NVSS at $f$=750 MHz \citep{Heeschen1964}, Molonglo Radio Telescope at $f$=408 MHz \citep{Large1981}, Culgoora circular array at $f$=60 and $f$=80 MHz (upper limits) \citep{Slee1995}, and Clark Lake Radio Telescope TPT array at $f$=57.5 MHz \citep{Israel1990}.

To determine the multiwavelength spectral characteristics of NGC 3521 over as wide a range as possible, this galaxy was observed with the Ukrainian T-shaped Radio Telescope (UTR-2) in the decameter range at $f$=28 and $f$=20 MHz in Jan-Feb 2022. UTR-2 is the world's largest decameter radio telescope (Kharkiv region, Ukraine) with a maximum effective area of $\approx$140,000 m$^2$, which is capable of operating in the frequency band from 8 to 32 MHz \citep{Konovalenko2016}. It was in active operation until Feb 2022 \citep{Albergaria2025}. 

\section{Aperture photometry for NGC 3521 from UV to radio cm range}
\label{sec:APT_UV_VLA}

\subsection{Photometric system and measurements}
\label{sec:phot_system}

Aperture photometry, which ensures consistent flux measurements across a wide range of wavelengths, particularly from the UV to the IR, poses challenges for nearby galaxies. We optimized our approach to encompass the full extent of NGC 3521 and derive flux densities across the UV-radio range within a standard aperture.
The zero-points were specified manually for each band: 22.5 mag for SDSS (u,g,r,i,z), 18.82 and 20.08 mag for GALEX (FUV, NUV), and 20.752, 19.569, 17.800, and 12.945 mag for WISE (W1–W4), respectively. The WISE magnitudes were additionally corrected from Vega to AB using the additive offsets of 2.699, 3.339, 5.174, and 6.620. For the FIR filters (Spitzer/MIPS and Herschel/PACS, SPIRE) and VLA data, the maps are provided in radiometric units (MJy sr$^{-1}$, Jy pix$^{-1}$, or Jy beam$^{-1}$) 
and were converted to mJy through integration over the aperture.

We then performed homogeneous multiwavelength aperture photometry for SDSS (u,g,r,i,z), GALEX (FUV, NUV), WISE (W1–W4), 
Spitzer/MIPS (24, 70, 160 $\mu$m), 
Herschel/PACS (70, 100, 160, $\mu$m), 
Herschel/SPIRE (250, 350, 500 $\mu$m), 
and VLA L-band data. A common elliptical aperture with semi-axes $a = 250\arcsec$ and $b = 120\arcsec$ was adopted to enclose the entire galaxy (see examples in Fig. \ref{fig:ngc3521_multiwavelength}). All photometric measurements were converted to flux densities on a common mJy scale.

\subsection{Photometric pipeline}

The photometric pipeline was built on top of publicly available Python astronomy libraries. In particular, we used stroPy for FITS I/O, WCS handling, unit conversions, and pixel–to–sky transformations \citep{Astropy2013, Astropy2018}, and Photutils for elliptical aperture photometry and mask generation \citep{Bradley2016Photutils}. Source detection and segmentation in the background-masked images were performed with SEP (the Python implementation of SExtractor; \citealt{Barbary2016SEP}), while foreground/source morphology refinements relied on functions from scipy.ndimage and scikit-image. Galactic extinction values were retrieved through astroquery.irsa (IRSA Dust service; \citealt{Astroquery2019}), and for multi-instrument image alignment we exploited reproject. 

Our own contribution was to integrate these components into a single end-to-end pipeline, to define uniform zero points and Vega$\rightarrow$AB corrections for all filters, and to implement band-dependent aperture masks and FIR/radio unit conversions to mJy (for more details see Appendix \ref{APT}), where flux density values conversion, background Subtraction, extinction correction, and uncertainty estimation are described. Examples of multiwavelength images of NGC 3521 used for aperture photometry and SED models are presented in Fig. \ref{fig:ngc3521_multiwavelength}\footnote{The mosaic of the SDSS image of NGC 3521 for further photometry analysis was performed by \url{https://reproject.readthedocs.io/en/stable/mosaicking.html}.} and resulting flux density values are in Table \ref{tab:NGC3521_fluxes}. 

\section{Aperture photometry of NGC 3521 in radio decameter range}
\label{section:APT_decameter}

A detailed description of the technical specifications and sensitivity measurement. and peculiarities of observations with UTR-2 telescope is provided in Appendix~\ref{sec:UTR2_tech_spec}.

\subsection{Data processing}

\begin{figure}
\centering
\includegraphics[width=\linewidth]{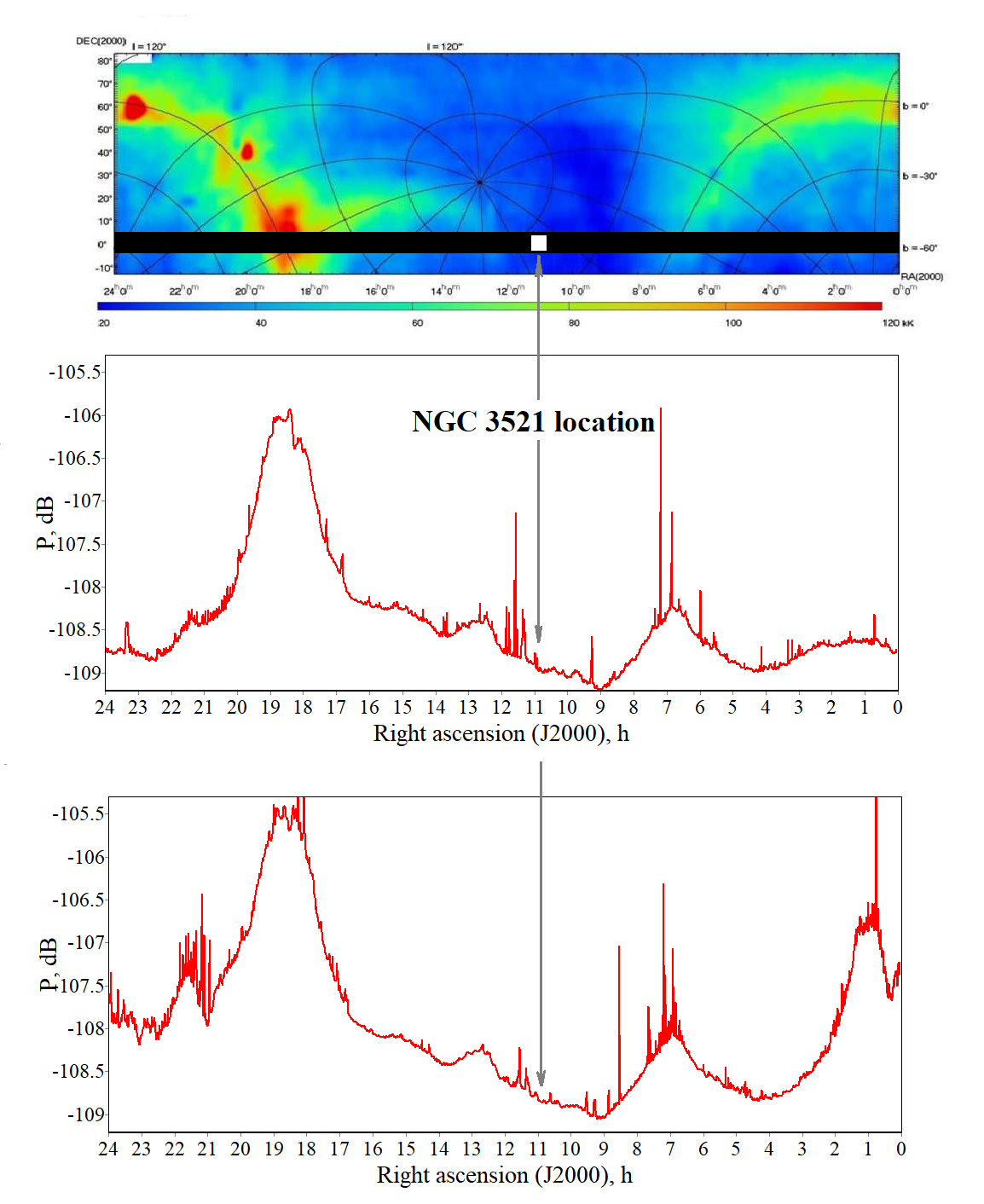}
\caption{
Upper: Map of non-thermal Galactic radio emission distribution (in terms of brightness temperatures $T_B$), obtained at UTR-2 for 20 MHz \citep{Sidorchuk2021}. The horizontal black strip schematically represents the Galaxy's intersection in the UTR-2 sky scans at DecJ = $0^\circ$. A white square indicates the position of NGC 3521.  Middle: Averaged scan for Jan 20-22, 2022 (daily sequence of total signal power changes at antenna output), 24-32 MHz band, integration time 30 s. Bottom: Averaged scan for Feb 3-5, 2022, 24-32 MHz band, integration time 30 s. 
}
\label{fig:map+2scans}
\end{figure}

Fig.~\ref{fig:map+2scans} (in the middle and bottom panels) shows the averaged two-day RA scans for DecJ = 0$^\circ$, obtained in sessions of Jan 20–22 and Feb 3–5, 2022, at the West–East antenna of UTR-2 radio telescope in 24–32 MHz band with integration time 30 s. The spatial resolution is $\approx$ $40'\times10^\circ$. The map at upper part of Fig.~\ref{fig:map+2scans} is given in RA-Dec terms. However, with some reservations, it can also be said that the resolution is about $10 \deg$ in galactic latitude in cases where the fixed antenna beam passes parallel to the galactic plane due to the daily rotation of the Earth. The significant difference between scans is due to markedly different levels of radio-frequency interference (Appendix~\ref{sec:UTR2_tech_spec}) during observations in Jan, 2022  (middle panel) and Feb, 2022 (bottom panel). Furthermore, the RFI peaks are observed at different times due to a daily time shift (4 minutes earlier each day) in the scan time for each RA position, which is particularly noticeable when the scans are separated by half a month.

From the upper part of Fig.~\ref{fig:map+2scans}, it is evident that the direction toward the NGC 3521 galaxy corresponds to areas of intermediate brightness temperatures of the Galactic background, of about $20{,}000$ K - $40{,}000$ K \citep{Sidorchuk2021}. In the middle and bottom panels of Fig. \ref{fig:map+2scans}, one can see numerous discrete sources that fell into the West-East antenna beam during the scanning of the celestial sphere at DecJ = 0$^\circ$. We note the distant, powerful sources – radio galaxies 3C348 (Hercules A, D $\simeq$ 700 Mpc) and 3C353 (D $\simeq$ 470 Mpc). Peaks are clearly recorded at RAJ = 16h 51m 08.1s and RAJ = 17h 20m 28.2s. The relative power of these peaks, after conversion into flux densities, corresponds to more than 100 Jy. These two radio sources were previously confidently observed at decameter wavelengths and included in the UTR-2 catalog \citep{Braude1979}.

\begin{figure}
    \centering
    \includegraphics[width=\linewidth]{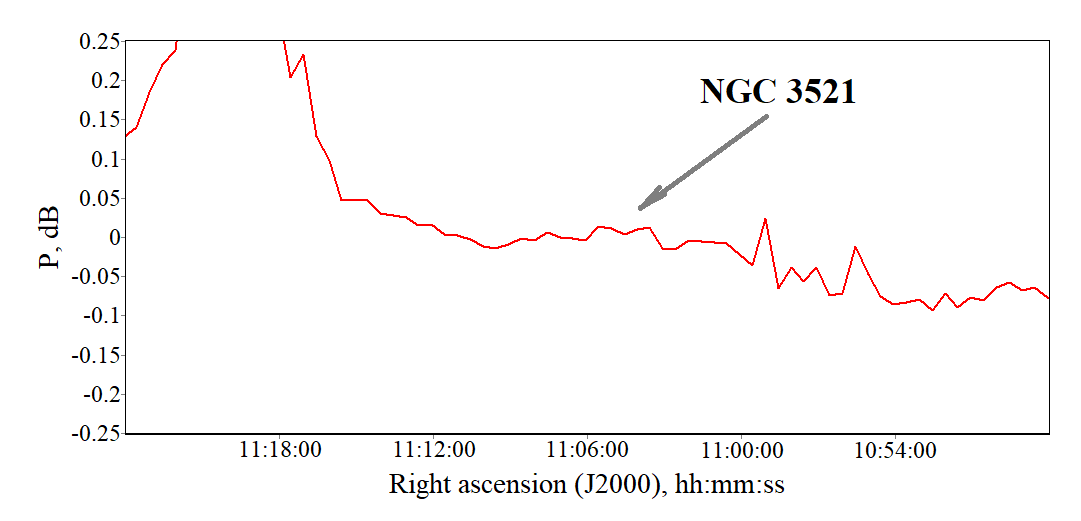}
    \caption{Part of the time sequence relative to the coordinates of NGC 3521 on the scan averaged over Jan 20–22 and Feb 3–5, 2022, obtained at the West-East antenna of the UTR-2 radio telescope for 24–32 MHz band, integration time 30 s.}
    \label{fig:NGC}
\end{figure}

\begin{figure}
    \centering
    \includegraphics[width=\linewidth]{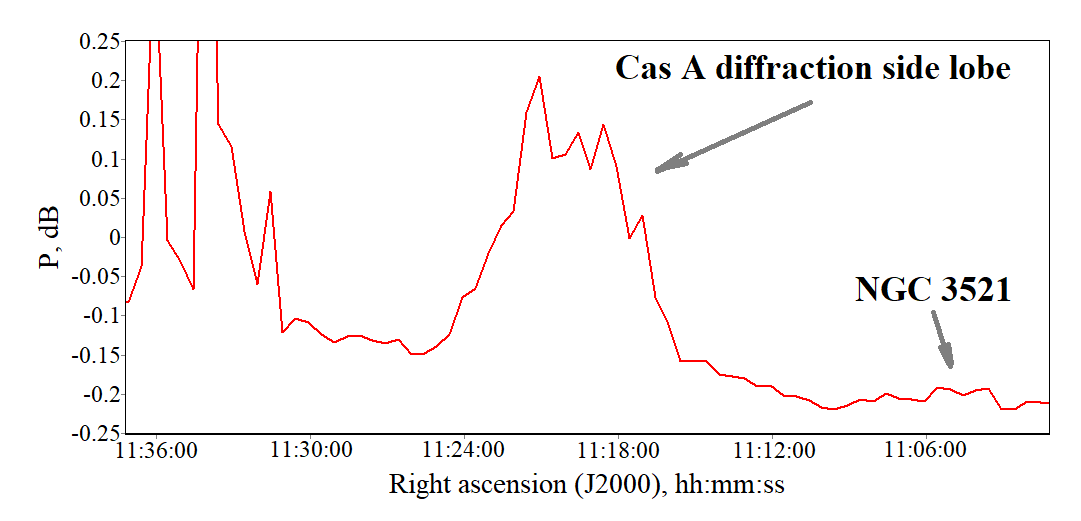}
    \caption{Part of the time sequence for the back diffraction side lobe of Cassiopeia A on the scan averaged over Jan 20–22 and Feb 3–5, 2022, obtained at the West-East antenna of the UTR-2 radio telescope for 24–32 MHz band, integration time 30 s.}
    \label{fig:CasA}
\end{figure}

Fig.~\ref{fig:NGC} shows a part of the averaged RA scan corresponding to NGC 3521 coordinates. From the obtained raw data, RA scans were implemented for the West-East antenna of UTR-2 (``knife'' beam, resolution by $RA \times Dec$ is about of $40' \times 10^\circ$) relative to the central frequency of 28 MHz (averaging band 8 MHz). The peak with a relative power of 0.022 dB corresponds to NGC 3521. Unfortunately, such a weak power signal-to-noise ratio does not meet the $3\sigma$ criterion. The observational data were analyzed, particularly in the 16-24 MHz band, but the level of fluctuations is much higher. The reason is that the scan distortion caused by narrow band, low-intensity RFI was included in the averaging. We also determined that the negative contribution to the obtained time sequences is made by the back diffraction side lobe of powerful radio emission from Cassiopeia A supernova remnant at the coordinate RAJ = 11h 19m 00s. It is received with a relative power of 0.36 dB at 28 MHz that corresponds to the flux density of 190.83 Jy {Fig.~\ref{fig:CasA}). This significantly affects the data quality in the scan. These reasons may explain why the NGC 3521 galaxy was not detected during the Decameter Sky Survey of the Northern Sky with the UTR-2 telescope in the late 1970s \citep{Braude1979}. 

To reach the $3\sigma$ criterion of detection of such a weak signal from NGC 3521, it was planned to follow observations 1) relative to the declination DecJ = $0^\circ$ during March, 2022; 2) with use of the high-quality data with the North-South antenna; 3) with implementation of the hardware multiplication of both antennas signal, which could allow obtain the data with high spatial resolution in the synthesized "pencil" beam mode. Unfortunately, these observations were interrupted \citep{Albergaria2025}. Additionally, despite the sufficiently high achieved sensitivity, the strong non-thermal Galactic background and the strong non-thermal radio emission of Cassiopeia A in the diffraction lobes require much greater sensitivity, which could be achieved by averaging a much larger number of daily scans. It would be advisable to use the data obtained at other decameter radio telescopes equipped with aperture synthesis and digital phasing systems (e.g., LOFAR LBA, NenuFAR, and others) for these purposes. 

\subsection{Flux density values of NGC 3521 in radio decameter range}

Due to the insufficient sensitivity for unambiguous detection of the signal from NGC 3521 in the decameter range for the reasons stated above, it is possible to estimate the upper limit on the flux density of NGC~3521 based on the relative power of peak 0.022 dB. The flux density is determined as
\begin{align}
S_{\nu} =\int_{\Omega} B_{\nu}(T)  d\Omega 
&= \frac{2 k T \nu^2}{c^2} \cdot \Omega \; [\mathrm{W}\cdot\mathrm{m}^{-2} \cdot \mathrm{Hz}^{-1}]
\label{eq:flux_dcm}
\end{align}

At decameter wavelengths, the antenna temperature $\Delta T_A$ recorded by the receiver is dominated by the brightness temperature of the Galactic background. The detected power, therefore, represents the excess of the Galactic background brightness temperature over the contribution from the discrete source. Converting 0.022 dB into linear units (1.00508) gives $\Delta T_A = (1.00508 - 1)\cdot T_B = 101.6$ K, which corresponds to a flux density of $S_{\nu} = 2k\Delta T_{A}\times A_{eff}^{-1} = 11.22$ Jy (in flux density units). Thereafter, the upper limit on the flux density of NGC 3521, obtained with UTR-2, for a frequency 28 MHz at $\Delta f = 8$ MHz and $\Delta \tau = 30$ s, will be 11.22 Jy. Substituting the obtained value of the upper limit on the flux density and the solid angle of NGC 3521 into formula \eqref{eq:flux_dcm}, we can estimate the brightness temperature of NGC 3521 through $B_\nu(T)$. The solid angle at a linear size of NGC 3521 of 25.9 kpc and a distance of 10.7 Mpc \citep{Warren2010} and an inclination angle of $ 72.7^{\circ}$ is $\Omega = 1.37 \cdot 10^{-6}$ sr. So, we obtain $B_\nu(T) = 8.19\times10^{-20}$~W $\cdot$ m$^{-2}$·Hz$^{-1}$ $\cdot$ sr$^{-1}$ and $T = 31.800$ K at 28 MHz.

The noise level for $\Delta f = 8$ MHz and $\Delta \tau = 30$ s is equal to $\sigma = 6.4~\cdot~10^{-5}$. We consider this value as 1$\sigma$ for 1 Jy. This 1$\sigma$ value for the obtained upper limit on the flux density at 28 MHz of 11.22 Jy would correspond to 0.00072 Jy (11.22 Jy $\pm$ 0.00072 Jy). However, these are purely statistical uncertainties that can be neglected. Much more significant are the systematic errors, including calibration uncertainties (up to 5\%) and baseline fluctuations in the scans (typically up to 5\%). Nevertheless, these values should not be taken strictly; these errors apply only to the upper limit of the flux density, which was determined by our measurements in the decameter range.

\section{Spectral energy distribution model of NGC 3521 from UV to radio decameter ranges}
\label{sec:sed_UV-Decameter} 

In this section, we derive physical properties of NGC 3521 from SED measurements using \textsc{CIGALE}. We emphasize that, unlike the observed photometric SED points, these inferred parameters are model-dependent and reflect the adopted set of templates and priors (e.g. star-formation history, attenuation law, dust emission model, and the energy-balance assumption). They help to translate overall SED shape and normalization into a compact set of physical descriptors (SFR, $M_\star$, $M_{\rm dust}$, etc.) that can be compared to the SED measurements performed by other authors and to the Milky Way data.

\subsection{Numerical simulations of the spectral energy distribution with CIGALE}
\label{sec:CIGALE}

To model and analyze the observed SED of NGC 3521, we used the \textsc{CIGALE} code \citep{Boquien2019}. We provide a detailed description of the selection of the baseline model and the choice of input parameters across the UV-radio cm range in \ref{sec:sed_UV-VLA}. The simulations were based on a combination of physically motivated modules, summarized in Table~\ref{tab:cigale_params_kharkiv}, along with the input parameters used to generate the model grid.

We chose a delayed star formation history with an optional burst or quenching episode
(sfhdelayedbq), expressed as
$\mathrm{SFR}(t)\propto t \exp(-t/\tau)$,
where $\tau$ is the e-folding timescale.
The stellar emission was computed using the stellar population synthesis models of \citep{Bruzual2003},
assuming a Chabrier initial mass function \citep{Chabrier2003} and approximately solar metallicity.
Nebular emission, including both line and continuum components, was incorporated following \citep{Inoue2011} and assuming zero escape fraction of Lyman continuum photons. Dust attenuation was described using the modified Calzetti starburst law
(dustatt\_modified\_starburst; \cite{Calzetti2000}, allowing for variations in the UV bump amplitude and in the power-law slope of the attenuation curve. The absorbed energy was re-emitted in the infrared following the dust emission models 
(dl2014; \cite{Draine2014}), which accounts for the emission of polycyclic aromatic hydrocarbons (PAHs) and large grains heated by a distribution of radiation field intensities.

The standard \textsc{CIGALE} radio module implements a single non-thermal power law anchored to the FIR–radio correlation via the $q_{\mathrm{IR}}$ parameter, assuming that the 1.4 GHz continuum is dominated by synchrotron emission with a fixed slope \citep{Helou1985, Boquien2019}. However, below $\nu \lesssim 100$ MHz, the integrated radio spectra of galaxies typically deviate from a pure power law. Namely, curvature and/or turnovers appear due to propagation and radiative-transfer effects in the ionized interstellar media (e.g., free–free absorption, possible synchrotron self-absorption). Observationally, global spectra show systematic low-frequency flattening and occasional turnovers, implying frequency-dependent opacity and multi-component synchrotron populations \citep{Condon1992, Lacki2013, Marvil2015, Chyzy2018}. To consistently model these effects within the SED energy-balance framework, we developed and implemented an extended radio module, radio\_extra\footnote{\url{https://github.com/OlenaKompaniiets/CIGALE_radio_extra_module}}, which retains the radio stock module but adds physically motivated low-frequency behavior.

Our radio\_extra module retains the star-forming (SF) synchrotron and the optional AGN power-law terms of the stock \textsc{CIGALE} radio module. On top of this baseline, we add: 

(i) a low-frequency (LF) component represented by a broken power law in $L_{\nu}$ with slopes $\alpha_{\rm lf}$ (below the break) and $\alpha_{\rm sf}$ (above the break). They are joined at a user-set break frequency $\nu_{\rm break}$. Its amplitude at the break is specified as a fraction $A_{\rm lf}$ of the SF $L_{\nu}$ at $\nu_{\rm break}$. This term mimics extended/aged ISM populations that preferentially shape the spectrum at $\nu \lesssim \nu_{\rm break}$ and reproduces the observed low-frequency curvature of integrated spectra \citep{Lacki2013, Marvil2015, Chyzy2018}. 

(ii) A frequency-dependent free–free-like absorption screen applied multiplicatively to \emph{all} radio components (SF+LF and, if presents, AGN), parameterized as $\tau_{\rm ff}(\nu)=\tau_{0,\rm ff} (\nu/\nu_{\rm ff})^{-\beta_{\rm ff}}$ with a canonical $\beta_{\rm ff}\simeq 2.1$ for homogeneous ionized gas. The emergent spectrum is attenuated by $\exp[-\tau_{\rm ff}(\nu)]$ \citep{Draine2011, RL79}. Internally, all components are computed in $L_{\nu}(\nu)$ on a wide grid (10 nm--100 m), converted to $L_{\lambda}(\lambda)$ for \textsc{CIGALE}. The SF terms are scaled by $L_{\rm dust}$.

\begin{table}
\centering
\caption{Parameter grid adopted for the CIGALE SED modeling from UV to decameter ranges}
\label{tab:cigale_params_kharkiv}
\begin{tabular}{ll}
\hline
\multicolumn{2}{l}{Star formation history: sfhdelayedbq} \\
\hline
$\tau_{\mathrm{main}}$ [Myr] & 1000, 2000, 4000 \\
$\mathrm{age_{main}}$ [Myr] & 7000 \\
$\mathrm{age_{bq}}$ [Myr] & 500, 1000 \\
$r_{\mathrm{SFR}}$ & 1.25, 1.5, 3.0 \\
\hline
\multicolumn{2}{l}{Stellar population synthesis: bc03} \\
\hline
IMF & Chabrier (1) \\
Metallicity $Z$ & 0.02 \\
Separation age [Myr] & 1000 \\
\hline
\multicolumn{2}{l}{Nebular emission: nebular} \\
\hline
$\log U$ & $-2.5$, $-3.0$ \\
Gas metallicity $Z_{\mathrm{gas}}$ & 0.02 \\
\hline
\multicolumn{2}{l}{Dust attenuation: dustatt\_modified\_starburst } \\
\hline
$E(B{-}V)_{\mathrm{lines}}$ & 0.5 \\
$E(B{-}V)_{\mathrm{factor}}$ & 0.44 \\
UV bump wavelength [nm] & 217.5 \\
UV bump width [nm] & 35.0 \\
UV bump amplitude & 3.0 \\
Power-law slope $\delta$ & $-0.4$, $-0.20$ \\
\hline
\multicolumn{2}{l}{Dust emission: dl2014} \\
$q_{\mathrm{PAH}}$ [\%] & 3.19, 4.58, 5.95 \\
$U_{\min}$ & 1, 2, 3, 5 \\
$\alpha$ & 2.0, 2.1, 2.2, 2.4 \\
$\gamma$ & 0.01, 0.05 \\
\hline
\multicolumn{2}{l}{Radio emission: radio\_extra} \\
\hline
FIR/radio $q$ (SF only), $q_{\mathrm{IR, SF}}$         & 2.3, 2.6, 2.9 \\
SF synch. slope, $\alpha_{\mathrm{sf}}$                  & 0.9 \\
LF amp. $\nu_{\mathrm{break}}$ (×SF), $A_{\mathrm{LF}}$& 1.4, 1.6, 2.0 \\
LF slope, $\alpha_{\mathrm{LF}}$                         & 1.5 \\
HF slope (extra), $\alpha_{\mathrm{SF}}$                 & 1.0, 1.2 \\
Break freq. (MHz), $\nu_{\mathrm{break}}$               & 57 \\
FF optical depth  $\nu_{\mathrm{ff}}$, $\tau_{\mathrm{ff}}$ & 1.0, 3.0, 5.0 \\
FF pivot freq. (MHz), $\nu_{\mathrm{ff}}$               & 100 \\
FF opacity slope, $\beta_{\mathrm{ff}}$                  & 2.1, 2.3, 2.5 \\
\hline
\end{tabular}
\end{table}

\subsection{Main parameters of NGC 3521 from spectral energy distribution}
\label{sec:sed_results}

The main parameters of the galaxy were estimated from the observational SED through comparison with a theoretical model. The total stellar luminosity is 
$L_{\star} = (7.59 \pm 0.38)\times10^{10} L_{\odot}$,
the other of the simple stellar population to be convolved with it, both using the values for the evolved stellar population of Table~1,
$L_{\star,\mathrm{old}} = (5.47 \pm 0.27)\times10^{10} L_{\odot}$
and the younger stellar component
$L_{\star,\mathrm{young}} = (2.12 \pm 0.11)\times10^{10} L_{\odot}$.
The integrated dust luminosity reaches
$L_{\mathrm{dust}} = (2.66 \pm 0.13)\times10^{10} L_{\odot}$,
and the associated dust mass is
$M_{\mathrm{dust}} = (1.26 \pm 0.14)\times10^{8} M_{\odot}$,
revealing a massive dust reservoir that can reprocess a large fraction of a stellar emission.
The sum of the stellar masses inferred from Bayesian analysis is
$M_{\star} = (5.99 \pm 0.30)\times10^{10} M_{\odot}$, old population in place
($M_{\star,\mathrm{old}} = (5.89 \pm 0.29)\times10^{10} M_{\odot}$).
The resulting star-formation rate from the delayed-$\tau$ star-formation history is
$\mathrm{SFR} = 1.65 \pm 0.08 M_{\odot} \mathrm{yr^{-1}}$.
in agreement with the main-sequence activity of massive late-type spirals.
Given the derived dust luminosity and mass,
The effective dust temperature is estimated using the standard scaling of the infrared luminosity-to-mass ratio for optically thin modified blackbody emission
\citep{Hildebrand1983,Draine2003,Casey2012} $L_{\mathrm{dust}}/M_{\mathrm{dust}} \propto T_{\mathrm{dust}}^{ 4+\beta}$, where $\beta = 2$ is adopted for the emissivity index of interstellar dust. 

The measured ratio 
$L_{\mathrm{dust}}/M_{\mathrm{dust}} \simeq 2.1\times10^{2} L_\odot/M_\odot$ 
corresponds to an effective dust temperature of $T_{\mathrm{dust}} \simeq 23 \mathrm{K}$. This value is typical for the diffuse dust in normal star-forming spiral galaxies.   In summary, these findings suggest that NGC 3521 harbors a massive stellar disk in which star formation rates remain moderate and dust content is well measured, in energy equilibrium with stellar radiation.

 \begin{figure}
    \centering
    \includegraphics[width=1.0\linewidth]{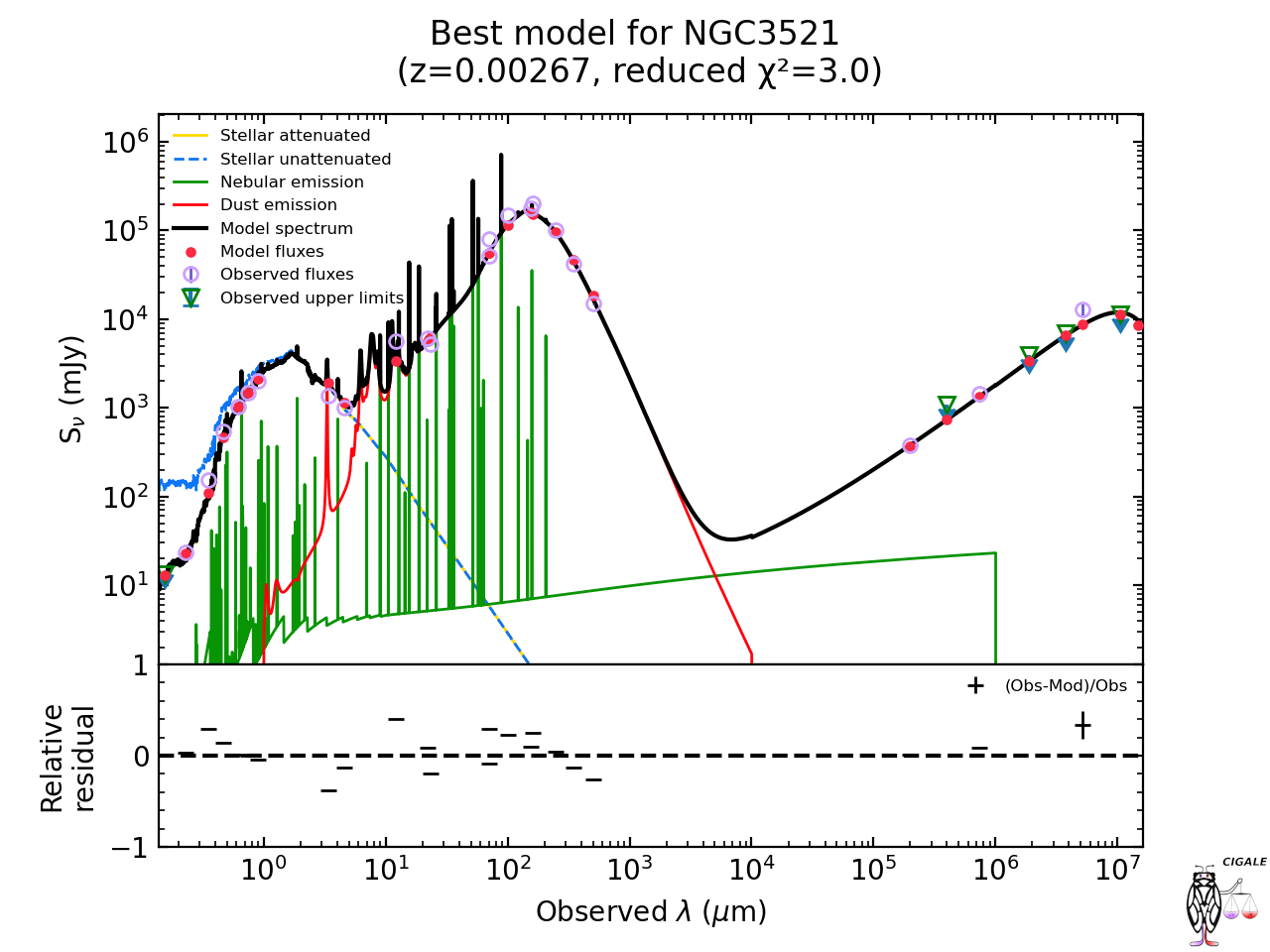}
    \caption{NGC 3521 SED from UV to decameter ranges}
    \label{fig:UV_to_decametr}
\end{figure}

In \textsc{CIGALE} \citep{Boquien2019}, physical properties are inferred from the marginal posterior PDFs obtained over the explored model grid with likelihood weights $w \propto \exp(-\chi^{2}/2)$. We report the expectation values of these marginal PDFs and their associated dispersions.

The posterior PDFs provide the primary diagnostic of parameter constraints. For the main quantities discussed in this work --- $\mathrm{SFR}$, $M_{\rm dust}$, $L_{\rm dust}$, $M_{\star}$, $L_{\star}$ and the corresponding old/young components --- the PDFs are single-peaked and maximize well within the explored parameter space, indicating that these parameters are meaningfully constrained by the UV-decameter SED coverage. The moderate widths of the $\mathrm{SFR}$ and $M_{\rm dust}$ PDFs are consistent with the expected coupling between recent star formation, attenuation, and dust heating/emission in star-forming spirals. In contrast, the mass-weighted stellar age ($t_{\rm M,\star}$) shows posterior probability accumulating toward the edge of the explored grid, implying that it is better interpreted as a one-sided constraint (a limit within the adopted model space) rather than as a tightly measured value.

As an additional validation check, we inspected the $\chi^{2}$ projections across the grid to identify potential alternative low-$\chi^{2}$ regions; the interpretation of constraints is based on the marginal posterior PDFs.

\section{Photometry of the central region of NGC 3521}
\label{sec:nucleus}

\begin{figure*}[ht]
\centering
\setlength{\tabcolsep}{4pt}
\begin{tabular}{ccc}
\includegraphics[width=.32\textwidth]{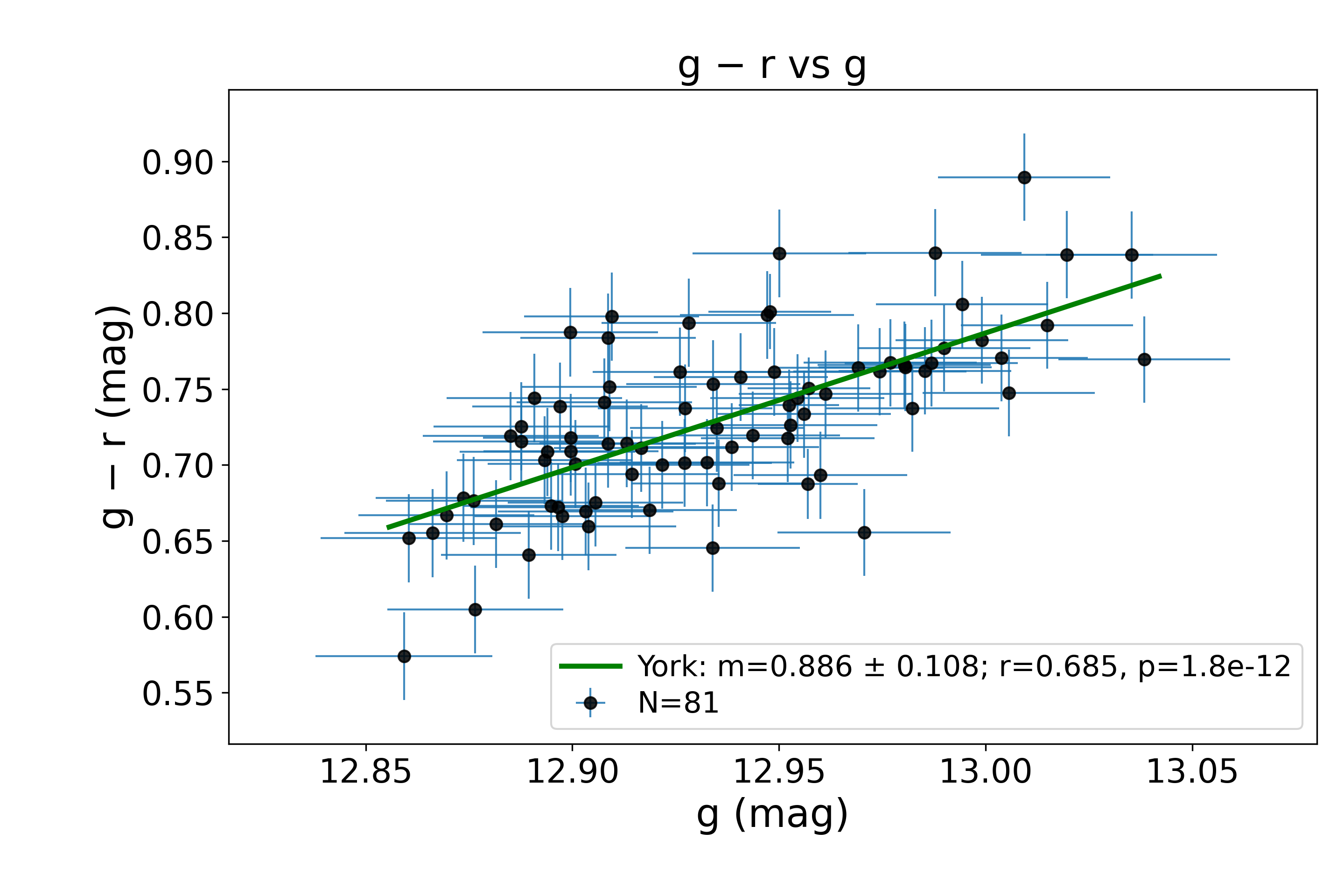} &
\includegraphics[width=.32\textwidth]{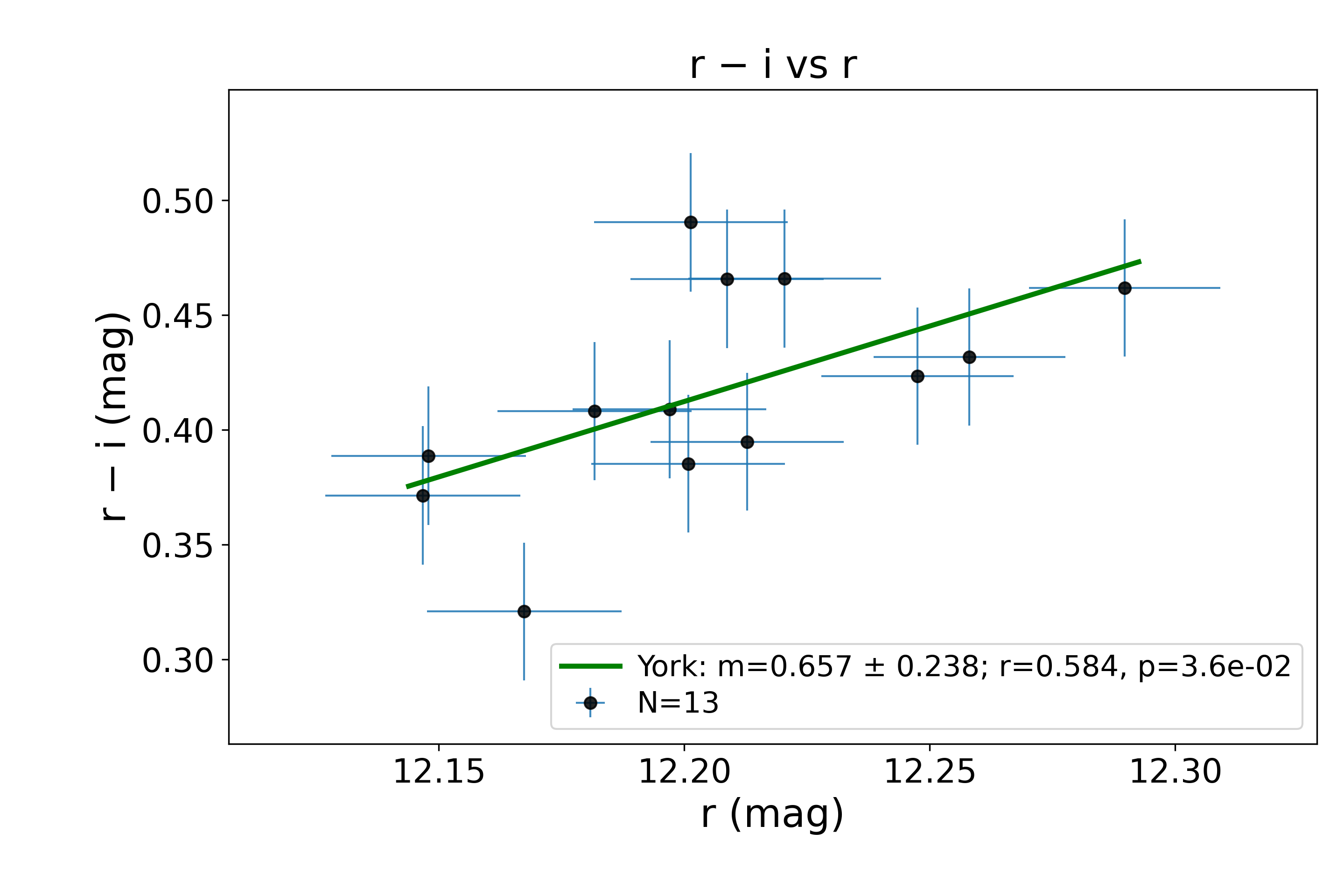} &
\includegraphics[width=.32\textwidth]{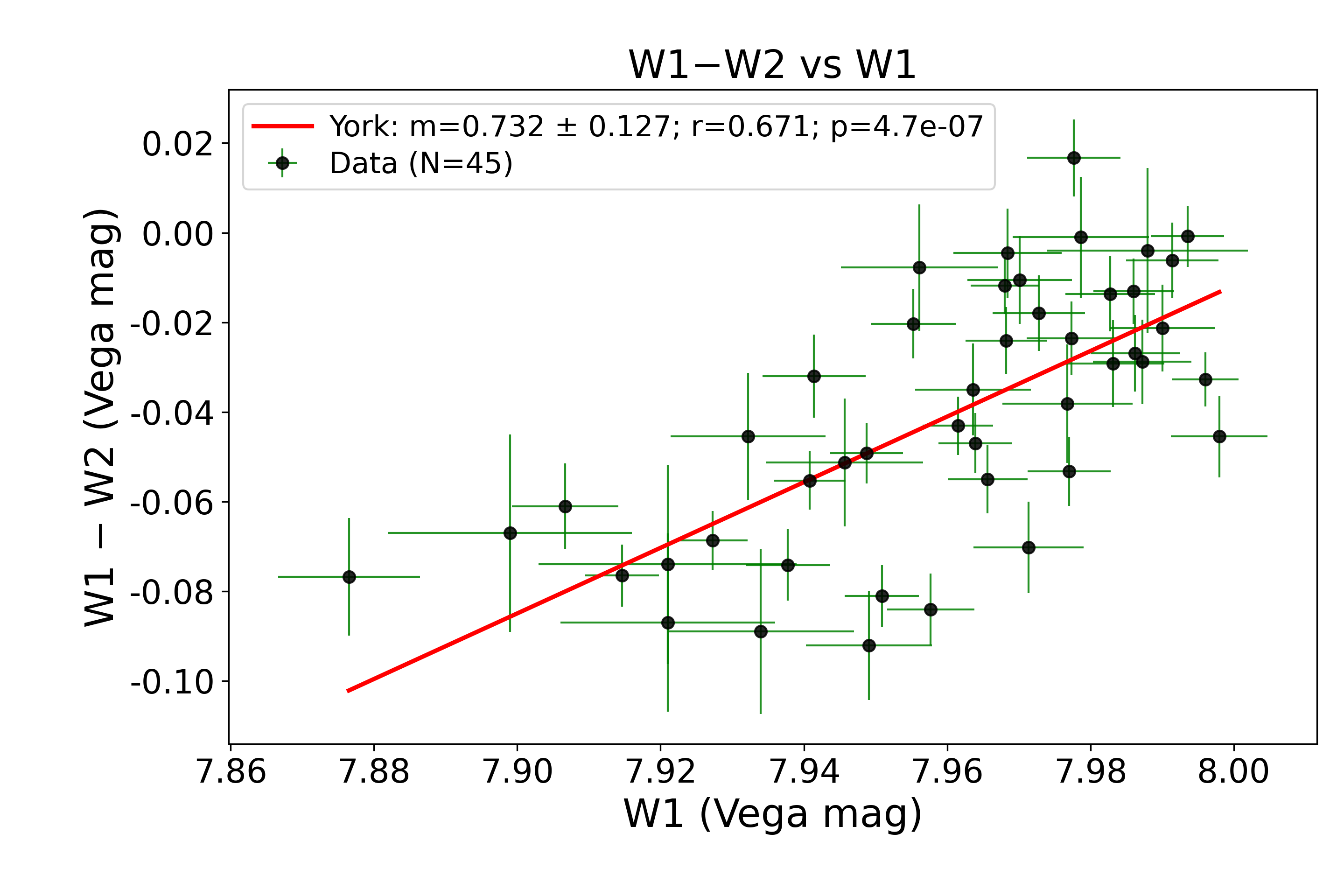} \\
\small (a) g-r vs g &
\small (b) r-i vs r &
\small (c) W1--W2 vs W1
\end{tabular}
\caption{Color--magnitude diagrams for NGC 3521. 
(a) g-r vs g: York slope $m_{\rm}= 0.886\pm0.108$, Pearson $r=0.685$, $p=1.8\times10^{-12}$.
(b) r-i vs r: $m_{\rm}= 0.657\pm0.238$, $r=0.584$, $p=3.6\times10^{-2}$.
(c) W1--W2 vs W1 (native Vega color): $m_{\rm}= 0.732\pm0.127$, $r=0.671$, $p=4.7\times10^{-7}$. All color--magnitude relations show a clear Bluer--When--Brighter (BWB) trend.}
\label{fig:cmd_row_twocol}
\end{figure*}

\subsection{X-ray data with Chandra}

NGC 3521 was observed by Chandra in the 0.3-8 keV range. \cite{Grier2011} exploited these data to search for a hidden nuclear X-ray activity in spiral galaxies, which are optically normal ones. They found no clearly defined X-ray peculiarities in this galaxy. One of the noteworthy features of NGC~3521 -- the presence of an ultraluminous X-ray (ULX) source, -- [SST2011] J110545.62+000016.2 -- was discovered by \cite{Heida2014} and further analyzed with VLT~X-shooter and Chandra data by \cite{Lopez2015}. This ULX is associated with an H~II region located $\sim138$~pc northeast of the nucleus, with an X-ray luminosity of $(1.9\pm0.8)\times10^{40}$~erg s$^{-1}$. 

The X-ray Eddington ratio for NGC 3521 was defined by \cite{Zhang2009} as $L_{0.3-8}/L_{Edd}$ = $2.2\times10^{-7}$. Given this value, $log(M_{SMBH}/M_{\odot }) = 6.86 \pm 0.58$ \citep{Davis2014}, and the LINER activity type \citep{Das2003, Pastoven2024}, we did not use Chandra data when constructing SED model. 

\subsection{Color--magnitude relations and spectral indices}

We analyzed active processes in the central region of NGC 3521, considering the photometric data obtained from the open archives of the Zwicky Transient Facility (ZTF) \citep{Masci2019ZTF} in three optical filters (g, r, i) and the NEOWISE data \citep{Mainzer2014NEOWISE} in two infrared channels (W1, W2). These data cover the periods 2018--2025 for ZTF and 2014--2024 for WISE.

After correcting for Galactic extinction, the multi-epoch photometry from ZTF (see \ref{sec:ZTF_optical}) and NEOWISE (see \ref{sec:NEOWISE_data}) was consolidated to mitigate inter-filter timing offsets. However, the binning schemes differ due to the distinct observing cadences. For ZTF, we computed nightly averages using a sliding window of $\Delta t \le 0.35$ d to obtain quasi-simultaneous g/r/i measurements. For NEOWISE, we derived nightly means from single-exposure photometry by grouping exposures into integer ``nights'' via
${\rm night\_id}=\lfloor{\rm MJD}+0.82\rfloor$,
where the offset was chosen to minimize splitting of dense exposure clusters across adjacent days.

Variability amplitudes were quantified with a robust percentile range (95th–5th percentiles), $\Delta m \equiv P_{95}-P_{05}$, which is less sensitive to outliers and occasional residual systematics than the full peak-to-peak spread. After the final quality screening and clipping, the optical bands show amplitudes $\Delta g=0.153$ mag, $\Delta r=0.144$ mag, and $\Delta i=0.120$ mag, while the infrared bands yield $\Delta W1=0.1608$ mag and $\Delta W2=0.1568$ mag. 

Paired nightly measurements were then used to construct three color--magnitude relations: g-r versus g (Fig.~\ref{fig:cmd_row_twocol}a), r-i versus r (Fig.~\ref{fig:cmd_row_twocol}b), and W1--W2 versus W1 (Fig.~\ref{fig:cmd_row_twocol}c). The uncertainty of each color index was computed via standard error propagation for independent measurement errors \citep{Bevington2003}. Because both axes of the color--magnitude diagrams contain measurement uncertainties, we applied York regression \citep{York2004}, which accounts for errors in both coordinates (including the covariance between magnitude and color). Overall, all three color--magnitude relations show a bluer-when-brighter (BWB) trend. The g-r vs g and W1--W2 vs W1 relations exhibit high Pearson correlation coefficients (0.69 and 0.67), while the r-i vs r relation shows a moderate correlation coefficient (0.58) based on fewer paired nights (Fig.~\ref{fig:cmd_row_twocol}).

\begin{table}[t]
\centering
\caption{Color--magnitude correlations and spectral parameters for NGC 3521.}

\label{tab:ztfwise}
\footnotesize
\setlength{\tabcolsep}{3pt}
\renewcommand{\arraystretch}{1.05}
\begin{tabular}{@{}lccccccc@{}}
\hline
Color & $m$ & r & $p$ & $K$ & $\tilde{C}$ & $\alpha_{\rm median}$ & ${\rm d}\alpha/{\rm d}(-X)$\\
\hline
\shortstack[l]{g--r\\vs g} 
& \shortstack[c]{0.89\\$\pm$0.11}
& 0.69
& $1.79\times10^{-12}$
& 0.319
& 0.73
& \shortstack[c]{2.28\\$\pm$0.02}
& \shortstack[c]{--2.78\\$\pm$0.34}\\[0.6ex]

\shortstack[l]{r--i\\vs r} 
& \shortstack[c]{0.66\\$\pm$0.24}
& 0.58
& $3.60\times10^{-2}$
& 0.224
& 0.41
& \shortstack[c]{1.82\\$\pm$0.06}
& \shortstack[c]{--2.93\\$\pm$1.06}\\[0.6ex]

\shortstack[l]{W1--W2\\vs W1} 
& \shortstack[c]{0.73\\$\pm$0.13}
&0.67
& $4.67\times10^{-7}$
& 0.344
& --0.68
& \shortstack[c]{--1.97\\$\pm$0.01}
& \shortstack[c]{--2.13\\$\pm$0.37}\\
\hline
\end{tabular}
\tablefoot{ Columns: $m$---York slope; $r,p$---Pearson coefficient and $p$-value; $K$---color--to--spectral-index factor; $\tilde{C}$---median color; $\alpha_{\rm median}$---median spectral index; ${\rm d}\alpha/{\rm d}(-X)$---slope with brightness. We adopt $F_\nu\propto\nu^{-\alpha}$ in all bands; median color W1--W2 is listed in AB color, while W1--W2 versus W1 (Fig.~\ref{fig:cmd_row_twocol}c) shown in native WISE Vega color.}
\end{table}

\section{Discussion}
\label{sec:discussion}

\subsection{Multiwavelength properties of NGC 3521}
\label{sec:discussionNGC3521}

As we noted in the Introduction, the SED models of NGC 3521 were constructed in various spectral ranges by \cite{Zibetti2011, Hunt2019, Erroz2019, Chang2020, Pastoven2024}. As well, the flux densities were obtained by \cite{Dale2012} in 3.6--500$\mu$m (Spitzer+Herschel data) using the dust model by \cite{Li2001} and KINGFISH FIR/sub-mm photometry; by \cite{Brown2014} who prepared the atlas of SEDs from UV to mid-IR (GALEX+SDSS+2MASS+Spitzer+WISE) using MAGPHYS software and distance to NGC 3521 as $D$ = 10.1 Mpc; by \cite{Pattle2023} from IR to sub-mm (Herschel+JCMT/SCUBA-2) for constructing of dust SED maps for nearby galaxies using $D$=12.42 Mpc for NGC 3521. Compared with these studies, we modeled the SED of NGC 3521 not only in a wider spectral range, from UV to radio decameter wavelengths, but also used aperture photometry across all ranges. The latter is essential because most of the aforementioned works did not consider aperture photometry and used different distance estimates for NGC 3521 (Table \ref{tab:NGC3521_massSFR}). The most meaningful discrepancies between the assumed distances are as follows (Column 1): almost 40 \% larger in the MUSE Atlas of Disks by \cite{Erroz2019}, where these authors have taken the distance from NED; almost 30 \% smaller in the work by \cite{Chang2020}, who have taken it from the JCMT Nearby Galaxies Legacy Survey by \cite{Wilson2012}. The stellar mass and SFR in work by \cite{Zibetti2011} were taken from \cite{Skibba2011} for distance 9.2 Mpc, which are based only on the optical g and $i$ bands and the NIR H bands using \cite{Zibetti2009} recipes. 

To provide a uniform and self-consistent approach to data processing, we used photometrically calibrated FITS images spanning the UV to radio cm ranges (see Section \ref{sec:APT_UV_VLA} and Appendix 2) and radio meter- and decameter-band flux densities on a uniform scale. As well, to compare our results with those mentioned in Column 1, we recalculated the SFR, stellar, and dust masses after correcting to the same distance scale ($D_0=10.7$ Mpc) using $Q(D_0)=Q(D) (D_0/D)^2$ as suggested by \cite{Kennicutt2011}. Such an approach was also applied by \cite{Hunt2019}. 

After rescaling to the common distance, the estimates of the global parameters show a better overall agreement. In particular, if the single most discrepant values are treated as method-dependent outliers, the remaining estimates are grouping within a relatively narrow range for all three quantities. For the SFR (Column~4), most values group around $\log \mathrm{SFR} \simeq 0.22$--$0.29$, while only the rescaled value by \citet{Skibba2011} is higher ($\log \mathrm{SFR}=0.46$). We note that these authors derived the SFR  independently from the H$\alpha$+$24 \mu$m combination. Therefore, this offset is more plausibly related to differences in the adopted SFR tracer and calibration, including different sensitivity to obscured star formation and to the timescale over which star formation is probed, rather than to a global inconsistency among the studies. The rescaled stellar mass estimates also agree reasonably well within $\sim$0.3  dex: $\log M_\star$ spans 10.33--10.90 (Column 2), where our Models~1--2 yield $\log M_\star\simeq 10.76$--10.78. For the dust mass, the rescaled values are remarkably consistent across independent analyses, with $\log M_{\rm dust}\approx 7.89$--8.12 (Column 3), and the estimate by \cite{Chang2020} becomes fully compatible.

A plausible explanation for the remaining outliers is that the published estimates are not based on identical observational constraints or on identical modeling assumptions. In particular, the stellar-mass estimates depend on whether one uses a color--mass-to-light calibration or full SED fitting. The dust-mass estimates are especially sensitive to the availability and weighting of FIR/sub-mm constraints, as well as to the adopted dust-emission prescription. The SFR estimates depend on the tracer itself,  UV+ far IR energy-balance SED fitting.

\begin{table*}
\centering
\caption{Estimates of $M_\star$, $M_{\rm dust}$ and SFR for NGC~3521
for the distance $D_0=10.7$ Mpc} 
\label{tab:NGC3521_massSFR}
\renewcommand{\arraystretch}{1.15}
\begin{tabularx}{\textwidth}{@{}X c c c@{}}
\toprule
\makecell[l]{Authors, spectral range / software / original values} &
\makecell[c]{$\log M_\star$} &
\makecell[c]{$\log M_{\rm dust}$} &
\makecell[c]{$\log \mathrm{SFR}$} \\
\midrule

\citet{Skibba2011}: $\log M_\star$ from $g-i$ colors and $H$-band luminosity using \citet{Zibetti2009}; $\log \mathrm{SFR}$ from H$\alpha$+$24 \mu$m (Sect.~4.4.1).~$D=9.2$ Mpc, $\log M_\star=10.52$, $\log \mathrm{SFR}=0.33$ &
10.65 & --- & 0.46 \\

\citet{Hunt2019} FUV to sub-mm (0.1--1000$\mu$m); multi-code SED fitting (CIGALE, GRASIL, MAGPHYS),~$D=11.2$  Mpc, $\log M_\star=10.68$, $\log M_{\rm dust}=7.93$, $\log \mathrm{SFR}=0.28$ &
10.64 & 7.89 & 0.24 \\

\citet{Erroz2019} optical IFS (MUSE) + ancillary UV--IR SEDs (MAD project),~$D=14.2$ Mpc, $\log M_\star=11.15$, $\log \mathrm{SFR}=0.47$ &
10.90 & --- & 0.22 \\

\citet{Chang2020} IR to sub-mm (Herschel+JCMT; global cold-dust properties),~$D=7.9$  Mpc, $\log M_{\rm dust}=7.65$ &
--- & 7.91 & --- \\

\citet{Pastoven2024} 135 nm to 21 cm ~(UV to radio), CIGALE (model A),~$D=10.7$  Mpc &
10.33 & 7.93 & 0.25 \\

This work, Model 1 (Appendix \ref{sec:sed_UV-VLA}) UV to radio cm, CIGALE+radio module,~$D=10.7$  Mpc &
10.76 & 7.99 & 0.29 \\

This work, Model 2, UV to radio decameter, CIGALE+radio\_extra module,~ $D=10.7$  Mpc  &
10.78 & 8.12 & 0.22 \\

\bottomrule
\end{tabularx}
\tablefoot{ A precise spectrophotometry of the SN 2024aecx event, which was recently recorded \citep{Stevance2024} and classified as a type IIb SN \citep{Andrews2024}, allowed \cite{Xi2026} to define the distance to NGC 3521 as 11.3 $\pm 1.1$ Mpc.}
\end{table*}

Let us compare the flux values adopted in our study (Table \ref{tab:NGC3521_fluxes}) with those in the works of \cite{Brown2014} and \cite{Pastoven2024}. It reveals a systematic underestimation across the UV (GALEX) to near-IR (WISE) wavelength range. The deviation from the fluxes obtained in our work ranges from $34$ to $66$~per~cent. This difference is expected because \cite{Pastoven2024} performed photometry using a rectangular aperture of $235\times56$~arcsec, whereas our measurements employ an elliptical aperture with semi-axes of $250\times120$~arcsec that encompasses the full extent of the galaxy (Fig. \ref{fig:ngc3521_multiwavelength}). We also obtained lower fluxes for Herschel/SPIRE~500~$\mu$m and Spitzer/MIPS~160~$\mu$m. This is consistent if we take into account that earlier photometry measurements were provided for a substantially larger aperture of $926\times455$~arcsec \citep{Dale2012}. All other flux values used by \cite{Brown2014} and \cite{Pastoven2024} agree with our measurements and lie within the confidence intervals derived in the present work.  

We reconstructed the baseline SED model in UV to radio 21 cm range by \citep{Pastoven2024} with minor adjustments to the grid parameters (Model 1 in Table \ref{tab:NGC3521_massSFR}, see Appendix \ref{sec:APT_UV_VLA} with Table \ref{tab:cigale_params} and Fig. \ref{fig:UV_to_VLA}) for comparison. It allowed us to reproduce the SED of NGC 3521 with high fidelity in this spectral domain. Extending it to the radio mater and decameter ranges resulted in more reliable estimates of the galaxy's physical properties (Model 2 in Table \ref{tab:NGC3521_massSFR}), most notably its star formation rate and smaller uncertainties than Model~1, The total stellar luminosities in both Models are similar, whereas the Model 2 gives a more consistent result between luminosities of old and young stellar populations, $L_{\star,\mathrm{old}} = (5.51 \pm 0.28)\times10^{10} L_{\odot}$ and $L_{\star,\mathrm{young}} = (2.14 \pm 0.11)\times10^{10} L_{\odot}$; respectively. The derived estimates of dust mass constitute approximately 1\% of the stellar mass, which is consistent with results from cosmological simulations of dust evolution in Milky Way galaxy analogs \citep{McKinnon2016}.

To place our results in the broader context of previous studies, let us follow the comparison of global parameters $M_\star$, $M_{\rm dust}$, and SFR from SEDs given in Table~\ref{tab:NGC3521_massSFR}. The range of stellar mass estimates is approximately 0.8 dex, ranging from $\log M_\star \sim 10.33$ to $11.15$. All studies agree on a massive, moderately star-forming spiral system. However, there are notable differences by \citet{Erroz2019}, who utilized UV–IR photometry from the MAD survey in conjunction with MUSE. We suggest that both the assumed distance (14.2 Mpc) and the specifics of stellar population synthesis in this range led to this result. For example, \cite{Mosenkov2019} working with the data in the Herschel band noted that surface density profiles for the dust mass have an obvious depletion in the inner galaxy region. They concluded that if this feature, together with the truncation in the galaxy outskirts, is not taken into account, the dust emission and dust mass density profiles look more or less exponential. Weak azimuthal variations in metallicity can play a definitive role \citep{Grasha2022}: the distribution of H~II regions exhibits enhanced or depleted oxygen abundance. We also note that the majority of authors report $\log M_{\rm dust} \sim 7.9$–$8.1$, and dust mass estimates are more consistent across studies. The inclusion of radio data through the radio$\_$extra module, which more successfully limits synchrotron contamination and cold dust emission, is the main reason why our Model 2 produces the highest dust mass ($\log M_{\rm dust} = 8.12$). \cite{Pastoven2024} and \cite{Hunt2019} analyzed similar spectral ranges with different CIGALE implementations and both found $\log M_{\rm dust} = 7.93$. \cite{Chang2020} resulted in a slightly lower dust mass value, $7.65$, but they considered missed warmer dust components, focusing only on the FIR to sub-mm range. 

The SFR values, which range from $\log \mathrm{SFR} = 0.22$ to $0.47$ ($\sim$1.6 to 3.0 M$_\odot$/yr), are likewise generally consistent except for a slightly higher result by \cite{Erroz2019}. In their earlier work \citep{Leroy2008} with UV (GALEX), IR (Spitzer), and kinematics (THINGS survey) data for estimation of the SFR, and the masses of the stellar and neutral gas components. They derived $\mathrm{SFR} \approx 2.1~M_{\odot} \mathrm{yr^{-1}}$, a stellar mass of $\sim 5\times10^{10} M_{\odot}$, and a neutral hydrogen mass of $\sim 1.4\times10^{10} M_{\odot}$.  \cite{Coccato2018}, who used archival ESO and Spitzer data for a spectroscopic decomposition of NGC~3521, performed a separate analysis of the bulge and disc in terms of their kinematics, stellar ages, and metallicities. They concluded that NGC~3521 has a complex star formation history with three major episodes: an old stellar population (> 7~Gyr), an intermediate-age component ($\approx$3~Gyr), and a young population (< 1~Gyr). The intermediate stellar population dominates both the galaxy's stellar mass and luminosity. 

Based on the results of our SED Model 2, the luminosity-weighted age of the stellar population is $5.00 \pm 0.5$~Gyr. Thus, while the sfhdelayed module does not reproduce the full complexity of the galaxy's star formation history, it successfully captures the formation of the dominant intermediate-age population together with the young stellar component. As for the gas component, \cite{Elson2014} discovered the presence of an anomalous HI region with a mass of $M_{H {I}}$ = 1.5 $\times 10^{9} M\odot$ (20 per cent of a total HI mass), analyzing VLA ``The HI Nearby Galaxy Survey''. This diffuse, slow-rotating halo gas component (25-125 km/s), with a height scale of about 3.5 kpc, is located in a thick disc that coincides with the inner regions, where the star formation rate is highest. These authors conclude that it serves as a "galactic fountain" depositing gas from the galaxy's disc into the halo.  

So, the estimates of SFR, stellar, and dust masses obtained from our SED modeling across UV to radio decameter range (Table \ref{tab:cigale_params_kharkiv}) provide a reliable characterization of the key parameters and star-forming history of NGC~3521. 

\subsection{Emission from the central region}
\label{sec:nuclear_activity}

As noted in Section~\ref{sec:nucleus}, we did not include Chandra data in the SED model for the X-ray range and analyzed the nuclear emission of NGC~3521 separately. In addition to the archival survey data, the NGC~3521 was previously monitored with the Ukrainian Zeiss-600 telescope at the Terskol Observatory during 2021--2022, including a 3-h r-band sequence obtained on 2022 February 11 \citep{Pastoven2024}. Using ZTF ($g,r,i$) and NEOWISE (W1, W2) multi-epoch photometry, we detected statistically significant color--magnitude correlations (Fig.~\ref{fig:cmd_row_twocol}; Table~\ref{tab:ztfwise}), demonstrating genuine variability of the central emission component over 2014--2025. 

In a highly inclined, dusty host with a composite H {\sc ii}/LINER nucleus (see Section~\ref{sec:Intro}), PSF-integrated photometry is expected to be strongly diluted by host starlight and affected by internal reddening, naturally producing a red mean continuum while enhancing bluer-when-brighter (BWB) trends when a compact blue variable component is presented. The median optical spectral indices, $\alpha_{gr}=2.28\pm0.02$ and $\alpha_{ri}=1.82\pm0.06$, confirm a red optical continuum within the $\sim$3\arcsec\ ZTF PSF, whereas the mid-IR spectral index $\alpha_{\rm MIR}=-1.97\pm0.01$ is close to the Rayleigh--Jeans limit ($\alpha=-2$ for $F_\nu\propto\nu^{-\alpha}$), implying that the integrated 3--5 $\mu$m emission within the 7.5\arcsec\ WISE PSF is dominated by stellar continuum rather than hot-dust (torus) emission (Table~\ref{tab:ztfwise}). The negative spectral-index responses to brightening, ${\rm d}\alpha/{\rm d}(-g)=-2.78\pm0.34$ and ${\rm d}\alpha/{\rm d}(-{\rm W1})=-2.13\pm0.37$, quantify the BWB behavior and support a scenario in which a weak, intrinsically bluer nuclear continuum varies atop a comparatively stable, reddened host background.

Similar behavior has been observed in the center of our Galaxy, around the SgrA SMBH. The highest near-IR brightness of Sgr A* over 20 years, with rapid flux variations up to 75-fold over two hours, was recorded during observations with the Keck telescope \citep{Do2019}. According to JWST observational data from 2023–2024, correlated fluctuations were detected at wavelengths of 2.1 and 4.8 $\mu$m, with a phase shift between them, indicating several emission mechanisms near the event horizon \citep{YusefZadeh2025}.

 \subsection {Twin SEDs of the Milky Way and NGC 3521}

The nearby NGC 3521 galaxy was first considered the MWA by \cite{McGaugh2016} based on the compositions of the MW and NGC 3521 rotation curves, and later by \cite{Pilyugin2019, Pilyugin2023} when they analyzed the metallicities of these galaxies. As we mentioned in the Introduction, NGC 3521 also shows other indicators suggesting an analogy with the Milky Way. We suggested that NGC~3521 and the MW should also exhibit a similar UV--to--radio decameter SED, and we explicitly test this using homogeneous aperture photometry.

The obtained SED parameters of NGC 3521 are comparable to those inferred for the Milky Way: $M_{\star} = 5.48^{+1.18}_{-0.94} \times 10^{10},M{\odot}$ \citep{Fielder2021}, $\mathrm{SFR} = 1.65 \pm 0.19,M_{\odot},\mathrm{yr}^{-1}$ \citep{LicquiaNewman2015ApJ}. The derived stellar mass and SFR are comparable to modern estimates for the Milky Way \citep{LicquiaNewman2015ApJ, Fielder2021}. Additionally, when comparing the positions of NGC 3521 and the Milky Way in the local cosmic web \citep{Kompaniiets2025}. The parameters of their local density environment are similar. The spectral type of the nuclear activity is sufficiently low to indicate that it is a LINER \citep{Das2003, Pastoven2024}. This is consistent with the expected type of nucleus activity of our Galaxy in the past \citep{Eckart2018, Ciurlo2025}. 

Thus, the close similarity of NGC 3521 to the Milky Way across numerous parameters — including those highlighted in our paper \citep{Vavilova2024} — makes it the most suitable candidate for constructing a reference SED of a Milky Way–like galaxy. A strict selection of Milky Way analogs according to these criteria, followed by detailed SED modeling, is essential for accurately reconstructing the Galaxy’s own SED. Such a study was performed by \cite{Fielder2021} with machine-learning techniques and a training sample of MWAs. However, their selection was based only on two global parameters — stellar mass and star-formation rate — which can bias the recovered SED of the Milky Way. Many galaxies share comparable stellar masses and SFRs but differ significantly in other key properties, e.g., their location in the cosmic web (denser environments, filaments, or voids), the presence or absence of a bar, and the level of nuclear activity. In turn, by applying isolation criteria, weak nuclear activity, and a small SMBH mass to achieve a more rigorous multi-parameter selection of MWAs \citep{Vavilova2024}, we minimize these biases.

We constructed an SED of NGC 3521 over a wider spectral range, incorporating observational data from radio meter to decameter wavelengths (Table \ref{tab:NGC3521_fluxes}, Fig. \ref{fig:UV_to_decametr}). 

The weak flux densities of NGC 3521 in the radio meter range were explained yet by \cite{Israel1990} as "the pervasive presence of a clumpy medium of well-mixed non-thermally emitting and thermally absorbing gas. For example, ionization of an interstellar medium with an electron temperature of 500 -- 1000 K and a clump density of 1 cm$^{-3}$ can be maintained by OB stars. Such a gas in the Milky Way should be located in the thick disc, mostly absent in the Galactic plane, and almost completely absent in the solar neighborhood".

To assess whether NGC~3521 can be considered a Milky Way galaxy-analogue or even twin by various multiwavelength properties, including in the decameter range, we need to estimate the flux density that our own Galaxy would have if placed at a distance of 10.7~Mpc. For the frequency 28~MHz and background brightness temperature of 20,000~K, 
$B_\nu(T) = 5.15\times10^{-20}$~W $\cdot$ m$^{-2}$·Hz$^{-1}$·sr$^{-1}$.  
The solid angle that our Galaxy would subtend at a distance of 10.7~Mpc, assuming a linear size of 32.4~kpc, is  $\Omega = 7.2\times10^{-6}$~sr. If we take into account the orientation at inclination angle of $ 72.7^{\circ}$, the solid angle will be $\Omega = 2.1\times10^{-6}$~sr. So, we obtain a flux density of 10.8 Jy for the Milky Way.

As can be seen, the upper limit on the flux density for NGC~3521 of 11.22~Jy at 28~MHz, and the expected flux density for our Galaxy of 10.8~Jy at 28~MHz if we place it at the same distance, are comparable and generally consistent. Thus, considering the estimated upper limits on the decameter flux density, the NGC~3521 galaxy can be regarded as similar to the Milky Way by flux density in the radio decameter range. This supports the classification of NGC~3521 as a Milky Way galaxy-analogue. Also, a comparison of the brightness temperatures of the Galactic background and NGC 3521 in the decameter range (20,000–30,000 K) further suggests that NGC 3521 lacks activity at its nucleus and is not a radio galaxy. Further studies of weak discrete radio sources at decameter wavelengths, with measured sensitivities up to several Jy, open a new window into the multiwavelength properties of MWAs and other galaxies.

\begin{figure}
    \centering
    \includegraphics[width=1.0\linewidth]{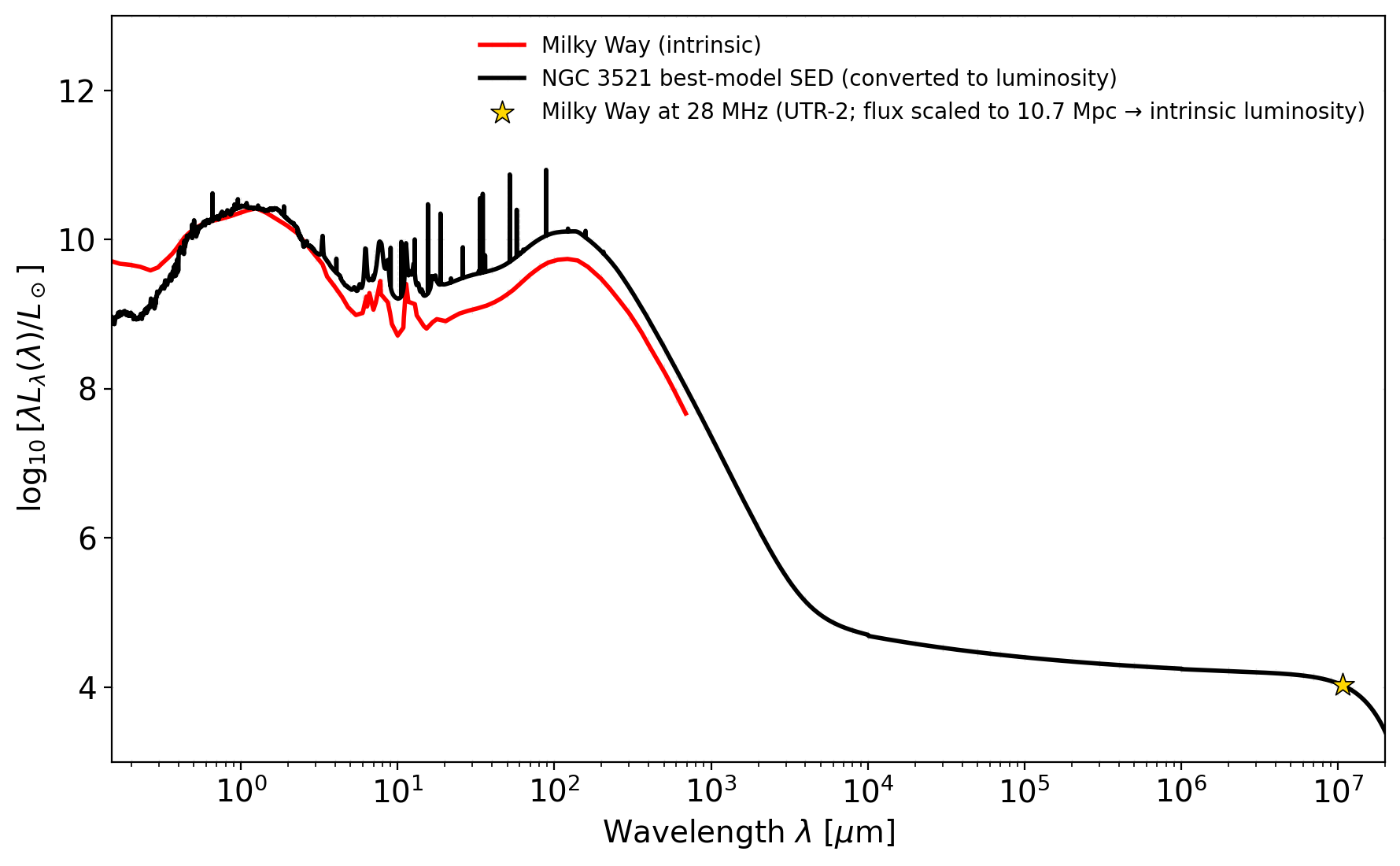}
    \caption{Multiwavelength SEDs of the Milky Way vs.\ NGC~3521 across UV to radio decameter ranges, shown as $\log_{10}\!\left[\lambda L_{\lambda}(\lambda)/L_{\odot}\right]$. Black curve: the best theoretical SED for NGC~3521 (Section~\ref{sec:CIGALE}), converted to intrinsic luminosity assuming $D=10.7 \mathrm{Mpc}$, and including 28 MHz UTR-2 measurement as a constraint. Red curve: the intrinsic SED of the Milky Way obtained by \cite{Natale2022} (originally provided as $L_{\nu}$) and recast to $\lambda L_{\lambda}/L_{\odot}$ without any distance scaling. Yellow star: Milky Way 28 MHz UTR-2 flux density scaled to $D=10.7 \mathrm{Mpc}$ and converted to intrinsic luminosity, plotted as $\lambda L_{\lambda}/L_{\odot}$.}
    \label{fig:MW-NGC3521_SED}
\end{figure}

\citet{Natale2022} derived an intrinsic, dust-corrected global SED of the Milky Way using a self-consistent 3D radiative-transfer (RT) model constrained by \emph{spatially resolved} all-sky surface-brightness data, rather than by a compilation of integrated fluxes. The geometry and amplitudes of the stellar and dust components were optimized by matching longitude and latitude profiles from the NIR to the sub-mm in a stepwise manner, exploiting the near-orthogonality of parameters that dominate different wavelength ranges (mid-IR for the compact star-forming component, sub-mm for the diffuse dust column, NIR for the old stellar disc and bulge). As they wrote, the resulting intrinsic SED is decomposed into old and young stellar components, as well as diffuse and clumpy dust emission. Within this framework, the UV output is dominated by the thin (young) stellar disc, while the optical is dominated by the old stellar disc plus bulge; the old stellar disc contributes $\sim 46\%$ of the total stellar luminosity. In the IR, the diffuse dust component dominates long wards of $50 \mu$m and accounts for $81\%$ of the total dust emission; overall, $(16\pm1)\%$ of the stellar luminosity is absorbed by dust and re-radiated in the IR/sub-mm, with young stellar populations (thin + inner thin discs) providing $71\%$ of the dust heating. The corresponding global parameters of the Milky Way are $\mathrm{SFR}=1.25\pm0.2 \mathrm{M_\odot yr^{-1}}$ (including $\mathrm{SFR}_{\rm in\text{-}tdisc}=0.25 \mathrm{M_\odot yr^{-1}}$), $M_\star=(4.9\pm0.3)\times10^{10} \mathrm{M_\odot}$, $\mathrm{sSFR}=(2.6\pm0.4)\times10^{-11} \mathrm{yr^{-1}}$, and $M_{\rm dust}=(4.78\pm0.06)\times10^{7} \mathrm{M_\odot}$.

We adopt the intrinsic SED model of the Milky Way from \citet{Natale2022} as a reference for comparison with our SED model of NGC~3521. To place both systems on the same footing, we compare them in intrinsic luminosity space, plotting the SEDs as $\log_{10}\!\left[\lambda L_{\lambda}(\lambda)/L_{\odot}\right]$ (Fig.~\ref{fig:MW-NGC3521_SED}). 

For the Milky Way, the reference model is provided as $L_{\nu}$ \citep{Natale2022}; we recast it to $\lambda L_{\lambda}=\nu L_{\nu}$ and normalize by $L_{\odot}$, without applying any distance scaling. 
For NGC~3521, we convert the best-fit model flux density $F_{\nu}$ to spectral luminosity via $L_{\nu}=4\pi D^{2}F_{\nu}$, adopting $D=10.7 \mathrm{Mpc}$, and then compute $\lambda L_{\lambda}=\nu L_{\nu}$ and $\lambda L_{\lambda}/L_{\odot}$. This representation allows as a direct comparison of the emitted power per logarithmic wavelength (or frequency) interval and facilitates comparisons with other galaxies without any rescaling to the distance of NGC~3521. We also included the Milky Way 28~MHz point in intrinsic luminosity by converting the UTR-2 flux density scaled to $D=10.7 \mathrm{Mpc}$ back to intrinsic luminosity using the same distance.

Figure~\ref{fig:MW-NGC3521_SED} demonstrates that the Milky Way and NGC~3521 have remarkably similar \emph{shapes} of their panchromatic SEDs, but systematically different \emph{amplitudes} in several key wavelength ranges. 
In the $\lambda L_{\lambda}/L_{\odot}$ representation, the two curves closely overlap across UV to the near-IR ($\lambda \sim 0.3{-}3~\mu$m), including the position and height of the stellar bump around $1~\mu$m. The overall shape of the dust and radio components is also similar: both SEDs exhibit a broad far-IR peak around $\lambda \sim 60{-}200~\mu$m and a smooth decline into the radio cm-- and decameter range. The Milky Way 28~MHz point lies very close to the extrapolated low-frequency tail of the NGC~3521 model. While the similarity of the MW and NGC3521 in terms of SED features and qualitative shape is indeed striking, there are notable differences.

In the mid- and far-IR, NGC~3521 is systematically brighter than the scaled Milky Way template, with stronger PAH features and a higher far-IR peak. A factor of $\gtrsim2$ in the dust peak corresponds to $\simeq0.3$ dex in luminosity normalization. In a broader population context, this is comparable to the intrinsic dispersion expected among local, massive star-forming discs: the width of the local star-forming ``main sequence'' is $\sim0.3$--0.45 dex from $M_\star\sim10^{10}$ to $10^{11}M_\odot$ \citep{Popesso2019}. The dust scaling relations for nearby star-forming galaxies show a similar $\sim$0.3--0.4 dex scatter in dust normalization, e.g., in $M_{\rm dust}/M_\star$) \citep{DeLooze2020}. Thus, while the offset is physically meaningful (reflecting a higher current dust luminosity and reprocessed output in NGC~3521), it is not exceptional relative to the expected scatter for galaxies with $L\sim10^{10}$--$10^{11}$. An additional, physically motivated caveat is that the Milky Way dust mass entering template-based comparisons may itself be underestimated by \cite{Natale2022}. 

For example, \cite{Ginolfi2018}, in their MW-sized halo modeling with the \textsc{GAMESH} pipeline, adopted an observed Milky Way dust mass of $M_{\rm dust}\sim1.5\times10^{8} M_\odot$. They found that AGB stars alone cannot build up more than $\sim3\times10^{7} M_\odot$ of dust, while SNe dominate the stellar dust production at all epochs. However, the matching $M_{\rm dust}\sim1.5\times10^{8} M_\odot$ with \emph{stellar sources only} has been required the assumption that the dust, formed in AGB/SN ejecta, is injected into the ISM without destruction by the SN reverse shock, which they regarded as unrealistic. Empirically, reverse-shock processing in well-studied remnants already removes $\sim10$--40\% of the initial dust mass \citep{Bocchio2016}. The scenario, in which reverse shock is a survival, was analyzed by \cite{Ginolfi2018}. These authors concluded that the net dust mass injected by stellar sources falls short of the Milky Way value by a factor of $\sim4$. In this context, part of the apparent MW--NGC~3521 far-IR mismatch could reflect systematics in the MW dust normalization adopted by the template, as well as genuine physical differences between the two galaxies.

The difference in the UV ranges is also due to the differing amounts of dust in the models of NGC 3521 and the MW. More dust attenuates stellar radiation, which in turn lowers the overall continuum of NGC 3521 relative to the Milky Way SED. At longer wavelengths, the radio continuum of NGC~3521 lies modestly above the scaled Milky Way SED over most of the radio cm range, again consistent with somewhat enhanced recent star formation and associated non-thermal emission. Even though the agreement at 28~MHz indicates that the two galaxies remain broadly comparable in their low-frequency radio output.

So, we may confidently count NGC~3521 as a near twin of the Milky Way, not only in its morphological type, structural parameters isolation and weak/absent nuclear activity \citep{Vavilova2024, Kompaniiets2025}, rotation curve \citep{McGaugh2016} and metallicity \citep{Pilyugin2023}, but also in its UV--to--radio SED shape and global properties (SFR, $M_\star$, $M_{\rm dust}$) within the expected scatter for Milky Way parameters. This result also supports our suggestion \citep{Vavilova2022} that SED shape, together with these global parameters, can serve as an additional indicator in the search for MW twins.

\section{Conclusions}
\label{sec:conclusions}

We determined the model SED of the spiral galaxy NGC 3521 in the widest range from UV to radio decameter wavelengths. In total, the SED measurements cover 27 photometric points, based on archival aperture imaging data from the UV to radio meter ranges and on our own observations in the decameter range. This provides comprehensive coverage of the key physical processes underlying the star formation history of NGC 3521, often considered the Milky Way twin.  

To provide this uniform observational data across the full wavelength range. We performed consistent aperture photometry from GALEX to VLA using a common elliptical isophotal aperture, producing integrated fluxes that are systematically higher in the UV--near-IR than earlier measurements based on smaller apertures \citep{Brown2014, Pastoven2024}. This homogenization reduces cross-band systematics and improves the robustness of the inferred physical parameters. To construct the SED over a broader range, we conducted observations at 28 MHz with the UTR-2 radio decameter telescope (Jan-Feb 2022), adopting a special technique and data processing. It allowed us to register an extremely weak signal and to estimate the upper limit on the flux density of 11.22 Jy. Moreover, we developed and implemented a dedicated radio module in the CIGALE software to ensure homogeneous scaling of data across the radiometer and decameter ranges, which were never used for the SED measurements of this galaxy. It is crucial because integrated galaxy spectra often show curvature and turnover below $\sim100$ MHz. 

We did not include Chandra X-ray data in the SED model of NGC 3521, given its LINER spectral type and weak nuclear activity. Instead, we analyzed central emission using ZTF and NEOWISE data of the 2014-2025 period. As a result, we found a clear bluer–when–brighter behavior of color–magnitude relations: the trend is detected in ZTF PSF-fit photometry, tracing the central $\lesssim3\arcsec$ region (as controlled by the seeing-quality cut), whereas in the MIR, a similar trend is present in the NEOWISE PSF-fit measurements (effective scale $\sim7.5\arcsec$), where the variability likely reflects a combination of the nuclear component and warm-dust emission from the inner region, see Fig.~\ref{fig:cmd_row_twocol}.

NGC 3521 is considered the Milky Way galaxy analog first and foremost by its main structural parameters, rotation curve, and metallicity. Additionally, given its LINER type of activity, small SMBH mass, location as an isolated system in the cosmic web, and well-sampled multiwavelength archival data, NGC 3521 is an excellent target for detailed SED modeling to compare its physical properties with those of the Milky Way (star and dust masses, star formation rate, star formation history). Our best baseline SED model gives $M_{\star}\simeq6\times10^{10} M_{\odot}$, ${\rm SFR}\simeq1.65 M_{\odot} {\rm yr^{-1}}$, and $M_{\rm dust}\simeq(1.3\pm0.1)\times10^{8} M_{\odot}$, with $T_{\rm dust}\sim23$ K. These values are consistent with independent multiwavelength estimates for NGC~3521 and with the baseline UV--radio cm model, while the extended wavelength coverage to radio meter and decameter range improves constraints on the star formation rate and dust mass.  The derived stellar mass and SFR are also comparable to modern estimates for the Milky Way. At the same time, in the specific MW--NGC~3521 comparison, the most noticeable residual offsets occur in the FUV and near the far-IR dust peak. These differences motivate future sample-based studies to investigate how the relative contributions of unobscured and dust-reprocessed emission relate to the evolutionary state of MW-like galaxies.

We demonstrate that the SED shapes and global parameters (SFR, stellar and dust masses) of our Galaxy and NGC~3521 are highly consistent, and conclude that they may serve as an additional indicator in the search for MWAs. In turn, it helps to extrapolate how MW properties appear to an extragalactic observer. 

\begin{acknowledgements}

This study was supported by the NRF of Ukraine (project 2023.03/0188). We consider our results in the decameter radio range to be a contribution to the "LOFAR 2.0 Ultra-Deep Observation" (LUDO) collaboration. Kompaniiets O.V. is sincerely grateful to Prof. Agnieszka Pollo and Prof. Katarzyna Małek (NCNR, Poland) for the introductory overview of CIGALE, and to NAWA for the PROM grant. We thank Prof. Johan Knapen and Prof. John Beckman (IAC, Spain) for fruitful seminars and discussions in March 2025. Kompaniiets O.V. expresses sincere gratitude to Prof. Michael Blanton (SDSS, USA) for useful remarks on the methodology for working with full-aperture SDSS data, which he provided in private e-mail correspondence \citep{Blanton}. The work by Junais at the IAC and visits by Kompaniiets O.V., Izviekova O.I and Dobrycheva D.V. to the IAC in June 2025 are co-funded by the European Union (MSCA Doctoral Network EDUCADO, GA 101119830 and Widening Participation, ExGal-Twin, GA 101158446). The authors thank the reviewer for useful comments that helped us to more fully present the results of our research. This work uses publicly available data from GALEX, SDSS, NEOWISE, ZTF, NRAO, and IRSA archives. The SAO/NASA ADS was very helpful in our research.

\end{acknowledgements}

\bibliographystyle{aa} 
\bibliography{aa58597-25} 

@ARTICLE{Andrews2024,
       author = {{Andrews}, J. and {Sand}, D.~J. and {Bostroem}, K.~A. and {Valenti}, S. and {Jha}, S. and {Lundquist}, M. and {Hoang}, E. and {Shrestha}, M. and {Dong}, Y. and {Janzen}, D. and {Hosseinzadeh}, G. and {Pearson}, J. and {Meza}, N. and {Metha}, D. and {Ravi}, A. and {Martas}, A. and {Subrayan}, B.},
        title = "{DLT40 Transient Classification Report for 2024-12-17}",
      journal = {Transient Name Server Classification Report},
         year = 2024,
        month = dec,
       volume = {2024-4943},
        pages = {1},
       adsurl = {https://ui.adsabs.harvard.edu/abs/2024TNSCR4943....1A}
}

@misc{Blanton,
  author = {Michael Blanton, SDSS},
  howpublished = {Private Communication},
  note = {Email to the authors, outlining suggections on the SDSS aperture photometry and zero points},
  year = {2025},
  month = {May},
  day = {15}
}

@techreport{Cutri2012WISE,
  author       = {Cutri, R.~M. and Wright, E.~L. and Conrow, T. and Fowler, J.~W. and Eisenhardt, P.~R.~M. and Grillmair, C. and Kirkpatrick, J.~D. and Masci, F. and McCallon, H. and Wheelock, S. and Fajardo-Acosta, S. and Yan, L. and Lake, S. and Mainzer, A.},
  year         = {2012},
  title        = {Explanatory Supplement to the WISE All-Sky Data Release Products},
  institution  = {IPAC, California Institute of Technology},
  address      = {Pasadena, CA},
  url          = {https://wise2.ipac.caltech.edu/docs/release/allsky/expsup/}
}

@techreport{Cutri2015NEOWISE,
  author       = {Cutri, R.~M. and Mainzer, A.~K. and Conrow, T. and Fowler, J.~W. and Bauer, J. and Kirkpatrick, J.~D. and Masci, F.~J. and Dailey, J. and Wright, E.~L.},
  year         = {2015},
  title        = {Explanatory Supplement to the NEOWISE Data Release Products},
  institution  = {IPAC, California Institute of Technology},
  address      = {Pasadena, CA},
  url          = {https://wise2.ipac.caltech.edu/docs/release/neowise/expsup/}
}

@ARTICLE{DESI2023,
       author = {{Moustakas}, John and {Lang}, Dustin and {Dey}, Arjun and {Juneau}, St{\'e}phanie and {Meisner}, Aaron and {Myers}, Adam D. and {Schlafly}, Edward F. and {Schlegel}, David J. and {Valdes}, Francisco and {Weaver}, Benjamin A. and {Zhou}, Rongpu},
        title = "{Siena Galaxy Atlas 2020}",
      journal = {ApJS},
         year = 2023,
        month = nov,
       volume = {269},
       number = {1},
          eid = {3},
        pages = {3},
          doi = {10.3847/1538-4365/acfaa2},
archivePrefix = {arXiv},
       eprint = {2307.04888},
 primaryClass = {astro-ph.GA},
       adsurl = {https://ui.adsabs.harvard.edu/abs/2023ApJS..269....3M}
}

@ARTICLE{DiValentino,
       author = {{Di Valentino}, Eleonora and {Said}, Jackson Levi and {Riess}, Adam and {Pollo}, Agnieszka and {Poulin}, Vivian and {G{\'o}mez-Valent}, Adri{\`a} and {Weltman}, Amanda and {Palmese}, Antonella and {Huang}, Caroline D. and {van de Bruck}, Carsten and {Saraf}, Chandra Shekhar and {Kuo}, Cheng-Yu and {Uhlemann}, Cora and {Grand{\'o}n}, Daniela and {Paz}, Dante and {Eckert}, Dominique and {Teixeira}, Elsa M. and {Saridakis}, Emmanuel N. and {Colg{\'a}in}, Eoin {\'O}. and {Beutler}, Florian and {Niedermann}, Florian and {Bajardi}, Francesco and {Barenboim}, Gabriela and {Gubitosi}, Giulia and {Musella}, Ilaria and {Banik}, Indranil and {Szapudi}, Istvan and {Singal}, Jack and {Cases}, Jaume Haro and {Chluba}, Jens and {Torrado}, Jes{\'u}s and {Mifsud}, Jurgen and {Jedamzik}, Karsten and {Said}, Khaled and {Dialektopoulos}, Konstantinos and {Herold}, Laura and {Perivolaropoulos}, Leandros and {Zu}, Lei and {Galbany}, Llu{\'\i}s and {Breuval}, Louise and {Visinelli}, Luca and {Escamilla}, Luis A. and {Anchordoqui}, Luis A. and {Sheikh-Jabbari}, M.~M. and {Lembo}, Margherita and {Dainotti}, Maria Giovanna and {Vincenzi}, Maria and {Asgari}, Marika and {Gerbino}, Martina and {Forconi}, Matteo and {Cantiello}, Michele and {Moresco}, Michele and {Benetti}, Micol and {Sch{\"o}neberg}, Nils and {Akarsu}, {\"O}zg{\"u}r and {Nunes}, Rafael C. and {Bernardo}, Reginald Christian and {Ch{\'a}vez}, Ricardo and {Anderson}, Richard I. and {Watkins}, Richard and {Capozziello}, Salvatore and {Li}, Siyang and {Vagnozzi}, Sunny and {Pan}, Supriya and {Treu}, Tommaso and {Irsic}, Vid and {Handley}, Will and {Giar{\`e}}, William and {Murakami}, Yukei and {Banihashemi}, Abdolali and {Poudou}, Ad{\`e}le and {Heavens}, Alan and {Kogut}, Alan and {Domi}, Alba and {Lenart}, Aleksander {\L}ukasz and {Melchiorri}, Alessandro and {Vadal{\`a}}, Alessandro and {Amon}, Alexandra and {Rivera}, Alexander Bonilla and {Reeves}, Alexander and {Zhuk}, Alexander and {Bonanno}, Alfio and {{\"O}vg{\"u}n}, Ali and {Pisani}, Alice and {Talebian}, Alireza and {Abebe}, Amare and {Aboubrahim}, Amin and {Gonz{\'a}lez Mor{\'a}n}, Ana Luisa and {Kov{\'a}cs}, Andr{\'a}s and {Lymperis}, Andreas and {Papatriantafyllou}, Andreas and {Liddle}, Andrew R. and {Paliathanasis}, Andronikos and {Borowiec}, Andrzej and {Yadav}, Anil Kumar and {Yadav}, Anita and {Sen}, Anjan Ananda and {William}, Anjitha John and {Davis}, Anne Christine and {Shajib}, Anowar J. and {Walters}, Anthony and {Lonappan}, Anto Idicherian and {Chudaykin}, Anton and {Capodagli}, Antonio and {da Silva}, Antonio and {De Felice}, Antonio and {Racioppi}, Antonio and {Oficial}, Araceli Soler and {Montiel}, Ariadna and {Favale}, Arianna and {Bernui}, Armando and {Velasco}, Arrianne Crystal and {Heinesen}, Asta and {Bakopoulos}, Athanasios and {Chatzistavrakidis}, Athanasios and {Khanpour}, Bahman and {Sathyaprakash}, Bangalore S. and {Zgirski}, Bartek and {L'Huillier}, Benjamin and {Famaey}, Benoit and {Jain}, Bhuvnesh and {Zhang}, Bing and {Karmakar}, Biswajit and {Dragovich}, Branko and {Thomas}, Brooks and {Correa}, Carlos and {Boiza}, Carlos G. and {Marques}, Catarina and {Escamilla-Rivera}, Celia and {Tzerefos}, Charalampos and {Zhang}, Chi and {De Leo}, Chiara and {Pfeifer}, Christian and {Lee}, Christine and {Venter}, Christo and {Gomes}, Cl{\'a}udio and {Roque De bom}, Clecio and {Moreno-Pulido}, Cristian and {Iosifidis}, Damianos and {Grin}, Dan and {Blixt}, Daniel and {Scolnic}, Dan and {Oriti}, Daniele and {Dobrycheva}, Daria and {Bettoni}, Dario and {Benisty}, David and {Fern{\'a}ndez-Arenas}, David and {Wiltshire}, David L. and {Sanchez Cid}, David and {Tamayo}, David and {Valls-Gabaud}, David and {Pedrotti}, Davide and {Wang}, Deng and {Staicova}, Denitsa and {Totolou}, Despoina and {Rubiera-Garcia}, Diego and {Milakovi{\'c}}, Dinko and {Pesce}, Dominic W. and {Sluse}, Dominique and {Borka}, Du{\v{s}}ko and {Yusofi}, Ebrahim and {Giusarma}, Elena and {Terlevich}, Elena and {Tomasetti}, Elena and {Vagenas}, Elias C. and {Fazzari}, Elisa and {Ferreira}, Elisa G.~M. and {Barakovic}, Elvis and {Dimastrogiovanni}, Emanuela and {Holm}, Emil Brinch and {Mottola}, Emil and {{\"O}z{\"u}lker}, Emre and {Specogna}, Enrico and {Brocato}, Enzo and {Jensko}, Erik and {Enriquez}, Erika Antonette and {Bhatia}, Esha and {Bresolin}, Fabio and {Avila}, Felipe and {Bouch{\`e}}, Filippo and {Bombacigno}, Flavio and {Anagnostopoulos}, Fotios K. and {Pace}, Francesco and {Sorrenti}, Francesco and {Lobo}, Francisco S.~N. and {Courbin}, Fr{\'e}d{\'e}ric and {Hansen}, Frode K. and {Sloan}, Greg and {Farrugia}, Gabriel and {Lynch}, Gabriel and {Garcia-Arroyo}, Gabriela and {Raimondo}, Gabriella and {Lambiase}, Gaetano and {Anand}, Gagandeep S. and {Poulot}, Gaspard and {Leon}, Genly and {Kouniatalis}, Gerasimos and {Nardini}, Germano and {Cs{\"o}rnyei}, G{\'e}za and {Galloni}, Giacomo},
        title = "{The CosmoVerse White Paper: Addressing observational tensions in cosmology with systematics and fundamental physics}",
      journal = {Physics of the Dark Universe},
         year = 2025,
        month = sep,
       volume = {49},
          eid = {101965},
        pages = {101965},
          doi = {10.1016/j.dark.2025.101965},
archivePrefix = {arXiv},
       eprint = {2504.01669},
 primaryClass = {astro-ph.CO},
       adsurl = {https://ui.adsabs.harvard.edu/abs/2025PDU....4901965D}
}

@article{Galex,
  author       = {Morrissey, P. and Conrow, T. and Barlow, T. A. and Small, T. and Seibert, M. and Wyder, T. K. and Budav{\'a}ri, T. and Arnouts, S. and Friedman, P. G. and others},
  title        = {The Calibration and Data Products of the Galaxy Evolution Explorer},
  journal      = {ApJS},
  year         = {2007},
  volume       = {173},
  pages        = {682--697},
  doi          = {10.1086/520512}
}

@article{H0DN,
  author  = {{H0DN Collaboration} and Casertano, S. and Anand, G. and Anderson, R. I. and Beaton, R. and Bhardwaj, A. and Blakeslee, J. P. and Boubel, P. and Breuval, L. and Brout, D. and Cantiello, M. and Cruz Reyes, M. and Cs{\"o}rnyei, G. and de Jaeger, T. and Dhawan, S. and Di Valentino, E. and Galbany, L. and Gil-Mar{\'\i}n, H. and Graczyk, D. and Huang, C. and Jensen, J. B. and Kervella, P. and Leibundgut, B. and Lengen, B. and Li, S. and Macri, L. and {\"O}z{\"u}lker, E. and Pesce, D. W. and Riess, A. and Romaniello, M. and Said, K. and Sch{\"o}neberg, N. and Scolnic, D. and Sicignano, T. and Skowron, D. M. and Uddin, S. A. and Verde, L. and Nota, A.},
  title   = {The Local Distance Network: a community consensus report on the measurement of the Hubble constant at 1\% precision},
  journal = {arXiv e-prints},
  year    = {2025},
  month   = oct,
  pages   = {arXiv:2510.23823},
  doi     = {10.48550/arXiv.2510.23823},
  eprint  = {2510.23823},
  archivePrefix = {arXiv},
  primaryClass  = {astro-ph.CO},
  adsurl  = {https://ui.adsabs.harvard.edu/abs/2025arXiv251023823H}
}

@article{Hersel,
  author       = {Kennicutt, R. C. Jr. and Calzetti, D. and Aniano, G. and Appleton, P. and Armus, L. and Beir{\~a}o, P. and Bolatto, A. D. and Brandl, B. and Crocker, A. and others},
  title        = {KINGFISH — Key Insights on Nearby Galaxies: A Far-Infrared Survey with Herschel},
  journal      = {PASP},
  year         = {2011},
  volume       = {123},
  pages        = {1347--1369},
  doi          = {10.1086/663818}
}

@article{NVSS,
  author       = {Condon, J. J. and Cotton, W. D. and Greisen, E. W. and Yin, Q. F. and Perley, R. A. and Taylor, G. B. and Broderick, J. J.},
  title        = {The NRAO VLA Sky Survey},
  journal      = {AJ},
  year         = {1998},
  volume       = {115},
  pages        = {1693--1716},
  doi          = {10.1086/300337}
}

@article{SDSS,
  author       = {Ahn, C. P. and Alexandroff, R. and Allende Prieto, C. and Anders, F. and Anderson, S. F. and Anderton, T. and Andrews, B. H. and others},
  title        = {The Ninth Data Release of the Sloan Digital Sky Survey: First Spectroscopic Data from the SDSS-III Baryon Oscillation Spectroscopic Survey},
  journal      = {ApJS},
  year         = {2012},
  volume       = {203},
  pages        = {21},
  doi          = {10.1088/0067-0049/203/2/21}
}

@article{Spitzer,
  author       = {Kennicutt, R. C. Jr. and Armus, L. and Bendo, G. and Calzetti, D. and Dale, D. A. and Draine, B. T. and Engelbracht, C. W. and Gordon, K. D. and Grauer, A. D. and others}, 
  title        = {SINGS: The SIRTF Nearby Galaxies Survey},
  journal      = {PASP},
  year         = {2003},
  volume       = {115},
  pages        = {928--952},
  doi          = {10.1086/376941}
}

@article{Wise,
  author       = {Wright, E. L. and Eisenhardt, P. R. M. and Mainzer, A. K. and Ressler, M. E. and Cutri, R. M. and Jarrett, T. and Kirkpatrick, J. D. and Padgett, D. and others},
  title        = {The Wide-field Infrared Survey Explorer (WISE): Mission Description and Initial On-orbit Performance},
  journal      = {AJ},
  year         = {2010},
  volume       = {140},
  pages        = {1868--1881},
  doi          = {10.1088/0004-6256/140/6/1868}
}

@ARTICLE{Albergaria2025,
       author = {{Albergaria}, Danilo and {Frantseva}, Kateryna and {Russo}, Pedro and {Babiichuk}, Svitlana and {Berezhna}, Oksana and {Denyshchenko}, Sofiia and {Dobrycheva}, Daria and {Kaydash}, Vadym and {Kompaniiets}, Olena and {Konovalenko}, Oleksander and {Kulinich}, Yurii and {Lukyanyk}, Igor and {Marsakova}, Vladyslava and {Novosyadlyj}, Bohdan and {Panko}, Elena and {Reshetnyk}, Volodymyr and {Slyusarev}, Ivan and {Sushch}, Iurii and {Tolstanova}, Ganna and {Vavilova}, Iryna and {Yankiv-Vitkovska}, Liubov and {Yatskiv}, Yaroslav and {Zakharenko}, Vyacheslav},
        title = "{Ukrainian wartime astronomy and its prospects}",
      journal = {Nature Astronomy},
         year = 2025,
        month = sep,
          doi = {10.1038/s41550-025-02657-w},
       adsurl = {https://ui.adsabs.harvard.edu/abs/2025NatAs.tmp..195A}
}

@article{Astropy2013,
author  = {Astropy Collaboration and Robitaille, T. P. and Tollerud, E. J. and Greenfield, P. and Droettboom, M. and Bray, E. and Aldcroft, T. and Davis, M. and Ginsburg, A. and Price-Whelan, A. M. and Kerzendorf, W. E. and Conley, A. and Crighton, N. and Barbary, K. and Muna, D. and Ferguson, H. and Grollier, F. and Parikh, M. M. and Nair, P. H. and G{\"u}nther, H. M. and Deil, C. and Woillez, J. and Conseil, S. and Kramer, R. and Turner, J. E. H. and Singer, L. and Fox, R. and Weaver, B. A. and Zabalza, V. and Edwards, Z. I. and Azalee Bostroem, K. and Burke, D. J. and Casey, A. R. and Crawford, S. M. and Dencheva, N. and Ely, J. and Jenness, T. and Labrie, K. and Lim, P. L. and Pierfederici, F. and Pontzen, A. and Ptak, A. and Refsdal, B. and Servillat, M. and Streicher, O.},
title   = {Astropy: A community Python package for astronomy},
journal = {A\&A},
volume  = {558},
pages   = {A33},
year    = {2013},
doi     = {10.1051/0004-6361/201322068}
}

@article{Astropy2018,
author  = {Astropy Collaboration and Price-Whelan, A. M. and Sip{\H{o}}cz, B. M. and G{\"u}nther, H. M. and Lim, P. L. and Crawford, S. M. and Conseil, S. and Shupe, D. L. and Craig, M. W.  and Dencheva, N. and Ginsburg, A. and VanderPlas, J. T. and Bradley, L. D. and P{\'e}rez-Su{\'a}rez, D. and De Val-Borro, M. and Aldcroft, T. L. and Cruz, K. L. and Robitaille, T. P. and Tollerud, E. J.},
title   = {The Astropy Project: Building an open-science project and status of the v2.0 core package},
journal = {AJ},
volume  = {156},
number  = {3},
pages   = {123},
year    = {2018},
doi     = {10.3847/1538-3881/aabc4f}
}

@ARTICLE{Astroquery2019,
       author = {{Ginsburg}, Adam and {Sip{\H{o}}cz}, Brigitta M. and {Brasseur}, C.~E. and {Cowperthwaite}, Philip S. and {Craig}, Matthew W. and {Deil}, Christoph and {Guillochon}, James and {Guzman}, Giannina and {Liedtke}, Simon and {Lian Lim}, Pey and {Lockhart}, Kelly E. and {Mommert}, Michael and {Morris}, Brett M. and {Norman}, Henrik and {Parikh}, Madhura and {Persson}, Magnus V. and {Robitaille}, Thomas P. and {Segovia}, Juan-Carlos and {Singer}, Leo P. and {Tollerud}, Erik J. and {de Val-Borro}, Miguel and {Valtchanov}, Ivan and {Woillez}, Julien and {Astroquery Collaboration} and {a subset of astropy Collaboration}},
        title = "{astroquery: An Astronomical Web-querying Package in Python}",
      journal = {\aj},
         year = 2019,
        month = mar,
       volume = {157},
       number = {3},
          eid = {98},
        pages = {98},
          doi = {10.3847/1538-3881/aafc33},
archivePrefix = {arXiv},
       eprint = {1901.04520},
 primaryClass = {astro-ph.IM},
       adsurl = {https://ui.adsabs.harvard.edu/abs/2019AJ....157...98G}
}

@article{Baldwin1981,
  author = {Baldwin, J. A. and Phillips, M. M. and Terlevich, R.},
  title = {Classification parameters for the emission-line spectra of extragalactic objects},
  journal = {PASP},
  volume = {93},
  pages = {5--19},
  year = {1981},
  doi = {10.1086/130766}
}

@ARTICLE{Banik2022,
       author = {{Banik}, Indranil and {Thies}, Ingo and {Truelove}, Roy and {Candlish}, Graeme and {Famaey}, Benoit and {Pawlowski}, Marcel S. and {Ibata}, Rodrigo and {Kroupa}, Pavel},
        title = "{3D hydrodynamic simulations for the formation of the Local Group satellite planes}",
      journal = {MNRAS},
         year = 2022,
        month = jun,
       volume = {513},
       number = {1},
        pages = {129-158},
          doi = {10.1093/mnras/stac722},
archivePrefix = {arXiv},
       eprint = {2204.09687},
 primaryClass = {astro-ph.GA},
       adsurl = {https://ui.adsabs.harvard.edu/abs/2022MNRAS.513..129B}
}

@article{Barbary2016SEP,
author  = {Barbary, K.},
title   = {SEP: Source Extractor as a library},
journal = {JOSS},
volume  = {1},
number  = {6},
pages   = {58},
year    = {2016},
doi     = {10.21105/joss.00058}
}

@book{Bevington2003,
  author    = {Bevington, Philip R. and Robinson, D. Keith},
  title     = {Data Reduction and Error Analysis for the Physical Sciences},
  edition   = {3},
  year      = {2003},
  publisher = {McGraw-Hill},
  address   = {New York},
  isbn      = {978-0-07-247227-1}
}

@ARTICLE{Boardman2020a,
       author = {{Boardman}, N. and {Zasowski}, G. and {Seth}, A. and {Newman}, J. and {Andrews}, B. and {Bershady}, M. and {Bird}, J. and {Chiappini}, C. and {Fielder}, C. and {Fraser-McKelvie}, A. and {Jones}, A. and {Licquia}, T. and {Masters}, K.~L. and {Minchev}, I. and {Schiavon}, R.~P. and {Brownstein}, J.~R. and {Drory}, N. and {Lane}, R.~R.},
        title = "{Milky Way analogues in MaNGA: multiparameter homogeneity and comparison to the Milky Way}",
      journal = {MNRAS},
         year = 2020,
        month = jan,
       volume = {491},
       number = {3},
        pages = {3672-3701},
          doi = {10.1093/mnras/stz3126},
archivePrefix = {arXiv},
       eprint = {1910.12896},
 primaryClass = {astro-ph.GA},
       adsurl = {https://ui.adsabs.harvard.edu/abs/2020MNRAS.491.3672B}
}

@ARTICLE{Boardman2020b,
       author = {{Boardman}, N. and {Zasowski}, G. and {Newman}, J.~A. and {Andrews}, B. and {Fielder}, C. and {Bershady}, M. and {Brinkmann}, J. and {Drory}, N. and {Krishnarao}, D. and {Lane}, R.~R. and {Mackereth}, T. and {Masters}, K. and {Stringfellow}, G.~S.},
        title = "{Are the Milky Way and Andromeda unusual? A comparison with Milky Way and Andromeda analogues}",
      journal = {MNRAS},
         year = 2020,
        month = nov,
       volume = {498},
       number = {4},
        pages = {4943-4954},
          doi = {10.1093/mnras/staa2731},
archivePrefix = {arXiv},
       eprint = {2009.02576},
 primaryClass = {astro-ph.GA},
       adsurl = {https://ui.adsabs.harvard.edu/abs/2020MNRAS.498.4943B}
}

@article{Boquien2019,
  author       = {Boquien, M. and Burgarella, D. and Roehlly, Y. and Corre, D. and L\'{e}rasle, M. and Ma\l{}ek, K. and Vidal-Garc{\'i}a, A.},
  year         = {2019},
  title        = {CIGALE: a Python Code Investigating Galaxy Emission},
  journal      = {A\&A},
  volume       = {622},
  pages        = {A103},
  doi          = {10.1051/0004-6361/201834156}
}

@article{Bocchio2016,
  author  = {Bocchio, M. and Marassi, S. and Schneider, R. and Bianchi, S. and Limongi, M. and Chieffi, A.},
  title   = {Dust grains from the heart of supernovae},
  journal = {A\&A},
  year    = {2016},
  volume  = {587},
  pages   = {A157},
  doi     = {10.1051/0004-6361/201527432}
}

@misc{Bradley2016Photutils,
       author = {{Bradley}, Larry and {Sipocz}, Brigitta and {Robitaille}, Thomas and {Tollerud}, Erik and {Deil}, Christoph and {Vin{\'\i}cius}, Z{\`e} and {Barbary}, Kyle and {G{\"u}nther}, Hans Moritz and {Bostroem}, Azalee and {Droettboom}, Michael and {Bray}, Erik and {Bratholm}, Lars Andersen and {Pickering}, T.~E. and {Craig}, Matt and {Pascual}, Sergio and {Greco}, Johnny and {Donath}, Axel and {Kerzendorf}, Wolfgang and {Littlefair}, Stuart and {Barentsen}, Geert and {D'Eugenio}, Francesco and {Weaver}, Benjamin Alan},
        title = "{Photutils: Photometry tools}",
 howpublished = {Astrophysics Source Code Library, record ascl:1609.011},
         year = 2016,
        month = sep,
          eid = {ascl:1609.011},
archivePrefix = {ascl},
       eprint = {1609.011},
       adsurl = {https://ui.adsabs.harvard.edu/abs/2016ascl.soft09011B}
}

@ARTICLE{Braude1979,
       author = {{Braude}, S. Ia. and {Megn}, A.~V. and {Sokolov}, K.~P. and {Tkachenko}, A.~P. and {Sharykin}, N.~K.},
        title = "{Decametric survey of discrete sources in the northern sky: V. Source catalogue in the range of declinations 0{\textdegree} to +10{\textdegree}}",
      journal = {Astrophys. Space Sci.},
         year = 1979,
        month = aug,
       volume = {64},
       number = {1},
        pages = {73-126},
          doi = {10.1007/BF00640035}
}

@article{Brown2014,
  author       = {Brown, M. J. I. and Armus, L. and Calvin, D. E. and et al.},
  title        = {An Atlas of Galaxy Spectral Energy Distributions from the Ultraviolet to the Mid-infrared},
  journal      = {ApJS},
  year         = {2014},
  volume       = {212},
  number       = {2},
  pages        = {18},
  doi          = {10.1088/0067-0049/212/2/18}
}

@article{Bruzual2003,
author  = {Bruzual, G. and Charlot, S.},
title   = {Stellar population synthesis at the resolution of 2003},
journal = {MNRAS},
year    = {2003},
volume  = {344},
pages   = {1000--1028}
}

@article{Burbidge1964,
author = {Burbidge, E. M. and Burbidge, G. R. and Prendergast, K. H.},
year = {1964},
title = {The Rotation and Mass of NGC 3521},
journal = {AJ},
volume = {139},
pages = {1058},
doi = {10.1086/147845}
}

@article{Calzetti2000,
author  = {Calzetti, D. and Armus, L. and Bohlin, R. C. and et al.},
title   = {The Dust Content and Opacity of Actively Star-forming Galaxies},
journal = {ApJ},
year    = {2000},
volume  = {533},
pages   = {682--695}
}

@ARTICLE{Cane1979,
       author = {{Cane}, H.~V.},
        title = "{Spectra of the non-thermal radio radiation from the galactic polar regions.}",
      journal = {MNRAS},
         year = 1979,
        month = nov,
       volume = {189},
        pages = {465-478},
          doi = {10.1093/mnras/189.3.465},
       adsurl = {https://ui.adsabs.harvard.edu/abs/1979MNRAS.189..465C}
}

@article{Cardelli1989,
author  = {Cardelli, J. A. and Clayton, G. C. and Mathis, J. S.},
title   = {The relationship between infrared, optical, and ultraviolet extinction},
journal = {ApJ},
volume  = {345},
pages   = {245--256},
year    = {1989},
doi     = {10.1086/167900}
}

@article{Casertano1991,
author = {Casertano, S. and van Gorkom, J. H.},
year = {1991},
title = {Declining rotation curves — The end of a conspiracy?},
journal = {AJ},
volume = {101},
pages = {1231--1241}
}

@article{Casey2012,
  author       = {Casey, C.M.},
  title        = {Far‑infrared Spectral Energy Distribution Fitting for Galaxies},
  journal      = {MNRAS},
  volume       = {425},
  number       = {4},
  pages        = {3094--3103},
  year         = {2012},
  doi          = {10.1111/j.1365-2966.2012.21455.x},
  url          = {https://doi.org/10.1111/j.1365-2966.2012.21455.x}
}

@ARTICLE{Chabrier2003,
       author = {{Chabrier}, Gilles},
        title = "{Galactic Stellar and Substellar Initial Mass Function}",
      journal = {PASP},
         year = 2003,
        month = jul,
       volume = {115},
       number = {809},
        pages = {763-795},
          doi = {10.1086/376392},
archivePrefix = {arXiv},
       eprint = {astro-ph/0304382},
 primaryClass = {astro-ph},
       adsurl = {https://ui.adsabs.harvard.edu/abs/2003PASP..115..763C}
}

@ARTICLE{Chang2020,
       author = {{Chang}, Zhengxue and {Zhou}, Jianjun and {Wilson}, Christine D. and {Esimbek}, Jarken and {Qiu}, Jianjie and {Li}, Dalei and {Zhou}, Minhua and {He}, Yuxin and {Ji}, Weiguang and {Tang}, Xindi and {Wu}, Gang and {Li}, Jun},
        title = "{Dissecting the Global Cold Dust Properties and Possible Submillimeter Excess of 13 Nearby Spiral Galaxies from the NGLS}",
      journal = {ApJ},
         year = 2020,
        month = sep,
       volume = {900},
       number = {1},
          eid = {53},
        pages = {53},
          doi = {10.3847/1538-4357/aba52f},
       adsurl = {https://ui.adsabs.harvard.edu/abs/2020ApJ...900...53C}
}

@article{Chyzy2018,
  author  = {Chy{\.z}y, K. T. and Jurusik, W. and Piotrowska, J. and et al.},
  year    = {2018},
  title   = {LOFAR MSSS: Flattening low-frequency radio continuum spectra of nearby galaxies},
  journal = {A\&A},
  volume  = {619},
  pages   = {A36},
  doi     = {10.1051/0004-6361/201833133}
}

@book{Ciurlo2025,
       author = {{Ciurlo}, Anna and {Morris}, Mark R.},
        title = "{Sagittarius A* -- The Milky Way Supermassive Black Hole}",
      publisher = {submited to Elsevier,  chapter for the Encyclopedia of Astrophysics (edited by I. Mandel, section editor S. McGee)},
         year = 2025,
        month = mar,
          eid = {arXiv:2503.20081},
        pages = { },
          doi = {10.48550/arXiv.2503.20081},
archivePrefix = {arXiv},
       eprint = {2503.20081},
 primaryClass = {astro-ph.GA},
       adsurl = {https://ui.adsabs.harvard.edu/abs/2025arXiv250320081C}
}

@article{Coccato2018,
author = {Coccato, L. and Fabricius, M. and Saglia, R. P. and others},
year = {2018},
title = {Spectroscopic decomposition of NGC 3521: unveiling the properties of the bulge and disc},
journal = {MNRAS},
volume = {477},
number = {2},
pages = {1958--1969},
doi = {10.1093/mnras/sty705}
}

@article{Condon1992,
  author  = {Condon, J. J.},
  year    = {1992},
  title   = {Radio Emission from Normal Galaxies},
  journal = {ARA\&A},
  volume  = {30},
  pages   = {575--611},
  doi     = {10.1146/annurev.aa.30.090192.003043}
}

@article{Do2019,
  author = {Do, T. and Witzel, G. and Gautam, A. K. and Chen, Z. and Ghez, A. M. and Morris, M. and Becklin, E. E. and Ciurlo, A. and Hosek, M. W. and Martinez, G. D. and Matthews, K. and Sakai, S. and Sch{\"o}del, R.},
  title = {Unprecedented Near-infrared Brightness and Variability of Sgr A*},
  journal = {ApJL},
  volume = {882},
  number = {2},
  year = {2019},
  pages = {L27},
  doi = {10.3847/2041-8213/ab38c3},
  url = {https://ui.adsabs.harvard.edu/abs/2019ApJ...882L..27D}
}

@ARTICLE{DeLooze2020,
       author = {{De Looze}, I. and {Lamperti}, I. and {Saintonge}, A. and {Rela{\~n}o}, M. and {Smith}, M.~W.~L. and {Clark}, C.~J.~R. and {Wilson}, C.~D. and {Decleir}, M. and {Jones}, A.~P. and {Kennicutt}, R.~C. and {Accurso}, G. and {Brinks}, E. and {Bureau}, M. and {Cigan}, P. and {Clements}, D.~L. and {De Vis}, P. and {Fanciullo}, L. and {Gao}, Y. and {Gear}, W.~K. and {Ho}, L.~C. and {Hwang}, H.~S. and {Micha{\l}owski}, M.~J. and {Lee}, J.~C. and {Li}, C. and {Lin}, L. and {Liu}, T. and {Lomaeva}, M. and {Pan}, H.-A. and {Sargent}, M. and {Williams}, T. and {Xiao}, T. and {Zhu}, M.},
        title = "{JINGLE - IV. Dust, H I gas, and metal scaling laws in the local Universe}",
      journal = {\mnras},
         year = 2020,
        month = aug,
       volume = {496},
       number = {3},
        pages = {3668-3687},
          doi = {10.1093/mnras/staa1496},
archivePrefix = {arXiv},
       eprint = {2006.01856},
 primaryClass = {astro-ph.GA},
       adsurl = {https://ui.adsabs.harvard.edu/abs/2020MNRAS.496.3668D}
}

@article{Dale2012,
  author       = {Dale, D. A. and Aniano, G. and Engelbracht, C. W. and et al.},
  title        = {Herschel Far-infrared and Submillimeter Photometry for the KINGFISH Sample of Nearby Galaxies},
  journal      = {ApJ},
  year         = {2012},
  volume       = {745},
  number       = {1},
  pages        = {95},
  doi          = {10.1088/0004-637X/745/1/95}
}

@article{Das2003,
author = {Das, M. and Teuben, P. J. and Vogel, S. N. and et al.},
year = {2003},
title = {Central Mass Concentration and Bar Dissolution in Nearby Spiral Galaxies},
journal = {ApJ},
volume = {582},
number = {1},
pages = {190--195},
doi = {10.1086/344480}
}

@ARTICLE{Davis2014,
       author = {{Davis}, Benjamin L. and {Berrier}, Joel C. and {Johns}, Lucas and {Shields}, Douglas W. and {Hartley}, Matthew T. and {Kennefick}, Daniel and {Kennefick}, Julia and {Seigar}, Marc S. and {Lacy}, Claud H.~S.},
        title = "{The Black Hole Mass Function Derived from Local Spiral Galaxies}",
      journal = {ApJ},
         year = 2014,
        month = jul,
       volume = {789},
       number = {2},
          eid = {124},
        pages = {124},
          doi = {10.1088/0004-637X/789/2/124},
archivePrefix = {arXiv},
       eprint = {1405.5876},
 primaryClass = {astro-ph.GA},
       adsurl = {https://ui.adsabs.harvard.edu/abs/2014ApJ...789..124D}
}

@inproceedings{Dettmar1993,
author = {Dettmar, R.-J. and Skiff, B. A.},
year = {1993},
title = {Declining rotation curves in interacting galaxies},
booktitle = {Evolution of Galaxies and their Environment},
pages = {251--252},
publisher = {NASA Ames Research Center}
}

@ARTICLE{deVaucouleurs1978,
       author = {{de Vaucouleurs}, G. and {Pence}, W.~D.},
        title = "{An outsider's view of the Galaxy: photometric parameters, scale lengths, and absolute magnitudes of the spheroidal and disk components of our Galaxy.}",
      journal = {AJ},
         year = 1978,
        month = oct,
       volume = {83},
        pages = {1163-1173},
          doi = {10.1086/112305},
       adsurl = {https://ui.adsabs.harvard.edu/abs/1978AJ.....83.1163D}
}

@ARTICLE{Dmytrenko2025a,
       author = {{Dmytrenko}, A.~M. and {Fedorov}, P.~N. and {Akhmetov}, V.~S. and {Velichko}, A.~B. and {Denyshchenko}, S.~I. and {Khramtsov}, V.~P. and {Vavilova}, I.~B. and {Dobrycheva}, D.~V. and {Sergijenko}, O. and O.~V. {Kompaniiets},and A.~A. {Vasylenko}},
        title = "{Spatial orientation and shape of the velocity ellipsoids of the Gaia DR3 giants and sub-giants in the Galactic plane}",
      journal = {MNRAS},
         year = 2025,
        month = sep,
       volume = {542},
       number = {3},
        pages = {2542-2559},
          doi = {10.1093/mnras/staf1408},
archivePrefix = {arXiv},
       eprint = {2412.18333},
 primaryClass = {astro-ph.GA},
       adsurl = {https://ui.adsabs.harvard.edu/abs/2025MNRAS.542.2542D}
}

@article{Dobrycheva2018,
author = {Dobrycheva, D. V. and Vavilova, I. B. and Melnyk, O. V. and Elyiv, A. A.},
year = {2018},
title = {Morphological type and color indices of the SDSS DR9 galaxies at 0.02 < z < 0.06},
journal = {Kinemat. Phys. Celesti. Bodies},
volume = {34},
number = {6},
pages = {290},
doi = {10.3103/S0884591318060028}
}

@ARTICLE{Dobrycheva2025,
       author = {{Dobrycheva}, D.~V. and {Hetmantsev}, O.~O. and {Vavilova}, I.~B. and {Shportko}, A. and {Gugnin}, O. and {Kompaniiets}, O.~V.},
        title = "{Discovery of the polar ring galaxies with deep learning}",
      journal = {\aap},
         year = 2025,
        month = oct,
       volume = {702},
          eid = {A258},
        pages = {A258},
          doi = {10.1051/0004-6361/202555052},
archivePrefix = {arXiv},
       eprint = {2505.05890},
 primaryClass = {astro-ph.GA},
       adsurl = {https://ui.adsabs.harvard.edu/abs/2025A&A...702A.258D}
}

@article{Draine2003,
  author       = {Draine, B. T.},
  title        = {Interstellar Dust Grains},
  journal      = {ARA\&A},
  volume       = {41},
  number       = {1},
  pages        = {241--289},
  year         = {2003},
  doi          = {10.1146/annurev.astro.41.011802.094840},
  url          = {https://doi.org/10.1146/annurev.astro.41.011802.094840}
}

@book{Draine2011,
  author    = {Draine, B. T.},
  year      = {2011},
  title     = {Physics of the Interstellar and Intergalactic Medium},
  publisher = {Princeton University Press},
  address   = {Princeton}
}

@article{Draine2014,
author  = {Draine, B. T. and Li, A.},
title   = {Infrared emission from interstellar dust. IV. The silicate-graphite-PAH model in the post-Spitzer era},
journal = {ApJ},
year    = {2014},
volume  = {780},
pages   = {172}
}

@ARTICLE{Eckart2018,
       author = {{Eckart}, A. and {Zajacek}, M. and {Parsa}, M. and {Fazeli}, E. Hosseini N. and {Busch}, G. and {Shahzamanian}, B. and {Subroweit}, M. and {Peissker}, F. and {Sabha}, N. and {Valencia-S.}, M. and {Horrobin}, M. and {Straubmeier}, C. and {Rost}, S. and {Borkar}, J. Schneeloch A. and {Karas}, V. and {Britzen}, S. and {Zensus}, A. and {Kamali}, F.},
        title = "{The Multifrequency Behavior of Sagittarius A*}",
      journal = {submited to proceedings of the XII workshop on 'Multifrequency Behavior of High Energy Cosmic Sources'},
         year = 2018,
        month = jun,
          eid = {arXiv:1806.00284 },
        pages = { },
          doi = {10.48550/arXiv.1806.00284},
 primaryClass = {astro-ph.HE},
       adsurl = {https://ui.adsabs.harvard.edu/abs/2018arXiv180600284E}
}

@article{Elson2014,
author = {Elson, E. C.},
year = {2014},
title = {An H I study of NGC 3521 — a galaxy with a slow-rotating halo},
journal = {MNRAS},
volume = {437},
number = {4},
pages = {3736--3749},
doi = {10.1093/mnras/stt2182}
}

@ARTICLE{Elyiv2020,
       author = {{Elyiv}, A.~A. and {Melnyk}, O.~V. and {Vavilova}, I.~B. and et al.},
        title = "{Machine-learning computation of distance modulus for local galaxies}",
      journal = {A\&A},
         year = 2020,
        month = mar,
       volume = {635},
          eid = {A124},
        pages = {A124},
          doi = {10.1051/0004-6361/201936883},
archivePrefix = {arXiv},
       eprint = {1910.07317},
 primaryClass = {astro-ph.CO}
}

@ARTICLE{Erroz2019,
       author = {{Erroz-Ferrer}, Santiago and {Carollo}, C. Marcella and {den Brok}, Mark and {Onodera}, Masato and {Brinchmann}, Jarle and {Marino}, Raffaella A. and {Monreal-Ibero}, Ana and {Schaye}, Joop and {Woo}, Joanna and {Cibinel}, Anna and {Debattista}, Victor P. and {Inami}, Hanae and {Maseda}, Michael and {Richard}, Johan and {Tacchella}, Sandro and {Wisotzki}, Lutz},
        title = "{The MUSE Atlas of Disks (MAD): resolving star formation rates and gas metallicities on <100 pc scales{\textdagger}}",
      journal = {MNRAS},
         year = 2019,
        month = apr,
       volume = {484},
       number = {4},
        pages = {5009-5027},
          doi = {10.1093/mnras/stz194},
archivePrefix = {arXiv},
       eprint = {1901.04493},
 primaryClass = {astro-ph.GA},
       adsurl = {https://ui.adsabs.harvard.edu/abs/2019MNRAS.484.5009E}
}

@inproceedings{Fabricius2015,
author = {Fabricius, M. H. and Coccato, L. and Bender, R. and others},
year = {2015},
title = {Regrowth of stellar disks in mature galaxies: The two component nature of NGC 7217 revisited with VIRUS-W},
booktitle = {IAU Proceedings},
volume = {309},
pages = {81--84},
doi = {10.1017/S1743921314009363}
}

@ARTICLE{Fedorov2025,
       author = {{Fedorov}, P.~N. and {Dmytrenko}, A.~M. and {Akhmetov}, V.~S. and {Velichko}, A.~B. and {Khramtsov}, V.~P. and {Denyshchenko}, S.~I. and {Vavilova}, I.~B. and {Dobrycheva}, D.~V. and {Sergijenko}, O. and {Vasylenko}, A.~A. and {Kompaniiets}, O.~V.},
        title = "{Galaxy rotation curve based on RGB stars from the Gaia DR3 catalogue}",
      journal = {submited to MNRAS},
         year = 2025,
        month = nov,
          eid = {arXiv:2511.22295},
        pages = { },
        archivePrefix = {arXiv},
       eprint = {2511.22295},
 primaryClass = {astro-ph.GA},
          doi = {10.48550/arXiv.2511.22295},
       adsurl = {https://ui.adsabs.harvard.edu/abs/2025arXiv251122295F}
}

@ARTICLE{Fielder2021,
       author = {{Fielder}, Catherine E. and {Newman}, Jeffrey A. and {Andrews}, Brett H. and {Zasowski}, Gail and {Boardman}, Nicholas F. and {Licquia}, Tim and {Masters}, Karen L. and {Salim}, Samir},
        title = "{Constraining the Milky Way's ultraviolet-to-infrared SED with Gaussian process regression}",
      journal = {MNRAS},
         year = 2021,
        month = dec,
       volume = {508},
       number = {3},
        pages = {4459-4483},
          doi = {10.1093/mnras/stab2618},
archivePrefix = {arXiv},
       eprint = {2106.14900},
 primaryClass = {astro-ph.GA},
       adsurl = {https://ui.adsabs.harvard.edu/abs/2021MNRAS.508.4459F}
}

@ARTICLE{Fraser2019,
       author = {{Fraser-McKelvie}, Amelia and {Merrifield}, Michael and {Arag{\'o}n-Salamanca}, Alfonso},
        title = "{From the outside looking in: what can Milky Way analogues tell us about the star formation rate of our own galaxy?}",
      journal = {MNRAS},
         year = 2019,
        month = nov,
       volume = {489},
       number = {4},
        pages = {5030-5036},
          doi = {10.1093/mnras/stz2493},
archivePrefix = {arXiv},
       eprint = {1909.01654},
 primaryClass = {astro-ph.GA},
       adsurl = {https://ui.adsabs.harvard.edu/abs/2019MNRAS.489.5030F}
}

@ARTICLE{Graham2007,
       author = {{Graham}, Alister W. and {Driver}, Simon P.},
        title = "{A Log-Quadratic Relation for Predicting Supermassive Black Hole Masses from the Host Bulge S{\'e}rsic Index}",
      journal = {ApJ},
         year = 2007,
        month = jan,
       volume = {655},
       number = {1},
        pages = {77-87},
          doi = {10.1086/509758},
archivePrefix = {arXiv},
       eprint = {astro-ph/0607378},
 primaryClass = {astro-ph},
       adsurl = {https://ui.adsabs.harvard.edu/abs/2007ApJ...655...77G}
}

@article{Grasha2022,
author = {Grasha, K. and Chen, Q. H. and Battisti, A. J. and others},
year = {2022},
title = {Metallicity, Ionization Parameter, and Pressure Variations of H II Regions in the TYPHOON Spiral Galaxies: NGC 1566, NGC 2835, NGC 3521, NGC 5068, NGC 5236, and NGC 7793},
journal = {ApJ},
volume = {929},
number = {2},
pages = {118},
doi = {10.3847/1538-4357/ac5ab2}
}

@ARTICLE{Grier2011,
       author = {{Grier}, C.~J. and {Mathur}, S. and {Ghosh}, H. and {Ferrarese}, L.},
        title = "{Discovery of Nuclear X-ray Sources in Sings Galaxies}",
      journal = {ApJ},
         year = 2011,
        month = apr,
       volume = {731},
       number = {1},
          eid = {60},
        pages = {60},
          doi = {10.1088/0004-637X/731/1/60},
archivePrefix = {arXiv},
       eprint = {1011.4295},
 primaryClass = {astro-ph.GA},
       adsurl = {https://ui.adsabs.harvard.edu/abs/2011ApJ...731...60G}
}

@article{Ginolfi2018,
  author  = {Ginolfi, M. and Graziani, L. and Schneider, R. and Marassi, S. and Valiante, R. and Dell'Agli, F. and Ventura, P. and Hunt, L. K.},
  title   = {Where does galactic dust come from?},
  journal = {MNRAS},
  year    = {2018},
  volume  = {473},
  number  = {4},
  pages   = {4538--4543},
  doi     = {10.1093/mnras/stx2572},
  url     = {https://doi.org/10.1093/mnras/stx2572}
}

@article{Haslbauer2024,
author = {Haslbauer, M. and Banik, I. and Kroupa, P. and others},
year = {2024},
title = {The Magellanic Clouds are very rare in the IllustrisTNG simulations},
journal = {Universe},
volume = {10},
number = {10},
pages = {385},
doi = {10.3390/universe10100385}
}

@ARTICLE{Heeschen1964,
       author = {{Heeschen}, D.~S. and {Wade}, C.~M.},
        title = "{A radio survey of galaxies}",
      journal = {AJ},
         year = 1964,
        month = may,
       volume = {69},
        pages = {277},
          doi = {10.1086/109269},
       adsurl = {https://ui.adsabs.harvard.edu/abs/1964AJ.....69..277H}
}

@article{Heida2014,
author = {Heida, M. and Jonker, P. G. and Torres, M. A. P. and others},
year = {2014},
title = {Near-infrared counterparts of ultraluminous X-ray sources},
journal = {MNRAS},
volume = {442},
number = {2},
pages = {1054--1067},
doi = {10.1093/mnras/stu928}
}

@article{Helou1985,
  author  = {Helou, G. and Soifer, B. T. and Rowan-Robinson, M.},
  year    = {1985},
  title   = {Thermal infrared and nonthermal radio: Remarkable correlation in disks of galaxies},
  journal = {ApJL},
  volume  = {298},
  pages   = {L7--L11},
  doi     = {10.1086/184556}
}

@article{Hildebrand1983,
  author       = {Hildebrand, R. H.},
  title        = {The determination of cloud masses and dust characteristics from submillimetre thermal emission},
  journal      = {QJRAS},
  year         = {1983},
  volume       = {24},
  pages        = {267--282}
}

@ARTICLE{Hunt2019,
       author = {{Hunt}, L.~K. and {De Looze}, I. and {Boquien}, M. and {Nikutta}, R. and {Rossi}, A. and {Bianchi}, S. and {Dale}, D.~A. and {Granato}, G.~L. and {Kennicutt}, R.~C. and {Silva}, L. and {Ciesla}, L. and {Rela{\~n}o}, M. and {Viaene}, S. and {Brandl}, B. and {Calzetti}, D. and {Croxall}, K.~V. and {Draine}, B.~T. and {Galametz}, M. and {Gordon}, K.~D. and {Groves}, B.~A. and {Helou}, G. and {Herrera-Camus}, R. and {Hinz}, J.~L. and {Koda}, J. and {Salim}, S. and {Sandstrom}, K.~M. and {Smith}, J.~D. and {Wilson}, C.~D. and {Zibetti}, S.},
        title = "{Comprehensive comparison of models for spectral energy distributions from 0.1 {\ensuremath{\mu}}m to 1 mm of nearby star-forming galaxies}",
      journal = {A\&A},
         year = 2019,
        month = jan,
       volume = {621},
          eid = {A51},
        pages = {A51},
          doi = {10.1051/0004-6361/201834212},
archivePrefix = {arXiv},
       eprint = {1809.04088},
 primaryClass = {astro-ph.GA},
       adsurl = {https://ui.adsabs.harvard.edu/abs/2019A&A...621A..51H}
}

@article{Inoue2011,
author  = {Inoue, A. K.},
title   = {Rest-frame ultraviolet-to-optical spectral properties of galaxies: Models and implications},
journal = {MNRAS},
year    = {2011},
volume  = {415},
pages   = {2920--2931}
}

@ARTICLE{Israel1990,
       author = {{Israel}, F.~P. and {Mahoney}, M.~J.},
        title = "{Low-Frequency Radio Continuum Evidence for Cool Ionized Gas in Normal Spiral Galaxies}",
      journal = {ApJ},
         year = 1990,
        month = mar,
       volume = {352},
        pages = {30},
          doi = {10.1086/168513},
       adsurl = {https://ui.adsabs.harvard.edu/abs/1990ApJ...352...30I}
}

@article{Karachentsev2022,
author = {Karachentsev, I. D. and Makarova, L. N. and Anand, G. S. and others},
year = {2022},
title = {Around the Spindle Galaxy: The Dark Halo Mass of NGC 3115},
journal = {AJ},
volume = {163},
number = {5},
pages = {234},
doi = {10.3847/1538-3881/ac5ab5}
}

@article{Karachentseva1973,
author = {Karachentseva, V. E.},
year = {1973},
title = {The Catalogue of Isolated Galaxies},
journal = {Astrofizicheskie Issledovaniya Byulletin},
volume = {8},
pages = {3--49}
}

@ARTICLE{Karachentseva2010,
       author = {{Karachentseva}, V.~E. and {Mitronova}, S.~N. and {Melnyk}, O.~V. and {Karachentsev}, I.~D.},
        title = "{Catalog of isolated galaxies selected from the 2MASS survey}",
      journal = {Astrophysical Bulletin},
         year = 2010,
        month = jan,
       volume = {65},
       number = {1},
        pages = {1-17},
          doi = {10.1134/S1990341310010013},
archivePrefix = {arXiv},
       eprint = {1005.3191},
 primaryClass = {astro-ph.CO},
       adsurl = {https://ui.adsabs.harvard.edu/abs/2010AstBu..65....1K}
}

@article{Kauffmann2003,
  author = {Kauffmann, G. and Heckman, T. M. and Tremonti, C. and Brinchmann, J. and Charlot, S. and White, S. D. M. and Ridgway, S. E. and Brinkmann, J. and Fukugita, M. and Hall, P. B. and Ivezi{\'c}, {\v{Z}}. and Richards, G. T. and Schneider, D. P.},
  title = {The host galaxies of active galactic nuclei},
  journal = {MNRAS},
  volume = {346},
  number = {4},
  pages = {1055--1077},
  year = {2003},
  doi = {10.1111/j.1365-2966.2003.07154.x}
}

@ARTICLE{Kennicutt2011,
       author = {{Kennicutt}, R.~C. and {Calzetti}, D. and {Aniano}, G. and {Appleton}, P. and {Armus}, L. and {Beir{\~a}o}, P. and {Bolatto}, A.~D. and {Brandl}, B. and {Crocker}, A. and {Croxall}, K. and {Dale}, D.~A. and {Donovan Meyer}, J. and {Draine}, B.~T. and {Engelbracht}, C.~W. and {Galametz}, M. and {Gordon}, K.~D. and {Groves}, B. and {Hao}, C.-N. and {Helou}, G. and {Hinz}, J. and {Hunt}, L.~K. and {Johnson}, B. and {Koda}, J. and {Krause}, O. and {Leroy}, A.~K. and {Li}, Y. and {Meidt}, S. and {Montiel}, E. and {Murphy}, E.~J. and {Rahman}, N. and {Rix}, H.-W. and {Roussel}, H. and {Sandstrom}, K. and {Sauvage}, M. and {Schinnerer}, E. and {Skibba}, R. and {Smith}, J.~D.~T. and {Srinivasan}, S. and {Vigroux}, L. and {Walter}, F. and {Wilson}, C.~D. and {Wolfire}, M. and {Zibetti}, S.},
        title = "{KINGFISH{\textemdash}Key Insights on Nearby Galaxies: A Far-Infrared Survey with Herschel: Survey Description and Image Atlas}",
      journal = {\pasp},
         year = 2011,
        month = dec,
       volume = {123},
       number = {910},
        pages = {1347},
          doi = {10.1086/663818},
archivePrefix = {arXiv},
       eprint = {1111.4438},
 primaryClass = {astro-ph.CO},
       adsurl = {https://ui.adsabs.harvard.edu/abs/2011PASP..123.1347K}
}

@article{Kewley2001,
  author = {Kewley, L. J. and Dopita, M. A. and Sutherland, R. S. and Heisler, C. A. and Trevena, J.},
  title = {Theoretical Modeling of Starburst Galaxies},
  journal = {ApJ},
  volume = {556},
  pages = {121--140},
  year = {2001},
  doi = {10.1086/321545}
}

@article{Knapik2000,
author = {Knapik, J. and Soida, M. and Dettmar, R.-J. and et al.},
year = {2000},
title = {Detection of spiral magnetic fields in two flocculent galaxies},
journal = {A\&A},
volume = {362},
pages = {910--920},
doi = {10.48550/arXiv.astro-ph/0009438}
}

@INPROCEEDINGS{Kompaniiets2023IAU,
       author = {Kompaniiets, O.~V. and Babyk, Iu. V. and Vasylenko, A.~A. and et al.},
        title = "{X-ray spectral and image spatial models of NGC 3081 with Chandra data}",
    booktitle = {The Predictive Power of Computational Astrophysics as a Discover Tool},
         year = 2023,
       series = {IAU Symposium},
       volume = {362},
        month = jan,
        pages = {100-104},
          doi = {10.1017/S1743921322001624},
       adsurl = {https://ui.adsabs.harvard.edu/abs/2023IAUS..362..100K}
}

@ARTICLE{Kompaniiets2024,
       author = {{Kompaniiets}, O.~V.},
        title = "{Multiwavelength properties of the low-redshift isolated galaxies with active nuclei modelled with CIGALE}",
      journal = {Space Sci.\&Technol.},
         year = 2023,
        month = jan,
       volume = {29},
       number = {5},
        pages = {88--98},
          doi = {10.15407/knit2023.05.088},
}

@ARTICLE{Kompaniiets2025,
       author = {{Kompaniiets}, O.~V. and {Kukhar}, O.~M. and {Vavilova}, I.~B. and {Dobrycheva}, D.~V. and {Fedorov}, P.~N. and {Dmytrenko}, A.~M. and {Khramtsov}, V.~P. and {Sergijenko}, O.~M. and {Vasylenko}, A.~A.},
        title = {Milky Way galaxy-analogs and isolated galaxies with bars: environmental density in the Local Volume},
      journal = {Space Sci.\&Technol.},
         year = 2025,
        volume = {31},
        number = {6},
        pages = {134--148},
        doi = {10.15407/knit2025.06.134}
}

@ARTICLE{Kompaniiets2025arxiv,
       author = {{Kompaniiets}, O.~V. and {Vasylenko}, A.~A. and {Vavilova}, I.~B.},
        title = "{The 2MIG isolated AGNs -- 2. X-ray general properties and peculiarities}",
      journal = {submited to MNRAS},
         year = 2025,
        month = jun,
          eid = {arXiv:2506.14348},
        pages = { },
        archivePrefix = {arXiv},
       eprint = {2506.14348},
 primaryClass = {astro-ph.HE},
          doi = {10.48550/arXiv.2506.14348},
       adsurl = {https://ui.adsabs.harvard.edu/abs/2025arXiv250614348K}
}

@article{Konovalenko2016,
author = {Konovalenko, A. and Sodin, L. and Zakharenko, V. and others},
title = {The modern radio astronomy network in Ukraine: UTR-2, URAN and GURT},
journal = {Experimental Astronomy},
volume = {42},
number = {1},
pages = {11--48},
year = {2016},
doi = {10.1007/s10686-016-9498-x}
}

@ARTICLE{Konovalenko2021,
       author = {{Konovalenko}, O.~O. and {Zakharenko}, V.~V. and {Lytvynenko}, L.~M. and {Ulyanov}, O.~M. and {Sidorchuk}, M.~A. and {Stepkin}, S.~V. and {Shepelev}, V.~A. and {Zarka}, P. and {Rucker}, H.~O. and {Lecacheux}, A. and {Panchenko}, M. and {Bruck}, Yu. M. and {Tokarsky}, P.~L. and {Bubnov}, I.~M. and {Yerin}, S.~M. and {Koliadin}, V.~L. and {Melnik}, V.~M. and {Kalinichenko}, M.~M. and {Stanislavsky}, O.~O. and {Dorovskyy}, V.~V. and {Khristenko}, O.~D. and {Shevchenko}, V.~V. and {Belov}, O.~S. and {Gridin}, A.~O. and {Antonov}, O.~V. and {Bovkun}, V.~P. and {Reznichenko}, O.~M. and {Bortsov}, V.~M. and {Kvasov}, G.~V. and {Ostapchenko}, L.~M. and {Shevchuk}, M.~V. and {Shevchenko}, V.~A. and {Yatskiv}, Ya. S. and {Vavilova}, I.~B. and {Braude}, I.~S. and {Shkuratov}, Y.~G. and {Ryabov}, V.~B. and {Pidgorny}, G.~I. and {Tymoshevsky}, A.~G. and {Lytvynenko}, O.~O. and {Galanin}, V.~V. and {Ryabov}, M.~I. and {Brazhenko}, A.~I. and {Vashchishin}, R.~V. and {Frantsuzenko}, A.~V. and {Koshovyy}, V.~V. and {Ivantyshyn}, C.~L. and {Lozinsky}, A.~B. and {Kharchenko}, B.~S. and {Vasylieva}, I.~Y. and {Kravtsov}, I.~P. and {Vasylkivsky}, Y.~V. and {Litvinenko}, G.~V. and {Mukha}, D.~V. and {Vasylenko}, N.~V. and {Shevtsova}, A.~I. and {Miroshnichenko}, A.~P. and {Kuhai}, N.~V. and {Sobolev}, Ya. M. and {Tsvyk}, N.~C.},
        title = "{The Founder of the Decameter Radio Astronomy in Ukraine Academician of Nas of Ukraine Semen Yakovych Braude is 110 Years Old: History of Creation and Development of the National Experimental Base for the Last Half Century}",
      journal = {RPRA},
         year = 2021,
        month = mar,
       volume = {26},
       number = {1},
        pages = {5-73},
          doi = {10.15407/rpra26.01.005},
       adsurl = {https://ui.adsabs.harvard.edu/abs/2021RPRA...26....5K}
}

@article{Kourkchi2017,
author = {Kourkchi, E. and Tully, R. B.},
year = {2017},
title = {Galaxy Groups Within 3500 km/s},
journal = {ApJ},
volume = {843},
number = {1},
pages = {16},
doi = {10.3847/1538-4357/aa76db}
}

@article{Lacki2013,
  author  = {Lacki, B. C.},
  year    = {2013},
  title   = {Interpreting the low-frequency radio spectra of starburst galaxies},
  journal = {MNRAS},
  volume  = {431},
  pages   = {3003--3027},
  doi     = {10.1093/mnras/stt372}
}

@ARTICLE{Large1981,
       author = {{Large}, M.~I. and {Mills}, B.~Y. and {Little}, A.~G. and {Crawford}, D.~F. and {Sutton}, J.~M.},
        title = "{The Molonglo Reference Catalogue of radio sources.}",
      journal = {MNRAS},
         year = 1981,
        month = feb,
       volume = {194},
        pages = {693-704},
          doi = {10.1093/mnras/194.3.693},
       adsurl = {https://ui.adsabs.harvard.edu/abs/1981MNRAS.194..693L}
}

@article{Leroy2008,
  author       = {Leroy, A. K. and Walter, F. and Brinks, E. and Bigiel, F. and de Blok, W. J. G. and Madore, B. and Thornley, M. D.},
  title        = {The Star Formation Efficiency in Nearby Galaxies: Measuring Where Gas Forms Stars Effectively},
  journal      = {AJ},
  year         = {2008},
  volume       = {136},
  number       = {6},
  pages        = {2782--2845},
  doi          = {10.1088/0004-6256/136/6/2782},
  eprint       = {arXiv:0810.2556},
  archivePrefix= {arXiv},
  primaryClass = {astro-ph}
}

@ARTICLE{Li2001,
       author = {{Li}, Aigen and {Draine}, B.~T.},
        title = "{Infrared Emission from Interstellar Dust. II. The Diffuse Interstellar Medium}",
      journal = {ApJ},
         year = 2001,
        month = jun,
       volume = {554},
       number = {2},
        pages = {778-802},
          doi = {10.1086/323147},
archivePrefix = {arXiv},
       eprint = {astro-ph/0011319},
 primaryClass = {astro-ph},
       adsurl = {https://ui.adsabs.harvard.edu/abs/2001ApJ...554..778L}
}

@ARTICLE{Licquia2015,
       author = {{Licquia}, Timothy C. and {Newman}, Jeffrey A. and {Brinchmann}, Jarle},
        title = "{Unveiling the Milky Way: A New Technique for Determining the Optical Color and Luminosity of Our Galaxy}",
      journal = {ApJ},
         year = 2015,
        month = aug,
       volume = {809},
       number = {1},
          eid = {96},
        pages = {96},
          doi = {10.1088/0004-637X/809/1/96},
archivePrefix = {arXiv},
       eprint = {1508.04446},
 primaryClass = {astro-ph.GA},
       adsurl = {https://ui.adsabs.harvard.edu/abs/2015ApJ...809...96L}
}

@article{LicquiaNewman2015ApJ,
  author       = {Licquia, Timothy C. and Newman, Jeffrey A.},
  title        = {Improved Estimates of the Milky Way's Stellar Mass and Star Formation Rate from Hierarchical Bayesian Meta-Analysis},
  journal      = {ApJ},
  year         = {2015},
  volume       = {806},
  number       = {1},
  pages        = {96},
  month        = jun,
  doi          = {10.1088/0004-637X/806/1/96},
  url          = {https://iopscience.iop.org/article/10.1088/0004-637X/806/1/96},
  eprint       = {1407.1078},
  archivePrefix= {arXiv},
  primaryClass = {astro-ph.GA},
  adsurl       = {https://ui.adsabs.harvard.edu/abs/2015ApJ...806...96L}
}

@ARTICLE{Lindner1995,
       author = {{Lindner}, U. and {Einasto}, J. and {Einasto}, M. and {Freudling}, W. and {Fricke}, K. and {Tago}, E.},
        title = "{The structure of supervoids. I. Void hierarchy in the Northern Local Supervoid.}",
      journal = {A\&A},
         year = 1995,
        month = sep,
       volume = {301},
        pages = {329},
          doi = {10.48550/arXiv.astro-ph/9503044},
archivePrefix = {arXiv},
       eprint = {astro-ph/9503044},
 primaryClass = {astro-ph},
       adsurl = {https://ui.adsabs.harvard.edu/abs/1995A&A...301..329L}
}

@article{Liu2011,
author = {Liu, G. and Koda, J. and Calzetti, D. and Fukuhara, M. and Momose, R.},
year = {2011},
title = {The Super-linear Slope of the Spatially Resolved Star Formation Law in NGC 3521 and NGC 5194 (M51a)},
journal = {ApJ},
volume = {735},
number = {1},
pages = {63},
doi = {10.1088/0004-637X/735/1/63}
}

@ARTICLE{Liang2023,
       author = {{Liang}, Shuang and {von der Linden}, Anja},
        title = "{Photometric calibration in u-band using blue halo stars}",
      journal = {\mnras},
         year = 2023,
        month = feb,
       volume = {519},
       number = {2},
        pages = {2281-2301},
          doi = {10.1093/mnras/stac3671},
archivePrefix = {arXiv},
       eprint = {2212.05135},
 primaryClass = {astro-ph.IM},
       adsurl = {https://ui.adsabs.harvard.edu/abs/2023MNRAS.519.2281L}
}

@article{Lopez2015,
author = {L{\'o}pez, K. M. and Jonker, P. G. and Heida, M.},
year = {2015},
title = {Discovery and analysis of a ULX nebula in NGC 3521},
journal = {MNRAS},
volume = {489},
number = {1},
pages = {1249--1264},
doi = {10.1093/mnras/stz2127}
}

@article{Mainzer2014NEOWISE,
  author  = {Mainzer, A. and Bauer, J. and Cutri, R. and et al.},
  year    = {2014},
  title   = {Initial Performance of the NEOWISE Reactivation Mission},
  journal = {ApJ},
  volume  = {792},
  number  = {1},
  pages   = {30},
  doi     = {10.1088/0004-637X/792/1/30}
}

@INPROCEEDINGS{Mast2006,
       author = {{Mast}, D. and {D{\'\i}az}, R.~J.},
        title = "{Stellar kinematics of spiral galaxies. NGC 2613, NGC 3521, M83}",
    booktitle = {RevMexAA Confer. Ser.},
         year = 2006,
       volume = {26},
        month = jun,
        pages = {193},
       adsurl = {https://ui.adsabs.harvard.edu/abs/2006RMxAC..26R.193M}
}

@article{Marvil2015,
  author  = {Marvil, J. and Owen, F. and Eilek, J.},
  year    = {2015},
  title   = {Integrated Radio Continuum Spectra of Galaxies},
  journal = {AJ},
  volume  = {149},
  pages   = {32},
  doi     = {10.1088/0004-6256/149/1/32}
}

@article{Masci2019ZTF,
  author  = {Masci, F. J. and Laher, R. R. and Rusholme, B. and et al.},
  year    = {2019},
  title   = {The Zwicky Transient Facility: Data Processing, Products, and Archive},
  journal = {PASP},
  volume  = {131},
  number  = {995},
  pages   = {018003},
  doi     = {10.1088/1538-3873/aae8ac}
}

@ARTICLE{Mazurenko2024,
       author = {{Mazurenko}, Sergij and {Banik}, Indranil and {Kroupa}, Pavel and {Haslbauer}, Moritz},
        title = "{A simultaneous solution to the Hubble tension and observed bulk flow within 250 h$^{-1}$ Mpc}",
      journal = {MNRAS},
         year = 2024,
        month = jan,
       volume = {527},
       number = {3},
        pages = {4388-4396},
          doi = {10.1093/mnras/stad3357},
archivePrefix = {arXiv},
       eprint = {2311.17988},
 primaryClass = {astro-ph.CO},
       adsurl = {https://ui.adsabs.harvard.edu/abs/2024MNRAS.527.4388M}
}

@article{McGaugh2016,
author = {McGaugh, S. S.},
year = {2016},
title = {The Surface Density Profile of the Galactic Disk from the Terminal Velocity Curve},
journal = {ApJ},
volume = {816},
number = {1},
pages = {42},
doi = {10.3847/0004-637X/816/1/42}
}

@ARTICLE{McKinnon2016,
       author = {{McKinnon}, Ryan and {Torrey}, Paul and {Vogelsberger}, Mark},
        title = "{Dust formation in Milky Way-like galaxies}",
      journal = {MNRAS},
         year = 2016,
        month = apr,
       volume = {457},
       number = {4},
        pages = {3775-3800},
          doi = {10.1093/mnras/stw253},
archivePrefix = {arXiv},
       eprint = {1505.04792},
 primaryClass = {astro-ph.GA},
       adsurl = {https://ui.adsabs.harvard.edu/abs/2016MNRAS.457.3775M}
}

@article{Melnyk2015,
author = {Melnyk, O. and Karachentseva, V. and Karachentsev, I.},
year = {2015},
title = {Star formation rates in isolated galaxies selected from the Two-Micron All-Sky Survey},
journal = {MNRAS},
volume = {451},
number = {2},
pages = {14},
doi = {10.1093/mnras/stv950}
}

@ARTICLE{Mosenkov2019,
       author = {{Mosenkov}, A.~V. and {Baes}, M. and {Bianchi}, S. and {Casasola}, V. and {Cassar{\`a}}, L.~P. and {Clark}, C.~J.~R. and {Davies}, J. and {De Looze}, I. and {De Vis}, P. and {Fritz}, J. and {Galametz}, M. and {Galliano}, F. and {Jones}, A.~P. and {Lianou}, S. and {Madden}, S.~C. and {Nersesian}, A. and {Smith}, M.~W.~L. and {Tr{\v{c}}ka}, A. and {Verstocken}, S. and {Viaene}, S. and {Vika}, M. and {Xilouris}, E.},
        title = "{Dust emission profiles of DustPedia galaxies}",
      journal = {A\&A},
         year = 2019,
        month = feb,
       volume = {622},
          eid = {A132},
        pages = {A132},
          doi = {10.1051/0004-6361/201833932},
archivePrefix = {arXiv},
       eprint = {1811.08923},
 primaryClass = {astro-ph.GA},
       adsurl = {https://ui.adsabs.harvard.edu/abs/2019A&A...622A.132M}
}

@ARTICLE{Muller2025,
       author = {{M{\"u}ller}, Oliver and {Jerjen}, Helmut and {Taibi}, Salvatore and {Heesters}, Nick and {Crosby}, Ethan and {Pawlowski}, Marcel S.},
        title = "{New dwarf galaxy candidates in the M106, NGC3521, and UGCA127 groups with the Hyper Suprime Camera}",
      journal = {Submitted to OJAp},
         year = 2025,
        month = apr,
          eid = {arXiv:2504.11608},
        pages = {arXiv:2504.11608},
          doi = {10.48550/arXiv.2504.11608},
archivePrefix = {arXiv},
       eprint = {2504.11608},
 primaryClass = {astro-ph.GA},
       adsurl = {https://ui.adsabs.harvard.edu/abs/2025arXiv250411608M}
}

@ARTICLE{Morrissey2007,
       author = {{Morrissey}, Patrick and {Conrow}, Tim and {Barlow}, Tom A. and {Small}, Todd and {Seibert}, Mark and {Wyder}, Ted K. and {Budav{\'a}ri}, Tam{\'a}s and {Arnouts}, Stephane and {Friedman}, Peter G. and {Forster}, Karl and {Martin}, D. Christopher and {Neff}, Susan G. and {Schiminovich}, David and {Bianchi}, Luciana and {Donas}, Jos{\'e} and {Heckman}, Timothy M. and {Lee}, Young-Wook and {Madore}, Barry F. and {Milliard}, Bruno and {Rich}, R. Michael and {Szalay}, Alex S. and {Welsh}, Barry Y. and {Yi}, Sukyoung K.},
        title = "{The Calibration and Data Products of GALEX}",
      journal = {\apjs},
         year = 2007,
        month = dec,
       volume = {173},
       number = {2},
        pages = {682-697},
          doi = {10.1086/520512},
archivePrefix = {arXiv},
       eprint = {0706.0755},
 primaryClass = {astro-ph},
       adsurl = {https://ui.adsabs.harvard.edu/abs/2007ApJS..173..682M}
}

@ARTICLE{Mutch2011,
       author = {{Mutch}, Simon J. and {Croton}, Darren J. and {Poole}, Gregory B.},
        title = "{The Mid-life Crisis of the Milky Way and M31}",
      journal = {ApJ},
         year = 2011,
        month = aug,
       volume = {736},
       number = {2},
          eid = {84},
        pages = {84},
          doi = {10.1088/0004-637X/736/2/84},
archivePrefix = {arXiv},
       eprint = {1105.2564},
 primaryClass = {astro-ph.CO},
       adsurl = {https://ui.adsabs.harvard.edu/abs/2011ApJ...736...84M}
}

@ARTICLE{Naidu2021,
       author = {{Naidu}, Rohan P. and {Conroy}, Charlie and {Bonaca}, Ana and {Zaritsky}, Dennis and {Weinberger}, Rainer and {Ting}, Yuan-Sen and {Caldwell}, Nelson and {Tacchella}, Sandro and {Han}, Jiwon Jesse and {Speagle}, Joshua S. and {Cargile}, Phillip A.},
        title = "{Reconstructing the Last Major Merger of the Milky Way with the H3 Survey}",
      journal = {ApJ},
         year = 2021,
        month = dec,
       volume = {923},
       number = {1},
          eid = {92},
        pages = {92},
          doi = {10.3847/1538-4357/ac2d2d},
archivePrefix = {arXiv},
       eprint = {2103.03251},
 primaryClass = {astro-ph.GA},
       adsurl = {https://ui.adsabs.harvard.edu/abs/2021ApJ...923...92N}
}

@ARTICLE{Natale2022,
       author = {{Natale}, Giovanni and {Popescu}, Cristina C. and {Rushton}, Mark and {Yang}, Ruizhi and {Thirlwall}, Jordan J. and {Pricopi}, Dumitru},
        title = "{A radiation transfer model for the Milky Way: II. The global properties and large-scale structure}",
      journal = {MNRAS},
         year = 2022,
        month = jan,
       volume = {509},
       number = {2},
        pages = {2339-2361},
          doi = {10.1093/mnras/stab2771},
archivePrefix = {arXiv},
       eprint = {2111.02180},
 primaryClass = {astro-ph.GA},
       adsurl = {https://ui.adsabs.harvard.edu/abs/2022MNRAS.509.2339N}
}

@article{Popesso2019,
  author  = {Popesso, P. and Concas, A. and Morselli, L. and Schreiber, C. and Rodighiero, G. and others},
  title   = {The main sequence of star-forming galaxies -- I. The local relation and its bending},
  journal = {MNRAS},
  year    = {2019},
  volume  = {483},
  number  = {3},
  pages   = {3213--3235},
  doi     = {10.1093/mnras/sty3230}
}

@article{Pastoven2024,
author = {Pastoven, O. S. and Kompaniiets, O. and Vavilova, I. and others},
title = {NGC 3521 as the Milky Way analogue: spectral energy distribution from UV to radio and photometric variability},
journal = {Space Sci. \& Technol.},
volume = {30},
number = {6},
pages = {67--83},
year = {2024},
doi = {10.15407/knit2024.06.067}
}

@ARTICLE{Pattle2023,
       author = {{Pattle}, Kate and {Gear}, Walter and {Wilson}, Christine D.},
        title = "{The JCMT nearby galaxies legacy survey: SCUBA-2 observations of nearby galaxies}",
      journal = {MNRAS},
         year = 2023,
        month = jun,
       volume = {522},
       number = {2},
        pages = {2339-2368},
          doi = {10.1093/mnras/stad652},
archivePrefix = {arXiv},
       eprint = {2302.14800},
 primaryClass = {astro-ph.GA},
       adsurl = {https://ui.adsabs.harvard.edu/abs/2023MNRAS.522.2339P}
}

@ARTICLE{Pilyugin2019,
       author = {{Pilyugin}, L.~S. and {Grebel}, E.~K. and {Zinchenko}, I.~A. and {Nefedyev}, Y.~A. and {V{\'\i}lchez}, J.~M.},
        title = "{Relations between abundance characteristics and rotation velocity for star-forming MaNGA galaxies}",
      journal = {A\&A},
         year = 2019,
        month = mar,
       volume = {623},
          eid = {A122},
        pages = {A122},
          doi = {10.1051/0004-6361/201834239},
archivePrefix = {arXiv},
       eprint = {1901.11001},
 primaryClass = {astro-ph.GA},
       adsurl = {https://ui.adsabs.harvard.edu/abs/2019A&A...623A.122P}
}

@ARTICLE{Pilyugin2023,
       author = {{Pilyugin}, L.~S. and {Tautvai{\v{s}}ien{\.{e}}}, G. and {Lara-L{\'o}pez}, M.~A.},
        title = "{Searching for Milky Way twins: Radial abundance distribution as a strict criterion}",
      journal = {A\&A},
         year = 2023,
        month = aug,
       volume = {676},
          eid = {A57},
        pages = {A57},
          doi = {10.1051/0004-6361/202346503},
archivePrefix = {arXiv},
       eprint = {2306.09854},
 primaryClass = {astro-ph.GA},
       adsurl = {https://ui.adsabs.harvard.edu/abs/2023A&A...676A..57P}
}

@article{Pulatova2015,
author = {Pulatova, N. G. and Vavilova, I. B. and Sawangwit, U. and others},
year = {2015},
title = {The 2MIG isolated AGNs. I. General and multiwavelength properties of AGNs and host galaxies in the northern sky},
journal = {MNRAS},
volume = {447},
number = {3},
pages = {2209--2223},
doi = {10.1093/mnras/stu2556}
}

@article{Pulatova2023,
author = {Pulatova, N. G. and Vavilova, I. B. and Vasylenko, A. A. and others},
year = {2023},
title = {Radio properties of the low-redshift isolated galaxies with active nuclei},
journal = {Kinemat. Phys. Celest. Bodies},
volume = {39},
number = {2},
pages = {47--72},
doi = {10.15407/kfnt2023.02.047}
}

@book{RL79,
  author    = {Rybicki, George B. and Lightman, Alan P.},
  title     = {Radiative Processes in Astrophysics},
  year      = {1979},
  publisher = {John Wiley \& Sons},
  address   = {New York}
}

@book{RybickiLightman1979,
  author    = {Rybicki, George B. and Lightman, Alan P.},
  title     = {Radiative Processes in Astrophysics},
  year      = {1979},
  publisher = {John Wiley \& Sons},
  address   = {New York},
  isbn      = {978-0-471-82759-7},
  url       = {https://ui.adsabs.harvard.edu/abs/1979rpa..book.....R}
}

@ARTICLE{Sawala2024,
       author = {{Sawala}, Till and {Delhomelle}, Jehanne and {Deason}, Alis J. and {Frenk}, Carlos S. and {H{\"a}kkinen}, Jenni and {Johansson}, Peter H. and {Keitaanranta}, Atte and {Rawlings}, Alexander and {Wright}, Ruby},
        title = "{No certainty of a Milky Way-Andromeda collision}",
      journal = {Nature Astronomy},
         year = 2025,
        month = aug,
       volume = {9},
        pages = {1206-1217},
          doi = {10.1038/s41550-025-02563-1},
archivePrefix = {arXiv},
       eprint = {2408.00064},
 primaryClass = {astro-ph.GA},
       adsurl = {https://ui.adsabs.harvard.edu/abs/2025NatAs...9.1206S}
}

@article{Schawinski2007,
  author = {Schawinski, K. and Thomas, D. and Sarzi, M. and Maraston, C. and Kaviraj, S. and Joo, S.-J. and Yi, S. K. and Silk, J.},
  title = {Observational evidence for AGN feedback in early-type galaxies},
  journal = {MNRAS},
  volume = {382},
  number = {4},
  pages = {1415--1431},
  year = {2007},
  doi = {10.1111/j.1365-2966.2007.12487.x}
}

@article{Schlafly2011,
author  = {Schlafly, E. F. and Finkbeiner, D. P.},
title   = {Measuring Reddening with Sloan Digital Sky Survey Stellar Spectra and Recalibrating SFD},
journal = {ApJ},
volume  = {737},
number  = {2},
pages   = {103},
year    = {2011},
doi     = {10.1088/0004-637X/737/2/103}
}

@article{Schlegel1998,
author  = {Schlegel, D. J. and Finkbeiner, D. P. and Davis, M.},
title   = {Maps of Dust Infrared Emission for Use in Estimation of Reddening and Cosmic Microwave Background Radiation Foregrounds},
journal = {ApJ},
volume  = {500},
pages   = {525--553},
year    = {1998},
doi     = {10.1086/305772}
}

@article{Sidorchuk2021,
author = {Sidorchuk, M. A. and Konovalenko, A. A. and Stepkin, S. V. and others},
title = {50 years of research in continuum at the UTR-2 radio telescope},
journal = {RPRA},
volume = {26},
number = {4},
pages = {287--313},
year = {2021},
doi = {10.15407/rpra26.04.287}
}

@ARTICLE{Skibba2011,
       author = {{Skibba}, Ramin A. and {Engelbracht}, Charles W. and {Dale}, Daniel and {Hinz}, Joannah and {Zibetti}, Stefano and {Crocker}, Alison and {Groves}, Brent and {Hunt}, Leslie and {Johnson}, Benjamin D. and {Meidt}, Sharon and {Murphy}, Eric and {Appleton}, Philip and {Armus}, Lee and {Bolatto}, Alberto and {Brandl}, Bernhard and {Calzetti}, Daniela and {Croxall}, Kevin and {Galametz}, Maud and {Gordon}, Karl D. and {Kennicutt}, Robert C. and {Koda}, Jin and {Krause}, Oliver and {Montiel}, Edward and {Rix}, Hans-Walter and {Roussel}, H{\'e}l{\`e}ne and {Sandstrom}, Karin and {Sauvage}, Marc and {Schinnerer}, Eva and {Smith}, J.~D. and {Walter}, Fabian and {Wilson}, Christine D. and {Wolfire}, Mark},
        title = "{The Emission by Dust and Stars of Nearby Galaxies in the Herschel KINGFISH Survey}",
      journal = {\apj},
         year = 2011,
        month = sep,
       volume = {738},
       number = {1},
          eid = {89},
        pages = {89},
          doi = {10.1088/0004-637X/738/1/89},
archivePrefix = {arXiv},
       eprint = {1106.4022},
 primaryClass = {astro-ph.CO},
       adsurl = {https://ui.adsabs.harvard.edu/abs/2011ApJ...738...89S}
}

@ARTICLE{Slee1995,
       author = {{Slee}, O.~B.},
        title = "{Radio sources observed with the Culgoora circular array.}",
      journal = {Aust. J. Phys.},
         year = 1995,
        month = jan,
       volume = {48},
        pages = {143-186},
          doi = {10.1071/PH950143},
       adsurl = {https://ui.adsabs.harvard.edu/abs/1995AuJPh..48..143S}
}

@ARTICLE{Stevance2024,
       author = {{Stevance}, H. and {Smith}, K.~W. and {Young}, D.~R. and {Nicholl}, M. and {Fulton}, M. and {McCollum}, M. and {Moore}, T. and {Weston}, J. and {Sheng}, X. and {Aamer}, A. and {Angus}, C.~R. and {Magill}, D. and {Ramsden}, P. and {Shingles}, L. and {Smartt}, S.~J. and {Srivastav}, S. and {Gillanders}, J. and {Cooper}, A.~J. and {Stoppa}, F. and {Rhodes}, L. and {Denneau}, L. and {Tonry}, J. and {Weiland}, H. and {Siverd}, R. and {Erasmus}, N. and {Koorts}, W. and {Jordan}, A. and {Suc}, V. and {Rest}, A. and {Chen}, T.~W. and {Stubbs}, C. and {Sommer}, J. and {Schmidt}, B.~P.},
        title = "{ATLAS24rkq (AT2024aecx): discovery of a young SN in NGC 3521 (11 Mpc)}",
      journal = {Transient Name Server AstroNote},
         year = 2024,
        month = dec,
       volume = {371},
        pages = {1},
       adsurl = {https://ui.adsabs.harvard.edu/abs/2024TNSAN.371....1S}
}

@BOOK{Tully1987,
       author = {{Tully}, R. Brent and {Fisher}, J. Richard},
        title = "{Atlas of Nearby Galaxies}",
         year = 1987,
       adsurl = {https://ui.adsabs.harvard.edu/abs/1987ang..book.....T}, 
      publisher = {Cambridge University Press}
}

@ARTICLE{Tuntipong2024,
       author = {{Tuntipong}, Sujeeporn and {van de Sande}, Jesse and {Croom}, Scott M. and {Barsanti}, Stefania and {Bland-Hawthorn}, Joss and {Brough}, Sarah and {Bryant}, Julia J. and {Casura}, Sarah and {Fraser-McKelvie}, Amelia and {Lawrence}, Jon S. and {Ristea}, Andrei and {Sweet}, Sarah M. and {Zafar}, Tayyaba},
        title = "{The SAMI galaxy survey: on the importance of applying multiple selection criteria for finding Milky Way analogues}",
      journal = {\mnras},
         year = 2024,
        month = oct,
       volume = {533},
       number = {4},
        pages = {4334-4359},
          doi = {10.1093/mnras/stae2042},
archivePrefix = {arXiv},
       eprint = {2408.12223},
 primaryClass = {astro-ph.GA},
       adsurl = {https://ui.adsabs.harvard.edu/abs/2024MNRAS.533.4334T}
}

@article{Vasylenko2020,
author = {Vasylenko, A. A. and Vavilova, I. B. and Pulatova, N. G.},
year = {2020},
title = {Isolated AGNs NGC 5347, ESO 438-009, MCG-02-04-090, and J11366-6002: Swift and NuSTAR joined view},
journal = {Astron. Nachrichten},
volume = {341},
number = {8},
pages = {801--811},
doi = {10.1002/asna.202013783}
}

@INCOLLECTION{Vavilova2020,
       author = {{Vavilova}, Irina and {Pakuliak}, Ludmila and {Babyk}, Iurii and {Elyiv}, Andrii and {Dobrycheva}, Daria and {Melnyk}, Olga},
        title = "{Surveys, Catalogues, Databases, and Archives of Astronomical Data}",
    booktitle = {Knowledge Discovery in Big Data from Astronomy and Earth Observation},
         year = 2020,
       editor = {{{\v{S}}koda}, Petr and {Adam}, Fathalrahman},
        pages = {57-102},
        publisher = {Elsevier},
          doi = {10.1016/B978-0-12-819154-5.00015-1},
       adsurl = {https://ui.adsabs.harvard.edu/abs/2020kdbd.book...57V}
}

@article{Vavilova2021,
author = {Vavilova, I. B. and Dobrycheva, D. V. and Vasylenko, M. Y. and others},
year = {2021},
title = {Machine learning technique for morphological classification of galaxies from the SDSS. I. Photometry-based approach},
journal = {A\&A},
volume = {648},
pages = {A122},
doi = {10.1051/0004-6361/202038981}
}

@article{Vavilova2022,
author = {Vavilova, I. B. and Khramtsov, V. and Dobrycheva, D. V. and others},
year = {2022},
title = {Machine learning technique for morphological classification of galaxies from SDSS. II. The image-based morphological catalogs of galaxies at 0.02 < z < 0.1},
journal = {Space Sci. \& Technol.},
volume = {28},
number = {1},
pages = {3--22},
doi = {10.15407/knit2022.01.003}
}

@article{Vavilova2024,
author = {Vavilova, I. B. and Fedorov, P. N. and Dobrycheva, D. V. and others},
year = {2024},
title = {An advanced approach for definition of the Milky Way galaxies-analogues},
journal = {Space Sci. \& Technol.},
volume = {30},
number = {4},
pages = {81--90},
doi = {10.15407/knit2024.04.081}
}

@article{VilaVilaro2015,
author = {Vila-Vilaro, B. and Cepa, J. and Zabludoff, A.},
year = {2015},
title = {The Arizona Radio Observatory Survey of Molecular Gas in Nearby Normal Spiral Galaxies I: The Data},
journal = {ApJS},
volume = {218},
number = {2},
pages = {28},
doi = {10.1088/0067-0049/218/2/28}
}

@article{Volvach2011,
author = {Volvach, A. E. and Volvach, L. N. and Kutkin, A. M. and others},
year = {2011},
title = {Multi-frequency studies of the non-stationary radiation of the blazar 3C 454.3},
journal = {Astronomy Reports},
volume = {55},
number = {7},
pages = {608--615},
doi = {10.1134/S1063772911070092}
}

@article{Walter2008,
author = {Walter, F. and Brinks, E. and de Blok, W. J. G. and others},
year = {2008},
title = {THINGS: The H I Nearby Galaxy Survey},
journal = {AJ},
volume = {136},
number = {6},
pages = {2563--2647},
doi = {10.1088/0004-6256/136/6/2563}
}

@article{Warren2010,
author = {Warren, B. E. and Wilson, C. D. and Israel, F. P. and others},
year = {2010},
title = {The James Clerk Maxwell Telescope Nearby Galaxies Legacy Survey. II. Warm Molecular Gas and Star Formation in Three Field Spiral Galaxies},
journal = {ApJ},
volume = {714},
number = {1},
pages = {571--588},
doi = {10.1088/0004-637X/714/1/571}
}

@ARTICLE{Wilson2012,
       author = {{Wilson}, C.~D. and {Warren}, B.~E. and {Israel}, F.~P. and {Serjeant}, S. and {Attewell}, D. and {Bendo}, G.~J. and {Butner}, H.~M. and {Chanial}, P. and {Clements}, D.~L. and {Golding}, J. and {Heesen}, V. and {Irwin}, J. and {Leech}, J. and {Matthews}, H.~E. and {M{\"u}hle}, S. and {Mortier}, A.~M.~J. and {Petitpas}, G. and {S{\'a}nchez-Gallego}, J.~R. and {Sinukoff}, E. and {Shorten}, K. and {Tan}, B.~K. and {Tilanus}, R.~P.~J. and {Usero}, A. and {Vaccari}, M. and {Wiegert}, T. and {Zhu}, M. and {Alexander}, D.~M. and {Alexander}, P. and {Azimlu}, M. and {Barmby}, P. and {Brar}, R. and {Bridge}, C. and {Brinks}, E. and {Brooks}, S. and {Coppin}, K. and {C{\^o}t{\'e}}, S. and {C{\^o}t{\'e}}, P. and {Courteau}, S. and {Davies}, J. and {Eales}, S. and {Fich}, M. and {Hudson}, M. and {Hughes}, D.~H. and {Ivison}, R.~J. and {Knapen}, J.~H. and {Page}, M. and {Parkin}, T.~J. and {Rigopoulou}, D. and {Rosolowsky}, E. and {Seaquist}, E.~R. and {Spekkens}, K. and {Tanvir}, N. and {van der Hulst}, J.~M. and {van der Werf}, P. and {Vlahakis}, C. and {Webb}, T.~M. and {Weferling}, B. and {White}, G.~J.},
        title = "{The JCMT Nearby Galaxies Legacy Survey {\textemdash} VIII. CO data and the L$_{CO(3-2)}$-L$_{FIR}$ correlation in the SINGS sample}",
      journal = {\mnras},
         year = 2012,
        month = aug,
       volume = {424},
       number = {4},
        pages = {3050-3080},
          doi = {10.1111/j.1365-2966.2012.21453.x},
archivePrefix = {arXiv},
       eprint = {1206.1629},
 primaryClass = {astro-ph.CO},
       adsurl = {https://ui.adsabs.harvard.edu/abs/2012MNRAS.424.3050W}
}

@ARTICLE{Xi2026,
       author = {{Xi}, Qiang and {Sun}, Ning-Chen and {Aguado}, David and {P{\'e}rez-Fournon}, Ismael and {Poidevin}, Fr{\'e}d{\'e}rick and {Jin}, Junjie and {Mao}, Yiming and {Niu}, Zexi and {Wang}, Beichuan and {Zhang}, Yu and {Misra}, Kuntal and {Janghel}, Divyanshu and {Maund}, Justyn R. and {Kumar}, Amit and {Tinyanont}, Samaporn and {Liu}, Liang-Duan and {Zhang}, Yu-Hao and {Ailawadhi}, Bhavya and {Dubey}, Monalisa and {Guo}, Zhen and {Gupta}, Anshika and {He}, Min and {Jain}, Dhruv and {Kar}, Debalina and {Li}, Wenxiong and {Lyman}, Joe D. and {Mu}, Haiyang and {Pranshu}, Kumar and {Sun}, Xinxiang and {Wang}, Lingzhi and {Yadav}, Sarvesh Kumar and {Zhao}, Yi-Han and {Zheng}, Jie and {Zhu}, Yinan and {L{\'o}pez Fern{\'a}ndez-Nespral}, David and {L{\'o}pez Oramas}, Alicia and {Wang}, Yanan and {Wiersema}, Klaas and {Liu}, Jifeng},
        title = "{SN 2024aecx: A Fast-evolving Type IIb Supernova with a Prominent Shock-cooling Peak}",
      journal = {\apj},
         year = 2026,
        month = feb,
       volume = {998},
       number = {1},
          eid = {98},
        pages = {98},
          doi = {10.3847/1538-4357/ae2d06},
archivePrefix = {arXiv},
       eprint = {2509.12343},
 primaryClass = {astro-ph.SR},
       adsurl = {https://ui.adsabs.harvard.edu/abs/2026ApJ...998...98X}
}

@ARTICLE{YusefZadeh2025,
       author = {{Yusef-Zadeh}, F. and {Bushouse}, H. and {Arendt}, R.~G. and {Wardle}, M. and {Michail}, J.~M. and {Chandler}, C.~J.},
        title = "{Nonstop Variability of Sgr A* Using JWST at 2.1 and 4.8 {\ensuremath{\mu}}m Wavelengths: Evidence for Distinct Populations of Faint and Bright Variable Emission}",
      journal = {\apjl},
         year = 2025,
        month = feb,
       volume = {980},
       number = {2},
          eid = {L35},
        pages = {L35},
          doi = {10.3847/2041-8213/ada88b},
archivePrefix = {arXiv},
       eprint = {2501.04096},
 primaryClass = {astro-ph.GA},
       adsurl = {https://ui.adsabs.harvard.edu/abs/2025ApJ...980L..35Y}
}

@article{York2004,
  author  = {York, D. and Evensen, N. M. and Mart{\'i}nez, M. L. and Delgado, J. D.},
  title   = {Unified equations for the slope, intercept, and standard errors of the best straight line},
  journal = {Am. J. Phys.},
  volume  = {72},
  number  = {3},
  pages   = {367--375},
  year    = {2004},
  doi     = {10.1119/1.1632486}
}

@article{Zakharenko2016,
author = {Zakharenko, V. and Konovalenko, A. and Sidorchuk, M. and others},
title = {Digital Receivers for Low-Frequency Radio Telescopes UTR-2, URAN, GURT},
journal = {JAI},
volume = {5},
number = {4},
year = {2016},
doi = {10.1142/S2251171716410105}
}

@article{Zeilinger2001,
author = {Zeilinger, W. W. and Vega Beltr{\'a}n, J. C. and Rozas, M. and Beckman, J. E. and Knapen, J. H.},
year = {2001},
title = {NGC 3521: Stellar Counter-Rotation Induced by a Bar Component},
journal = {Ap\&SS},
volume = {276},
number = {2/4},
pages = {643--650},
doi = {10.1023/A:1017548101623}
}

@ARTICLE{Zhang2009,
       author = {{Zhang}, Wei Ming and {Soria}, Roberto and {Zhang}, Shuang Nan and et al.},
        title = "{A Census of X-ray Nuclear Activity in Nearby Galaxies}",
      journal = {ApJ},
         year = 2009,
        month = jul,
       volume = {699},
       number = {1},
        pages = {281-297},
          doi = {10.1088/0004-637X/699/1/281},
archivePrefix = {arXiv},
       eprint = {0904.1091},
 primaryClass = {astro-ph.HE},
       adsurl = {https://ui.adsabs.harvard.edu/abs/2009ApJ...699..281Z}
}

@ARTICLE{Zibetti2009,
       author = {{Zibetti}, Stefano and {Charlot}, St{\'e}phane and {Rix}, Hans-Walter},
        title = "{Resolved stellar mass maps of galaxies - I. Method and implications for global mass estimates}",
      journal = {\mnras},
         year = 2009,
        month = dec,
       volume = {400},
       number = {3},
        pages = {1181-1198},
          doi = {10.1111/j.1365-2966.2009.15528.x},
archivePrefix = {arXiv},
       eprint = {0904.4252},
 primaryClass = {astro-ph.CO},
       adsurl = {https://ui.adsabs.harvard.edu/abs/2009MNRAS.400.1181Z}
}

@ARTICLE{Zibetti2011,
       author = {{Zibetti}, Stefano and {Groves}, Brent},
        title = "{Resolved optical-infrared spectral energy distributions of galaxies: universal relations and their break-down on local scales}",
      journal = {MNRAS},
         year = 2011,
        month = oct,
       volume = {417},
       number = {2},
        pages = {812-834},
          doi = {10.1111/j.1365-2966.2011.19286.x},
archivePrefix = {arXiv},
       eprint = {1106.2165},
 primaryClass = {astro-ph.CO},
       adsurl = {https://ui.adsabs.harvard.edu/abs/2011MNRAS.417..812Z}
}

\begin{appendix}
\section{Technical specifications and low-frequency sensitivity measurements of NGC 3521 with the UTR-2 radio decameter telescope }
\label{sec:UTR2_tech_spec}

\subsection{Technical specifications}

UTR-2 (Ukrainian T-shaped Radio telescope, second model) is located 30 km away from Chuhuiv city, Kharkiv region, Ukraine. Its exploitation started in 1971, and for fifty years \citep{Konovalenko2021}, until the end of February 2022, this instrument worked continuously on many scientific programs\footnote{Until the UTR-2 was significantly damaged as a result of the occupation of the Chuhuiv district of the Kharkiv region during the Russian military aggression against Ukraine. After the end of hostilities, it is planned to significantly modernize the UTR-2 with the help of the European astronomical community \citep{Albergaria2025}}. 

UTR-2 consists of two mutually orthogonal multi-element phased antenna arrays, arranged in the shape of the letter "T" (a Mills cross-type configuration without one (fourth) arm). The larger array, North-South, with dimensions of 1800×60 m and consisting of 1440 elements (half-wave dipoles of the Nadenenko system), is located along the meridian and is responsible for declination resolution ($\sim12^\circ\times20'$ at 25 MHz). Its effective area is $\approx$100,000 m$^2$. The smaller array, West-East, with dimensions of 900×60 m and consisting of 600 elements, is located along the parallel and is responsible for right-ascension resolution ($\sim40'\times12^\circ$ at 25 MHz), with an effective area of $\approx$40,000 m$^2$. Observations could be performed separately on each antenna array, as well as in the mode of their signal multiplication. The latter allows the synthesis of a pencil beam with a size of $25'\times25'$ at 25 MHz \citep{Konovalenko2016}.

\subsection{Sensitivity measurements}

Decameter data for NGC 3521 were obtained during a highly sensitive sky survey in a wide frequency band pass relative to the declination DecJ = 0$^\circ$ (sessions from January 20 to 22 and from February 3 to 5, 2022). The UTR-2 radio telescope beam was fixedly oriented in relation to this declination (azimuth $A = 0^\circ$, elevation angle $\varepsilon \sim 40^\circ$). The daily rotation of Earth provided the complete coverage by scanning the celestial sphere areas within the right ascensions RA from 0$^h$ to 24$^h$, crossing the Galactic plane twice a day. Signal registration, pre-processing, and data recording were performed using DSP-Z digital multi-channel spectrum analyzers \citep{Zakharenko2016} over the frequency range 8.25-33 MHz, with a frequency resolution of 4 kHz and a time resolution of 0.1 s. The corresponding spectra, which illustrate in particular the radio frequency interference (RFI) situation, are shown in Fig.~\ref{fig:RFI_spectra}.

\begin{figure}
    \centering
    \includegraphics[width=\linewidth]{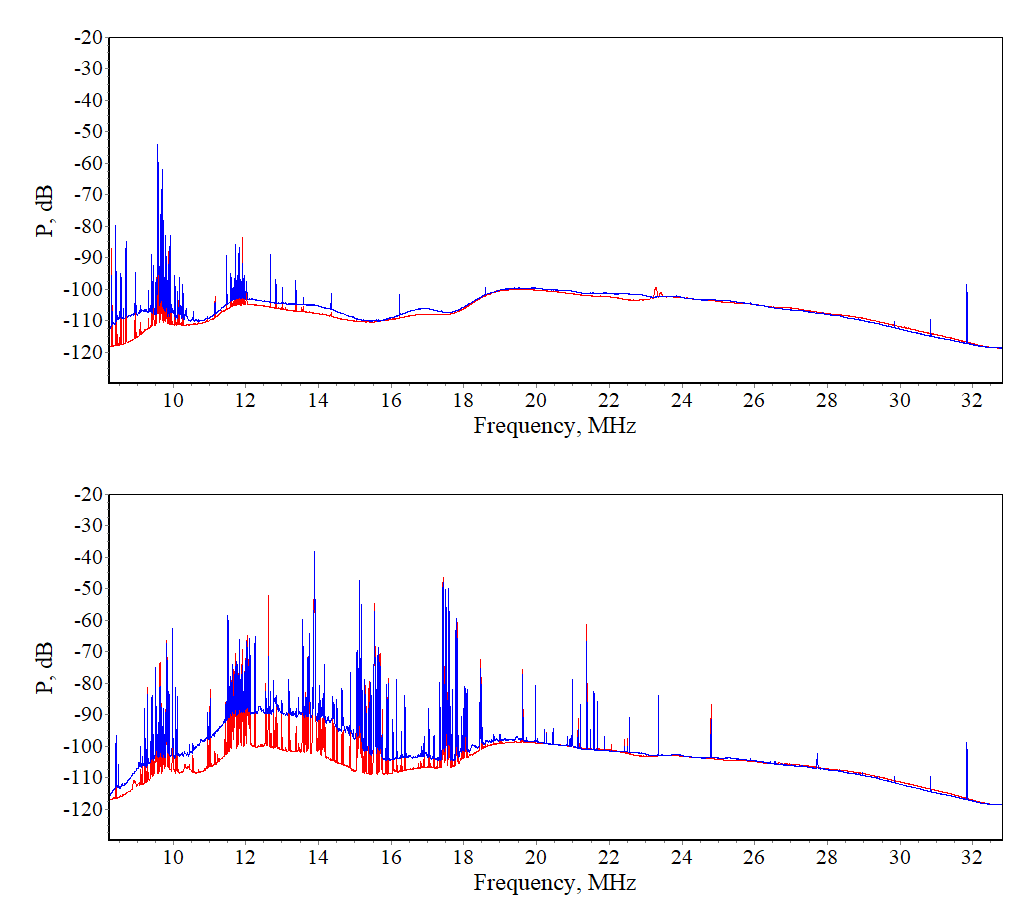}
    \caption{Spectra obtained during observations at UTR-2 radio telescope in the scanning mode relative to DecJ = 0º. Red line – signal from North – South antenna; Blue line – signal from West – East antenna.  Upper - Spectrum obtained on January 22, 2022 at 00:34:05 (UTC) (culmination time for NGC 3521); Lower - Spectrum obtained on January 22, 2022 at 11:00:00 (UTC).}
    \label{fig:RFI_spectra}
\end{figure}

Let’s estimate the expected flux density from the NGC 3521 galaxy in the decameter range. NGC 3521 is a spiral galaxy with an intermediate level of star formation. It is not a radio galaxy and reveals at decameter wavelengths due to its own synchrotron radio emission. The flux density from NGC 3521 at high frequencies (1.4 GHz) is about 374.77 mJy (Table \ref{tab:NGC3521_fluxes}). Converting this value to decameter-range frequencies (for example, 20 MHz ($S \propto \nu^{-\alpha}$, where $\nu$ is the frequency, $\alpha$ is the spectral index, which for spiral galaxies is 0.6–0.8), we obtain a value of $\approx$10 Jy. But it should be noted that the non-thermal Galactic background, against which the search for the signal from NGC 3521 may be performed, has brightness temperatures a thousand times higher at decameter wavelengths. In the case of NGC 3521, the detection of such a weak extragalactic signal from NGC 3521 is extremely challenging.

To estimate the sensitivity required for detecting a weak signal from an object such as NGC 3521 with the UTR-2 radio telescope, we consider that the system temperature in the decameter range is dominated by the Galactic background, whose brightness temperature is about 43,000 K at 20 MHz \citep{Cane1979}. The corresponding sensitivity, expressed in terms of the brightness temperature, can be estimated as $\Delta T = T_B/\sqrt{\Delta f  \Delta \tau}$, where $T_B$ is the brightness temperature of the Galactic background, $\Delta f$ is the receiver bandwidth, and $\Delta \tau$ is the integration time in seconds. To increase sensitivity, $\Delta f$ and $\Delta \tau$ should be as large as possible. Based on analysis of the spectra shown in Fig.~\ref{fig:RFI_spectra}, it was determined that the cleanest data, free from strong and low-intensity RFI, were obtained for frequencies above 25 MHz. To achieve a maximum $\Delta f$, its value can be chosen as 8 MHz for the 24--32 MHz band (with a central frequency of 28 MHz). 

To reduce the level of fluctuations, it is advisable to select an integration time $\Delta \tau$ corresponding to the beam size in the narrow cross-section along which the scan is performed---in this case, not exceeding 30 s (based on the beam size of about $40'$ by right ascension). Substituting $T_B = 43{,}000$ K for 20 MHz and $T_B = 20{,}000$ K for 28 MHz \citep{Cane1979, Sidorchuk2021}, $\Delta f = 8$ MHz, and $\Delta \tau = 30$ s into the previous equation, we obtain $\Delta T = 1.94$ K for 20 MHz and $\Delta T = 1.29$ K for the chosen central frequency of 28 MHz. Then, this value can be converted into sensitivity in terms of flux density as $\Delta S = 2k \Delta T/A_{\text{eff}}$, where $k = 1.38\times10^{-23} \mathrm{J K^{-1}}$ is the Boltzmann constant and $A_{\text{eff}}$ is the effective area of the antenna in square meters. For the UTR-2 West–East array at an elevation angle of $\varepsilon \approx 40^\circ$, the effective area is about 35,000 m$^2$ at 20 MHz and 25,000 m$^2$ at 28 MHz. Accordingly, the resulting sensitivity values are $\Delta S_{20} = 0.15$ Jy and $\Delta S_{28} = 0.14$ Jy. At a bandwidth of $\Delta f = 8$ MHz and an integration time of $\Delta \tau = 30$ s, these sensitivities are sufficient to detect the expected signal from NGC 3521, while averaging all available scans ($N$) would improve the signal-to-noise ratio by a factor of $\sqrt{N}$.

\begin{figure}
    \centering
    \includegraphics[width=\linewidth]{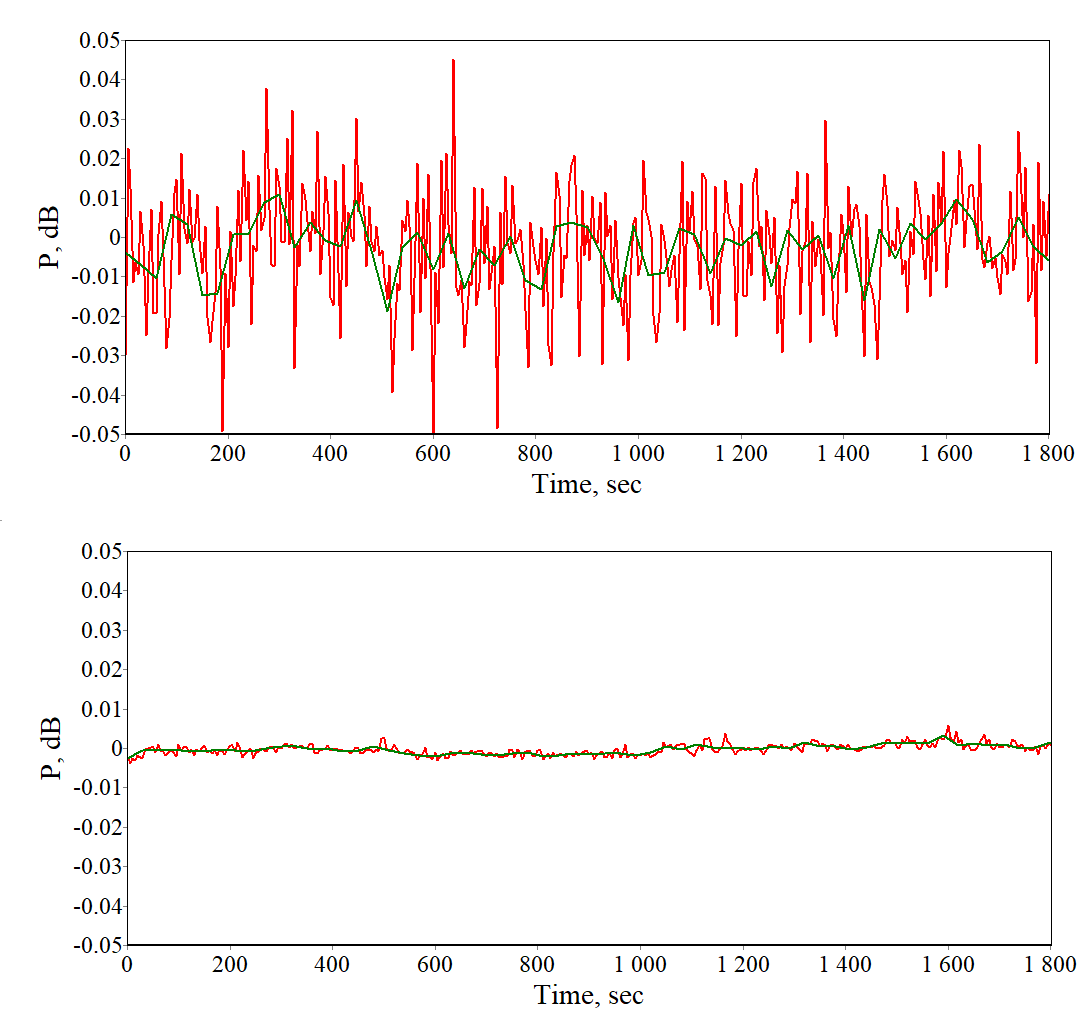}
    \caption{30-minute time sequence at 20 MHz built on the data of calibration signal recording from the reference noise generator at a level corresponding to the maximum power level of the Galactic background. Above: $\Delta f = 10$ kHz – the red line corresponds to $\Delta \tau = 5$ s, the green line corresponds to $\Delta \tau = 30$ s. Below: $\Delta f = 8$ MHz – the red line corresponds to $\Delta \tau = 5$ s, the green line corresponds to $\Delta \tau = 30$ s.}
    \label{fig:calibr_dcm}
\end{figure}

The calibration time sequences shown in Fig.~\ref{fig:calibr_dcm} illustrate the effect of increasing $\Delta f$ and $\Delta \tau$ on the noise level. For $\Delta f = 10$ kHz and $\Delta \tau = 5$ s the $\sigma = 1 / \sqrt{\Delta f  \Delta \tau}=4.5~\cdot~10^{-3}$. For $\Delta f = 10$ kHz and $\Delta \tau = 30$ s the $\sigma = 1.8~\cdot~10^{-3}$. For $\Delta f = 8$ MHz and $\Delta \tau = 5$ s the $\sigma = 1.6~\cdot~10^{-4}$. For $\Delta f = 8$ MHz and $\Delta \tau = 30$ s the $\sigma = 6.4~\cdot~10^{-5}$. This sensitivity, when expressed in dB, fully corresponds to the relative powers obtained during calibration with a noise generator (Fig.~\ref{fig:calibr_dcm}).

\subsection{Peculiarities of observations} 

Observations in the decameter range ($\sim 10$--$30 \mathrm{MHz}$) are limited by the bright and structured Galactic background that sets the system temperature, by strong and variable radio-frequency interference (RFI) from non-cosmic signals, and by ionospheric distortions. We used UTR-2 in drift-scan mode with the beam fixed at $\mathrm{Dec}_J = 0^\circ$ and exploited Earth's rotation to sweep RA, recording wide-band spectra with DSP-Z analyzers ($4 \mathrm{kHz}$ frequency and $0.1 \mathrm{s}$ time resolution). Sensitivity was improved by selecting the cleanest sub-band ($\gtrsim 25 \mathrm{MHz}$), adopting a large effective bandwidth ($\Delta f = 8 \mathrm{MHz}$) and an integration time matched to the narrow beam cross-section ($\Delta \tau \approx 30 \mathrm{s}$), which maximizes $\Delta f \Delta \tau$ and lowers the radiometer noise to $\Delta T \simeq T_B/\sqrt{\Delta f \Delta \tau} \sim 1$--$2 \mathrm{K}$.

Using the effective area of the West--East array for the observed elevation angle, this yields a flux-density sensitivity of $\Delta S \approx 2k\Delta T/A_{\rm eff} \approx 0.14$--$0.15 \mathrm{Jy}$, consistent with calibration sequences using a reference noise source and with SEFD estimates. Additional accuracy is achieved by averaging multiple daily scans ($\mathrm{S/N} \propto \sqrt{N}$), carefully excising RFI, and accounting for side-lobe contamination (e.g., Cassiopeia~A) that can bias the baseline. Planned hardware multiplication of the signals from the North--South and West--East arrays (synthesized ``pencil-beam'' mode) would further suppress confusion and enhance the detectability of faint extragalactic sources such as NGC 3521.

\section{Data processing for flux density values of NGC 3521 from UV to Radio cm ranges}
\label{APT}

\subsection{Aperture and PSF corrections}

For all ranges, we assumed a circular Gaussian point-spread function (PSF) with FWHM values corresponding to the nominal angular resolution of each instrument. To correct for the fraction of light falling outside the adopted elliptical aperture, we computed the encircled energy (EE) fraction as
\begin{equation}
{\rm EE}(r) = 1 - \exp\!\left[-\tfrac{1}{2}\left(\frac{r}{\sigma}\right)^{2}\right],
\qquad \sigma = \frac{{\rm FWHM}}{2.355},
\end{equation}
where r is the equivalent radius of the aperture, $r = \sqrt{ab}$.
The aperture–corrected flux is then
\begin{equation}
F_{\nu}^{\rm apcor} = \frac{F_{\nu}}{{\rm EE}(r)}.
\end{equation}
For the radio cm data, we adopted ${\rm EE}=1$, since the aperture fully encloses the source.

\subsection{Flux conversion and background subtraction}

Background levels were estimated locally using a sigma–clipped median algorithm on PSF-matched images, excluding the entire science aperture from the background mesh to avoid flux losses. The background-subtracted image is then defined as $I' = I - B$, where $I$ is the original image intensity and $B$ is the local background model. Flux densities were computed from the total background-subtracted signal within the aperture as $S = \sum_{\rm ap} I'$, where the summation extends over all unmasked pixels inside the aperture. The conversion from instrumental counts to physical flux densities done with the adopted zero-points\footnote{ For GALEX see - \cite{Morrissey2007}; for SDSS: \url{https://www.sdss4.org/dr17/algorithms/magnitudes/} and \cite{Liang2023}; for WISE: \url{https://irsa.ipac.caltech.edu/data/WISE/docs/release/All-Sky/expsup/sec4_4h.html}} and flux calibration factors defined for each filter (Section~\ref{sec:phot_system}). For the optical and ultraviolet ranges, this relation follows the standard AB magnitude convention, while for the far-infrared and radio cm images, the conversion depends on the maps' native radiometric units, as described below.

The AB mag and zero-points, in mJy, were taken into account in the calculations of optical and UV photometric parameters, as follows from eq. \ref{eq:phot}
\begin{equation}
\label{eq:phot}
m = -2.5\log_{10} S + {\rm ZP}, \qquad 
F_{\nu} [{\rm mJy}] = 10^{-0.4(m - 8.90)} \times 10^{3},
\end{equation}
where ${\rm ZP}$ is the adopted photometric zero-point, and the constant 8.90 encodes the AB zero-point for Jy.

Photometric data for the FIR and Radio cm ranges, in radiometric units, were calculated using eq. \ref{eq:beam},\ref{eq:radio_phot}. Let $\Omega_{\rm pix}$ is the pixel solid angle (sr), and $\Omega_{\rm beam}$ is the beam solid angle derived from the restoring beam parameters (${\rm BMAJ, BMIN}$):
\begin{equation}
\label{eq:beam}
\Omega_{\rm beam} = 2\pi\sigma_x\sigma_y, \qquad
\sigma = \frac{{\rm FWHM}}{2.355}.
\end{equation}
Then the flux density in mJy is given by the following equations:
\begin{align}
\label{eq:radio_phot}
{\rm MJy sr^{-1}}: \quad &
F_{\nu}[{\rm mJy}] = \left(\sum_{\rm ap} I'_{\rm MJy/sr}\right)\Omega_{\rm pix} 10^{9},\\
{\rm Jy pix^{-1}}: \quad & 
F_{\nu}[{\rm mJy}] = \left(\sum_{\rm ap} I'_{\rm Jy/pix}\right)10^{3},\\
{\rm Jy beam^{-1}}: \quad & 
{\rm pix/beam} = \frac{\Omega_{\rm beam}}{\Omega_{\rm pix}},
F_{\nu}[{\rm mJy}] = 
\frac{\sum_{\rm ap} I'_{\rm Jy/beam}}{\rm (pix/beam)} 10^{3}.
\end{align}

\subsection{Extinction correction}

The Milky Way extinction was corrected using $E(B{-}V)$ values obtained from the IRSA Dust maps via the astroquery.irsa interface 
\citep{Astroquery2019}, scaled by a factor of 0.86 to match the 
 recalibration \citep{Schlafly2011} of the dust map \citep{Schlegel1998}. So, with $A_V = 3.1 E(B{-}V)_{\rm eff}$ and the  reddening law \citep{Cardelli1989} applied for $\lambda \le 3.3 \mu$m, the extinction-corrected flux density is $A_{\lambda} = {\rm CCM89}(\lambda, A_V, R_V=3.1)$ and $F_{\nu}^{\rm corr} = F_{\nu}^{\rm apcor}\times 10^{0.4A_{\lambda}}$.

Extinction corrections were neglected for $\lambda > 3.3 \mu$m, as well as for all data in FIR and radio cm ranges.

\subsection{Uncertainty estimation}

The background noise per aperture was calculated as $\sigma_{\rm ap} = f_{\rm corr} \sigma_{\rm bkg} \sqrt{N_{\rm pix}}$, where $\sigma_{\rm bkg}$ is the standard deviation of the local background, $N_{\rm pix}$ is the number of unmasked pixels inside the aperture, and $f_{\rm corr}$ is a correlation factor accounting for PSF broadening and repixelization effects. For PACS, SPIRE, and VLA data, we used empirically calibrated $f_{\rm corr}$ values ranging from 1.1 to 1.8.

For fluxes derived from magnitudes, we used equation \ref{eq:unc_mag}:
\begin{equation}
\label{eq:unc_mag}
\delta m = \frac{2.5}{\ln 10} \frac{\sigma_{\rm ap}}{S}, \qquad
\frac{\delta F_{\nu}}{F_{\nu}} = \frac{\ln 10}{2.5} \delta m.
\end{equation}
The propagated uncertainties after aperture and extinction corrections calculated by eq. \ref{eq:unc_ext}
\begin{equation}
\label{eq:unc_ext}
\delta F_{\nu}^{\rm apcor} = 
\frac{\delta F_{\nu}}{{\rm EE}(r_{\rm eq})}, \qquad
\delta F_{\nu}^{\rm corr} = 
\delta F_{\nu}^{\rm apcor} 10^{0.4A_{\lambda}}.
\end{equation}
The results of aperture photometry from UV to radio cm ranges are given in Table \ref{tab:NGC3521_fluxes}.

\section{Baseline model construction for UV to radio cm spectral energy distribution}
\label{sec:sed_UV-VLA}

We adopted the baseline SED model developed in our previous preliminary UV--to--radio (cm) fitting \cite{Pastoven2024}. Here we apply the same model to the fluxes derived from aperture photometry. The baseline model and the corresponding grid parameters are summarized in Table~\ref{tab:cigale_params}.

The total stellar luminosity is
$L_{\star} = (7.53 \pm 0.38)\times10^{10} L_{\odot}$,
and for the evolved population of stars
$L_{\star,\mathrm{old}} = (6.28 \pm 0.31)\times10^{10} L_{\odot}$.
The integrated dust luminosity reaches
$L_{\mathrm{dust}} = (2.78 \pm 0.14)\times10^{10} L_{\odot}$,
and the corresponding dust mass is
$M_{\mathrm{dust}} = (9.8 \pm 2.1)\times10^{7} M_{\odot}$,
which unveils a massive dust reservoir that reradiate much of the stellar emission at infrared wavelengths. The sum of the stellar mass obtained from a Bayesian analysis is 2.1 ± 0.8.
$M_{\star} = (5.73 \pm 0.29)\times10^{10} M_{\odot}$,
old stellar component dominated
($M_{\star,\mathrm{old}} = 5.73 \pm 0.29\times10^{10} M_{\odot}$),
This indicates that the stellar population of NGC 3521 is dominated by an evolved disk population. The mass-weighted stellar age is
$t_{\star,\mathrm{mw}} = 5.00 \pm 0.01~\mathrm{Gyr}$,
signifying a more evolved stellar component, as that would be characteristic for metal-rich spirals in the local Universe. The corresponding star-formation rate from the delayed-$\tau$ star-formation history is
$\mathrm{SFR} = 1.93 \pm 0.10 M_{\odot} \mathrm{yr^{-1}}$,
which is in agreement with a main-sequence activity of massive late-type spiral galaxies. The attenuation parameters of the SED fitting are $A_{B,90} = 0.845 \pm 0.016~\mathrm{mag}$,
$A_{V,90} = 0.621 \pm 0.017~\mathrm{mag}$,
and 
$A_{\mathrm{FUV}} = 2.62 \pm 0.03~\mathrm{mag}$,
which are typical of galaxies with intermediate inclination. Above all, these results imply that NGC 3521 hosts a huge stellar disk with an enhanced dust component, where the current star formation is distinctly calm. The system appears to be close to energy equilibrium between stellar and dust radiation.

For the determination of whether our physical parameters are robust, we examined the distributions of reduced $\chi^{2}$s saved for each physical parameter. All parameters investigated exhibit a clear minimum at a single solution concentration; therefore, the best-fit model can be considered statistically unique within the covered grid. No secondary minima or extended plateaus are observed, indicating that the degeneracies among parameters (stellar and dust masses, SFR) are small. While the wider distributions in $\mathrm{SFR}$ and $L_{\star,\mathrm{old}}$ can be attributed to the correlation between recent SFR activity and dust attenuation among spiral galaxies. These tests show that the selected \textsc{CIGALE} configuration has adequate grid resolution and that the inferred parameters are well-constrained and physically motivated.

\begin{figure}
    \centering
    \includegraphics[width=0.75\linewidth]{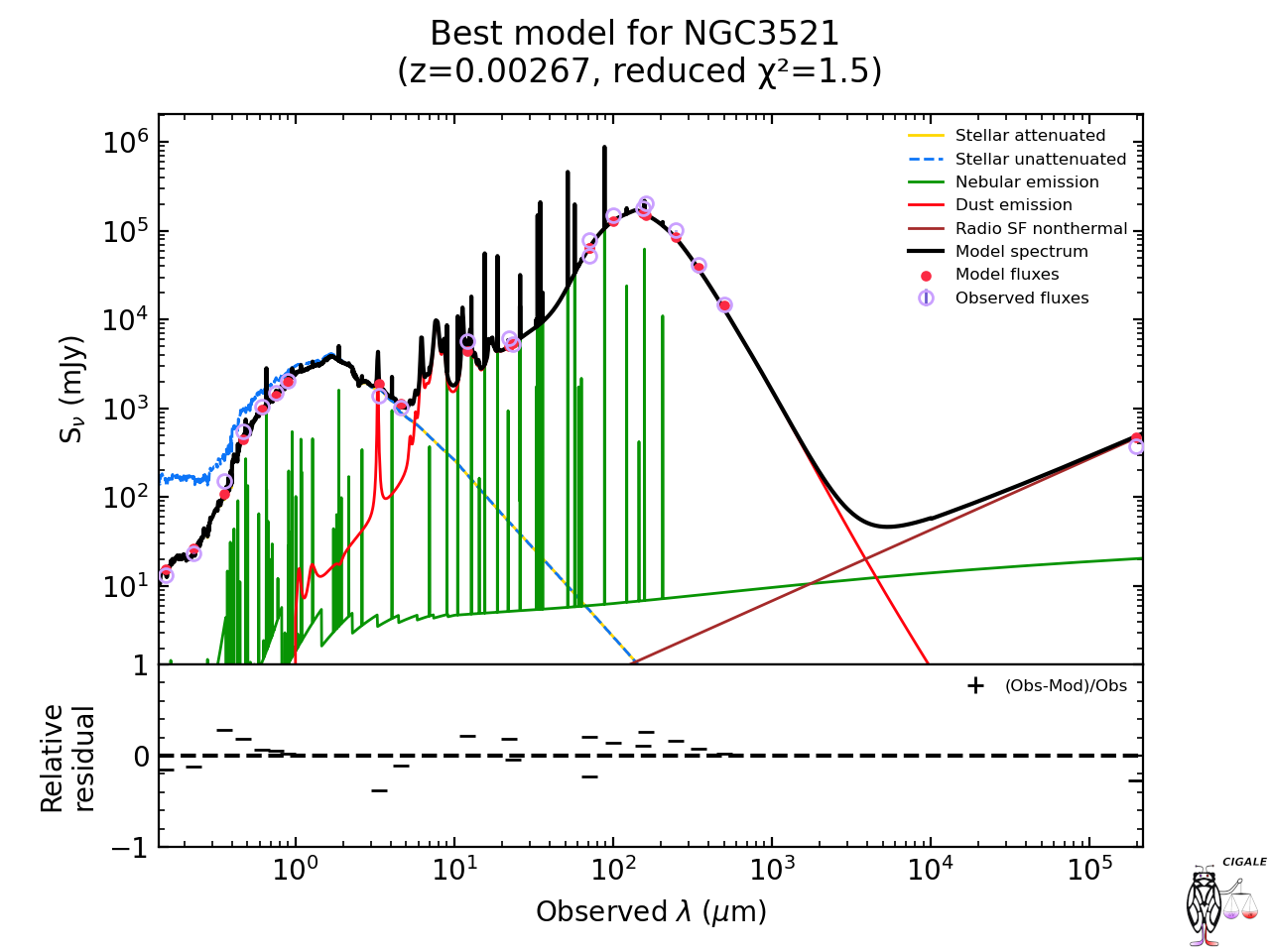}
    \caption{NGC 3521 SED from UV to radio cm ranges}
    \label{fig:UV_to_VLA}
\end{figure}

\begin{table}
\centering
\caption{Parameter grid adopted for the SED modeling.}
\label{tab:cigale_params}
\begin{tabular}{ll}
\hline
\multicolumn{2}{l}{Star formation history: sfhdelayedbq} \\
\hline
$\tau_{\mathrm{main}}$ [Myr] & 1000, 2000, 4000, 6000 \\
$\mathrm{age_{main}}$ [Myr] & 7000, 8000 \\
$\mathrm{age_{bq}}$ [Myr] & 75, 100, 150, 200 \\
$r_{\mathrm{SFR}}$ & 1.0, 1.1, 1.5, 2.5 \\
\hline
\multicolumn{2}{l}{Stellar population synthesis: bc03} \\
\hline
Initial Mass Function & Chabrier (1) \\
Metallicity $Z$ & 0.02, 0.05 \\
\hline
\multicolumn{2}{l}{Nebular emission: nebular} \\
\hline
$\log U$ & $-3.0$, $-2.5$ \\
Gas metallicity $Z_{\mathrm{gas}}$ & 0.02, 0.03 \\
\hline
\multicolumn{2}{l}{Dust attenuation: dustatt\_modified\_starburst} \\
\hline
$E(B{-}V)_{\mathrm{lines}}$ & 0.5 \\
$E(B{-}V)_{\mathrm{factor}}$ & 0.44 \\
UV bump amplitude & 3.0 \\
UV bump width [nm] & 35.0 \\
UV bump wavelength [nm] & 217.5 \\
Power-law slope $\delta$ & $-0.4$, $-0.2$ \\
\hline
\multicolumn{2}{l}{Dust emission: dl2014}\\
\hline
$q_{\mathrm{PAH}}$ [\%] & 3.19, 3.9, 4.58 \\
$U_{\min}$ & 0.5, 1, 2, 3, 5, 10, 20 \\
$\alpha$ & 2.0, 2.1, 2.2, 2.4 \\
$\gamma$ & 0.01, 0.05, 0.1 \\
\hline
\multicolumn{2}{l}{Radio emission: radio} \\
$q_{\mathrm{IR, SF}}$ & 2.58 \\
$\alpha_{\mathrm{SF}}$ & 0.8 \\
\hline
\end{tabular}
\end{table}

\section{Emission from the central region: ZTF and NEOWISE data preparation }

\subsection{Optical data with ZTF}
\label{sec:ZTF_optical}

ZTF provides pipeline PSF-fit photometry with a typical seeing of 2.0--2.2\arcsec\ (FWHM) and a plate scale of $\sim 1.01\arcsec$/pix \citep{Masci2019ZTF}.  

To mitigate epoch-to-epoch PSF/seeing systematics, we use the ZTF pipeline PSF-fit magnitudes and attach the per-epoch seeing (FWHM) from the exposure-level metadata. Because NGC~3521 is spatially extended, seeing variations can modulate the host-galaxy contribution even for PSF-fit measurements. We therefore apply a seeing-quality cut (FWHM $\leq 3.00\arcsec$), rejecting 123/633 measurements (19.4\%) and retaining 510 epochs (median FWHM $2.26\arcsec$, 95th percentile $2.88\arcsec$). After this cut, we apply standard QC (maximum magnitude uncertainty and positional offset) and robust MAD-based global clipping in each band ($K_{\rm global}=2.5 \sigma_{\rm MAD}$), yielding 454 clean measurements.
To verify that the inferred color--magnitude relations are not driven by seeing, we (i) repeat the full pipeline using stricter cuts (FWHM $\leq 2.70\arcsec$ and $\leq 2.50\arcsec$) and (ii) perform a residual-magnitude test in which a linear magnitude--seeing term is removed in each band before repeating the nightly color--magnitude analysis. All slopes remain consistent within uncertainties (Table~\ref{tab:ztf_cmd_seeing_robust}). Since the results are consistent within uncertainties across the tested thresholds, we adopt FWHM $\leq 3.00\arcsec$ as our fiducial cut to maximize the number of paired measurements and minimize the regression uncertainties.

The effective wavelengths and transmission curves were obtained from the SVO Filter Profile Service\footnote{\url{https://svo2.cab.inta-csic.es/theory/fps/index.php?id=Palomar/ZTF.i_fil&&mode=browse&gname=Palomar&gname2=ZTF}}:
$\lambda_{\rm eff}(g)=4746.48$ \AA, $\lambda_{\rm eff}(r)=6366.38$ \AA, $\lambda_{\rm eff}(i)=7829.03$ \AA. 

We adopted the standard power-law convention $F_{\nu}\propto\nu^{-\alpha}$ \citep{RybickiLightman1979} consistently across the optical and MIR bands. For any two bands $X$ and $Y$ (AB magnitudes $m_X,m_Y$; effective wavelengths $\lambda_X,\lambda_Y$), the color $(X-Y)\equiv m_X-m_Y$ is related to the spectral index by
\begin{equation}
\alpha = \frac{(X-Y)}{K_{XY}}, \qquad
K_{XY} = 2.5\log_{10}\!\left(\frac{\nu_X}{\nu_Y}\right)
       = 2.5\log_{10}\!\left(\frac{\lambda_Y}{\lambda_X}\right),
\label{eq:alpha_color_relation}
\end{equation}
where $\nu=c/\lambda$. Using the ZTF effective wavelengths we obtain color--to--index factors
\begin{equation}
K_{gr}=2.5\log_{10}\!\left(\frac{\lambda_r}{\lambda_g}\right)=0.3188,
\qquad
K_{ri}=2.5\log_{10}\!\left(\frac{\lambda_i}{\lambda_r}\right)=0.2245.
\label{eq:K_values_gr_ri}
\end{equation}

\subsection{Infrared data from NEOWISE}
\label{sec:NEOWISE_data}
Infrared photometry in the W1 (3.35 $\mu$m) and W2 (4.60 $\mu$m) bands was obtained from the single-exposure NEOWISE Source Table (neowiser\_p1bs\_psd) hosted by the NASA/IPAC Infrared Science Archive (IRSA)\footnote{\url{https://irsa.ipac.caltech.edu}} \citep{Mainzer2014NEOWISE}. These wavelengths correspond to the effective filter response of the WISE detectors as defined in the WISE All-Sky Release Explanatory Supplement \citep{Cutri2012WISE}.

The NEOWISE measurements used here are the pipeline profile-fit magnitudes (w1mpro, w2mpro) from individual exposures (native Vega system). Given the WISE PSF (FWHM $\approx 6\arcsec$), each measurement inevitably combines the nuclear emission with a significant fraction of the inner stellar/dust disc \citep{Cutri2012WISE}. The profile-fit photometry is obtained by PSF fitting within a circular region of radius $7.5\arcsec$ for non-saturated W1/W2 detections ($1.25\times{\rm FWHM}$), while the fitting radius is enlarged when the source core is saturated \citep[Sec.~IV.4.c.iii]{Cutri2012WISE}. For our target, the single-exposure catalog diagnostics confirm the non-saturated regime for all queried measurements (w1fitr=w2fitr=$7.5\arcsec$ and w1sat=w2sat=0 throughout) \citep{Cutri2015NEOWISE}. We queried IRSA with a cone search of radius $5\arcsec$ around the galaxy center and applied standard quality filtering (SNR thresholds, cc\_flags, qual\_frame) followed by robust global clipping in each band and epoch-based averaging.

For the spectral-index analysis, we convert W1 and W2 to AB using the standard offsets from the WISE All-Sky Release Explanatory Supplement
(Section~IV.4.h.i, Table~3; \citealt{Cutri2012WISE})\footnote{\url{https://irsa.ipac.caltech.edu/data/WISE/docs/release/All-Sky/expsup/sec4_4h.html}}:
\begin{equation}
{\rm W1}_{\rm AB} = {\rm W1}_{\rm Vega} + 2.699, \qquad
{\rm W2}_{\rm AB} = {\rm W2}_{\rm Vega} + 3.339,
\end{equation}
so that $({\rm W1}-{\rm W2})_{\rm AB}=({\rm W1}-{\rm W2})_{\rm Vega}-0.64$. Adopting $F_{\nu}\propto\nu^{-\alpha}$, we define
\begin{equation}
K_{\rm W1W2}=2.5 \log_{10}\!\left(\frac{\nu_{\rm W1}}{\nu_{\rm W2}}\right)=0.3441
\end{equation}

with the effective wavelengths from the SVO Filter Profile Service\footnote{\url{https://svo2.cab.inta-csic.es/theory/fps/index.php?mode=browse&gname=WISE&asttype=}}
 $\lambda_{\rm eff}(W1)=33526.00$ \AA, $\lambda_{\rm eff}(W2)=46028.00$ \AA, and compute $\alpha_{\rm MIR}=({\rm W1}-{\rm W2})_{\rm AB}/K_{\rm W1W2}$ and ${d\alpha}/{d(-{\rm W1})}=-m_{\rm color}/K_{\rm W1W2}$.
\begin{table*}
\centering
\caption{Robustness of ZTF nightly color--magnitude relations to the seeing-quality cut and to the residual-magnitude test.}
\label{tab:ztf_cmd_seeing_robust}
\small
\setlength{\tabcolsep}{6pt}
\begin{tabular}{l l c c c c c}
\hline
Seeing cut & Variant & $N_{\rm pairs}$ &
$m$ &
$\alpha_{\rm median}$ &
${\rm d}\alpha/{\rm d}(-X)$ &
r \\
\hline
\multicolumn{7}{c}{g-r vs g (nightly)}\\
\hline
$\le 3.00''$ & RAW   & 81 & 0.89$\pm$0.11 & 2.28$\pm$0.02 & $-2.78\pm 0.34$ & 0.69 \\
$\le 3.00''$ & RESID & 81 & 0.83$\pm$0.11 & 2.30$\pm$0.03 & $-2.59\pm 0.35$ & 0.65 \\
$\le 2.70''$ & RAW   & 63 & 0.85$\pm$0.12 & 2.28$\pm$0.03 & $-2.66\pm 0.37$ & 0.68 \\
$\le 2.70''$ & RESID & 63 & 0.83$\pm$0.12 & 2.28$\pm$0.03 & $-2.59\pm 0.38$ & 0.66 \\
$\le 2.50''$ & RAW   & 44 & 0.83$\pm$0.14 & 2.31$\pm$0.04 & $-2.59\pm 0.44$ & 0.65 \\
$\le 2.50''$ & RESID & 44 & 0.82$\pm$0.15 & 2.32$\pm$0.04 & $-2.56\pm 0.46$ & 0.64 \\
\hline
\multicolumn{7}{c}{r-i vs r (nightly)}\\
\hline
$\le 3.00''$ & RAW   & 13 & 0.66$\pm$0.24 & 1.82$\pm$0.06 & $-2.93\pm 1.06$ & 0.58 \\
$\le 3.00''$ & RESID & 13 & 0.64$\pm$0.27 & 1.84$\pm$0.07 & $-2.85\pm 1.18$ & 0.55 \\
$\le 2.70''$ & RAW   & 10 & 0.61$\pm$0.31 & 1.85$\pm$0.09 & $-2.74\pm 1.37$ & 0.53 \\
$\le 2.70''$ & RESID & 10 & 0.64$\pm$0.27 & 1.85$\pm$0.10 & $-2.85\pm 1.18$ & 0.56 \\
$\le 2.50''$ & RAW   &  8 & 0.46$\pm$0.15 & 1.85$\pm$0.07 & $-2.04\pm 0.68$ & 0.65 \\
$\le 2.50''$ & RESID &  8 & 0.54$\pm$0.19 & 1.84$\pm$0.10 & $-2.40\pm 0.86$ & 0.68 \\
\hline
\end{tabular}
\tablefoot{ Seeing cut --- maximum allowed PSF FWHM in arcseconds; Variant---analysis type (RAW: standard pipeline magnitudes after seeing+QC+MAD cleaning; RESID: residual-magnitude test where, in each band, a linear magnitude--seeing term is removed and the analysis is repeated on the residuals); $N_{\rm pairs}$---number of paired nightly colour measurements used in the regression;$m$---York slope; $\alpha_{\rm median}$ --- median spectral index; ${\rm d}\alpha/{\rm d}(-X)$--- spectral slope with brightness;  r --- Pearson coefficient.}
\end{table*}

\end{appendix}
\end{document}